\def\huawei{0}

\if\huawei1
\documentclass[acmsmall,authordraft,nonacm]{acmart}
\else
\documentclass[acmsmall]{acmart}
\fi

\usepackage{algorithm}
\usepackage{algpseudocode}

\usepackage{xspace}
\usepackage{listings}

\usepackage{multirow}

\usepackage{array}
\usepackage{lipsum}

\newcommand{\APPR}{ATUA\xspace}
\newcommand{\AutAut}{\APPR}

\newcommand{\appdiff}{AppDiff\xspace}

\newcommand{\RF}{$\mathcal{L}$\xspace}

\newcommand{\R}[1]{\underline{#1}}
\newcommand{\FIXME}[1]{\textcolor{black}{#1}}

\newcommand{\CHANGED}[1]{\textcolor{black}{#1}}
\definecolor{byzantine}{rgb}{0.74, 0.2, 0.64}
\newcommand{\CHANGEDNEW}[1]{\textcolor{black}{#1}}
\newcommand{\CHANGEDOCT}[1]{\textcolor{black}{#1}}
\newcommand{\CHANGEDNOV}[1]{\textcolor{black}{#1}}

\newcommand{\DELETE}[1]{}

\usepackage{marginnote}
\usepackage[backgroundcolor=white,bordercolor=blue,linecolor=blue]{todonotes}

\newcommand{\JMR}[2]{\textcolor{black}{#2}}
\newcommand{\JMRCHANGE}[1]{\textcolor{black}{#1}}

\newcommand{\JMRTWO}[2]{#2}

\newcommand{\CHANGETWO}[1]{#1}

\AtBeginDocument{%
  \providecommand\BibTeX{{%
    \normalfont B\kern-0.5em{\scshape i\kern-0.25em b}\kern-0.8em\TeX}}}

\setcopyright{acmcopyright}
\acmJournal{TOSEM}
\acmYear{2021} \acmVolume{1} \acmNumber{1} \acmArticle{1} \acmMonth{1} \acmPrice{15.00}\acmDOI{10.1145/3502297}

\lstdefinelanguage{json}{
    basicstyle=\scriptsize\ttfamily,
    numbers=left,
    numberstyle=\scriptsize,
    stepnumber=1,
    numbersep=8pt,
    showstringspaces=false,
    breaklines=true,
    frame=lines,
    backgroundcolor=\color{white},
    literate=
     *{0}{{{\color{black}0}}}{1}
      {1}{{{\color{black}1}}}{1}
      {2}{{{\color{black}2}}}{1}
      {3}{{{\color{black}3}}}{1}
      {4}{{{\color{black}4}}}{1}
      {5}{{{\color{black}5}}}{1}
      {6}{{{\color{black}6}}}{1}
      {7}{{{\color{black}7}}}{1}
      {8}{{{\color{black}8}}}{1}
      {9}{{{\color{black}9}}}{1}
      {:}{{{\color{darkgray}{:}}}}{1}
      {,}{{{\color{darkgray}{,}}}}{1}
      {\{}{{{\color{darkgray}{\{}}}}{1}
      {\}}{{{\color{darkgray}{\}}}}}{1}
      {[}{{{\color{darkgray}{[}}}}{1}
      {]}{{{\color{darkgray}{]}}}}{1}
      {/}{{{\color{darkgray}{/}}}}{1},
}

\begin{document}

\title{Automated, Cost-effective, and Update-driven App Testing}

\if\huawei1
\subtitle{Working draft about Huawei/SnT partnership project results on \emph{Automated Testing of Software Upgrades for Android Systems}.}
\fi

\author{Chanh Duc Ngo}
\affiliation{%
  \institution{SnT Centre, University of Luxembourg}
  \streetaddress{JFK 29}
  \city{Luxembourg}
  \country{Luxembourg}}
\email{chanh-duc.ngo@uni.lu}

\author{Fabrizio Pastore}
\affiliation{%
  \institution{SnT Centre, University of Luxembourg}
  \streetaddress{JFK 29}
  \city{Luxembourg}
  \country{Luxembourg}}
\email{fabrizio.pastore@uni.lu}

\author{Lionel Briand}
\affiliation{%
  \institution{SnT Centre, University of Luxembourg}
  \streetaddress{JFK 29}
  \city{Luxembourg}
  \country{Luxembourg}}
  \affiliation{%
  \institution{School of EECS, University of Ottawa}
  \city{Ottawa}
  \country{Canada}}
\email{lionel.briand@uni.lu}

\begin{abstract}

Apps' pervasive role in our society led to the definition of test automation approaches to ensure their dependability. However, state-of-the-art approaches tend to generate large numbers of test \CHANGED{inputs} and are unlikely to achieve more than 50\% method coverage.

In this paper, 
we propose a strategy to achieve significantly higher coverage of the code affected by updates with a much smaller number of test \CHANGED{inputs}, thus alleviating the test oracle problem.
 
More specifically, we present \APPR, a model-based approach that synthesizes App models with static analysis, integrates a dyna\-mically-refined state abstraction function and combines  complementary testing strategies, including (1) coverage of the model structure, (2) coverage of the App code, (3) random exploration, and (4) coverage of dependencies identified through information retrieval. Its model-based strategy enables \APPR to generate a small set of inputs that exercise only the code affected by the updates. In turn, this makes common test oracle solutions more cost-effective as they tend to involve human effort.
 
A large empirical evaluation, conducted with \FIXME{72} App versions belonging to nine popular Android Apps, has shown that \APPR is more effective and less effort intensive than state-of-the-art approaches when testing App updates.

\end{abstract}



\maketitle

\section{Introduction}

The business-critical role played by \CHANGED{software applications for mobile devices (Apps)} in our society~\cite{Deloitte:US:2018} has led to the development of dedicated techniques for their automated testing~\cite{Linares:ICSME:2017}. 
Since most of the code in an App concerns the handling of input values and events, test automation approaches automatically generate sequences of events and input values (hereafter, input sequences) that simulate the use of the App under test in its deployed environment. 
These approaches mainly differ with respect to the strategy used to create input sequences, such as random, evolutionary, and model-based approaches relying either on static or dynamic information~\cite{Linares:ICSME:2017}.

Unfortunately, state-of-the-art automated App testing techniques show limited code coverage capabilities, thus indicating they are unlikely to exercise all the features of the App under test. For example, they typically exercise about half of the methods implemented by commercial apps~\cite{Wang:EmpStudy:2018}. 
\CHANGED{As a result, all methods and instructions that are not automatically tested should be exercised by manually implemented test cases, an expensive task that may delay the App release.}
Also, though existing techniques show a degree of  complementarity~\cite{Wang:EmpStudy:2018}, state-of-the-art approaches do not attempt to integrate them to achieve better results. 

Existing testing approaches do not account for the high release frequency of a typical App's lifecycle, which are usually driven by marketing strategies aiming at increasing visibility~\cite{Mcilroy:FrequentlyUpdatedApps:ESE:2016,Calciati:NewAppReseases:MSR:2018,Alvarez:WAMA:2019}. 
\CHANGED{As a result, existing work does not include effective means of prioritizing the testing of modified or newly introduced features and are thus not addressing one of the major needs of App developers. However, this is an important requirement for any testing strategy as exercising all the features of an App in each release is enormously wasteful.}
Existing work on testing App upgrades is limited to the selection of subsets of events that may trigger modified code~\cite{Sharma:QADroid:ISSTA:2019} or the selection of regression test cases~\cite{Choi:DetReduce:ICSE:2018}. This is, however, not adequate when, to start with, available test cases do not exercise all the new and modified features of the software.

\CHANGED{Finally, the current body of work does not address the \emph{oracle problem}~\cite{Barr2015,Linares:ICSME:2017,Rubinov:AndroidAppTestingSurvey:2018}. 
More precisely, testing techniques cannot discover functional failures beyond crashes and the manual verification of the App outputs is difficult due the large number of inputs they exercise~\cite{Choi:DetReduce:ICSE:2018}.
However, in the context of frequent App updates, with a test input generation strategy that effectively exercises updated features, it is conceivable to address the oracle problem 
by relying on dedicated strategies to minimize test inputs. 
Failures affecting unchanged features (i.e., regressions) can be automatically detected by 
comparing the output of different App versions for a same input~\cite{Bogdan:1998,Shamshiri:2013,Pastore:2014,Gao:GUIRegressionTesting:ASE:2015}, whereas  
the output of new and modified features can instead be verified, at reduced costs, by relying on internal or external crowdsourcing~\cite{Pastore:CrowdOracles}. Nevertheless, such solutions are only practical if the number of test inputs is kept down to a reasonable number.}

\CHANGED{Keeping the number of test inputs to the strict minimum is important to minimize human intervention since it may be required when executing the same inputs on different software versions, e.g., to adapt input sequences to changes in the GUI~\cite{Li:Atom:2017,Pan:GUIrepair:2019}. 
Further, screenshots of the results must be visualized after every input.
Unfortunately, state-of-the-art App testing approaches generate large test suites, while test suite reduction approaches require to perform runtime monitoring of the App, which slows down execution and diminishes test automation effectiveness ~\cite{Choi:DetReduce:ICSE:2018}.}

\CHANGEDNEW{In summary, to address the limitations above, we aim to achieve the following two objectives:
(O1) maximize the number of updated methods and their instructions that are automatically exercised within practical test execution time, and 
(O2) generate a significantly reduced set of inputs, compared to state-of-art approaches, thus decreasing human effort.}

\CHANGEDNEW{To achieve the two objectives above, it is necessary to integrate multiple analysis strategies. Objective O1 can be effectively  achieved by means of static analysis, to determine updated features (e.g., through the identification of updated methods\cite{Sharma:QADroid:ISSTA:2019}) and identify the inputs that may trigger a specific feature (e.g., the input that leads to a particular Window)~\cite{yang-jase18}. 
Unfortunately, static analysis alone may not enable the effective testing of Apps; indeed, they typically rely on APIs dedicated to input handling that are hardly processed by static analysis tools, as discussed in related work  ~\cite{Rubinov:AndroidAppTestingSurvey:2018}. Random exploration is thus required to discover, at runtime, inputs that may trigger a potentially large subset of modified methods.
Unfortunately, random exploration might be particularly inefficient and conflict with objective O2 (e.g., it may require thousands of inputs to exercise features that depend on specific App states). For this reason it is necessary to determine which inputs bring the App into distinct program states by relying on dynamically-refined state abstraction functions~\cite{Gu:APE:ICSE:2019} and by identifying dependencies among App features (e.g., to determine that an option in the settings page enables a specific feature).}

In this paper we present \emph{\APPR\footnote{Atua is also the name of spirits in Polynesia, https://maoridictionary.co.nz/word/494} (Automated Testing of Updates for Apps)}, the first approach that integrates multiple test strategies to efficiently use the test budget 
\JMR{2.1}{and achieve the two above-mentioned objectives.
The rationale followed by \APPR is that Apps can be cost-effectively tested by combining static and dynamic program analysis
to select the inputs that exercise updated methods, our test targets.
Also, given the complexity of Apps, testing should be performed incrementally, by focusing first on objectives that are easier to achieve. For this reason, \APPR works in three phases: (1) it exercises all the features that may trigger modified methods (e.g., submitting a registration form that is processed by an updated method), (2) to maximize coverage in the presence of data-dependencies, it exercises updated features with diverse input values (e.g., a diverse set of values in a form), (3) to maximize coverage in the presence of state-dependencies, it exercises related features (e.g., submit a registration form after changing language settings).}
\APPR implements a model-based approach that 
integrates a \emph{dynamically-refined state abstraction function} and
complementary testing strategies, including (1) coverage of the \emph{model structure}, (2) coverage of the \emph{App code}, (3) \emph{random} exploration, and (4) coverage of \emph{dependencies} among App windows.

\APPR generates models 
of the App under test by combining static and dynamic program analysis.
It extends static program analysis approaches~\cite{yang-jase18} to automatically generate extended window transition graphs (EWTG), i.e., finite state machines that capture which inputs trigger window transitions and updated methods.
Also, it introduces a state abstraction function that 
refines the states of the EWTGs to
capture differences in the user interface that are not detected by means of static analysis (e.g., the presence of dynamically disabled buttons).
The state abstraction function is automatically refined to eliminate or, when not possible, reduce non-determinism while minimizing the number of abstract states.

To automatically exercise Apps, \APPR relies on the generated EWTGs to identify the sequences of inputs that trigger the execution of updated methods. When there are discrepancies between the EWTGs and the observed behavior, random exploration is used to refine the former. 
Code coverage is used to identify the methods that require additional testing effort. Finally, using information retrieval techniques~\cite{Toffola:MiningInputs:ASE:2017}, \APPR identifies dependencies between App windows that may prevent the execution of certain methods.

We assume that, for every software version, engineers are interested in testing the updated methods only. 
However, the general principles behind \APPR can easily be adopted also for other ways of characterising change, e.g., based on impact analysis~\cite{Ryder:2001}.
Indeed, other criteria for selecting target methods are straightforward to integrate into \APPR.

An empirical evaluation conducted with nine popular, commercial Apps shows that, 
compared to state-of-the-art approaches  (i.e., DM2~\cite{Borges-Droidmate2-ASE-2018}, APE~\cite{Gu:APE:ICSE:2019}, and Monkey~\cite{monkey}),
\APPR leads to reduced test costs.
Indeed, it generates less than 70\%, 4\%, and 2\% of the inputs generated by DM2, APE, and Monkey, respectively.
By automatically exercising, on average, 2.6 instructions belonging to updated methods for every generated test inputs, \APPR is the most cost-effective approach.
Further, on average, \APPR, for a same test execution budget (e.g., 1 hour test execution time), 
\JMRCHANGE{improves  the method and instruction coverage achieved by the second best, state-of-the-art approach by at least 10\%}.

The paper is structured as follows. 
Section~\ref{sec:background} introduces background technologies.
Section~\ref{sec:approach} provides the technical details of the proposed approach.  
Section~\ref{sec:empirical} reports on the results of our empirical evaluation.  
Section~\ref{sec:related} discusses related work.
Section~\ref{sec:conclusion} concludes the paper.

\section{Background}
\label{sec:background}

\begin{figure*}
\begin{minipage}{.49\textwidth}
	\vspace{11mm}
	\includegraphics[width=7cm]{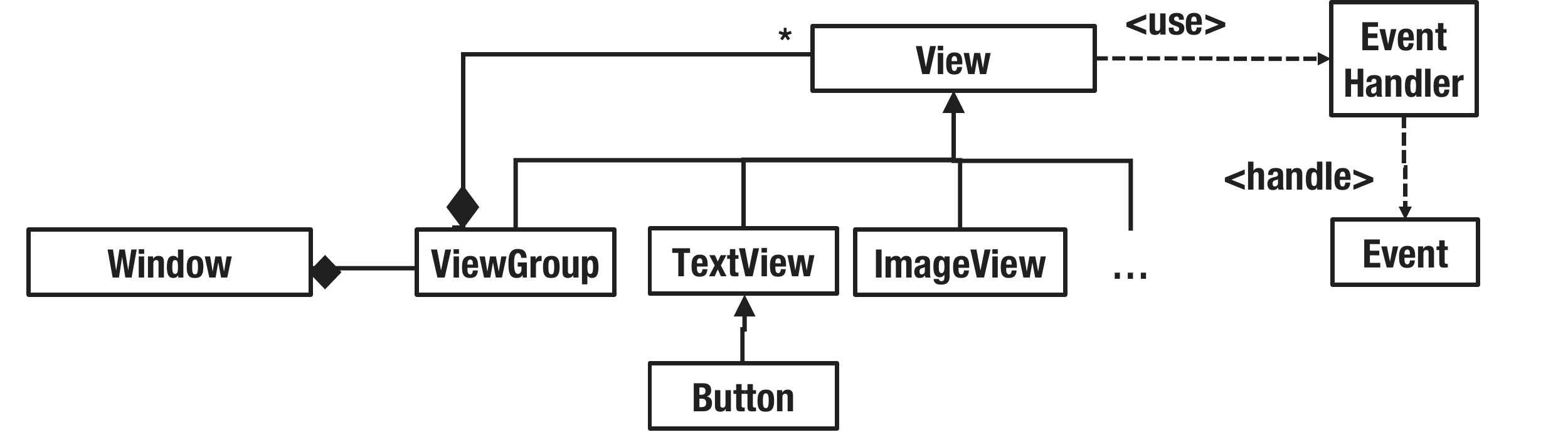}
      \caption{Overview of the Android Apps's GUI architectural components.}
      \label{fig:android:arch}
\end{minipage}\hspace{2mm}
\begin{minipage}{.49\textwidth}
       \includegraphics[width=6.5cm]{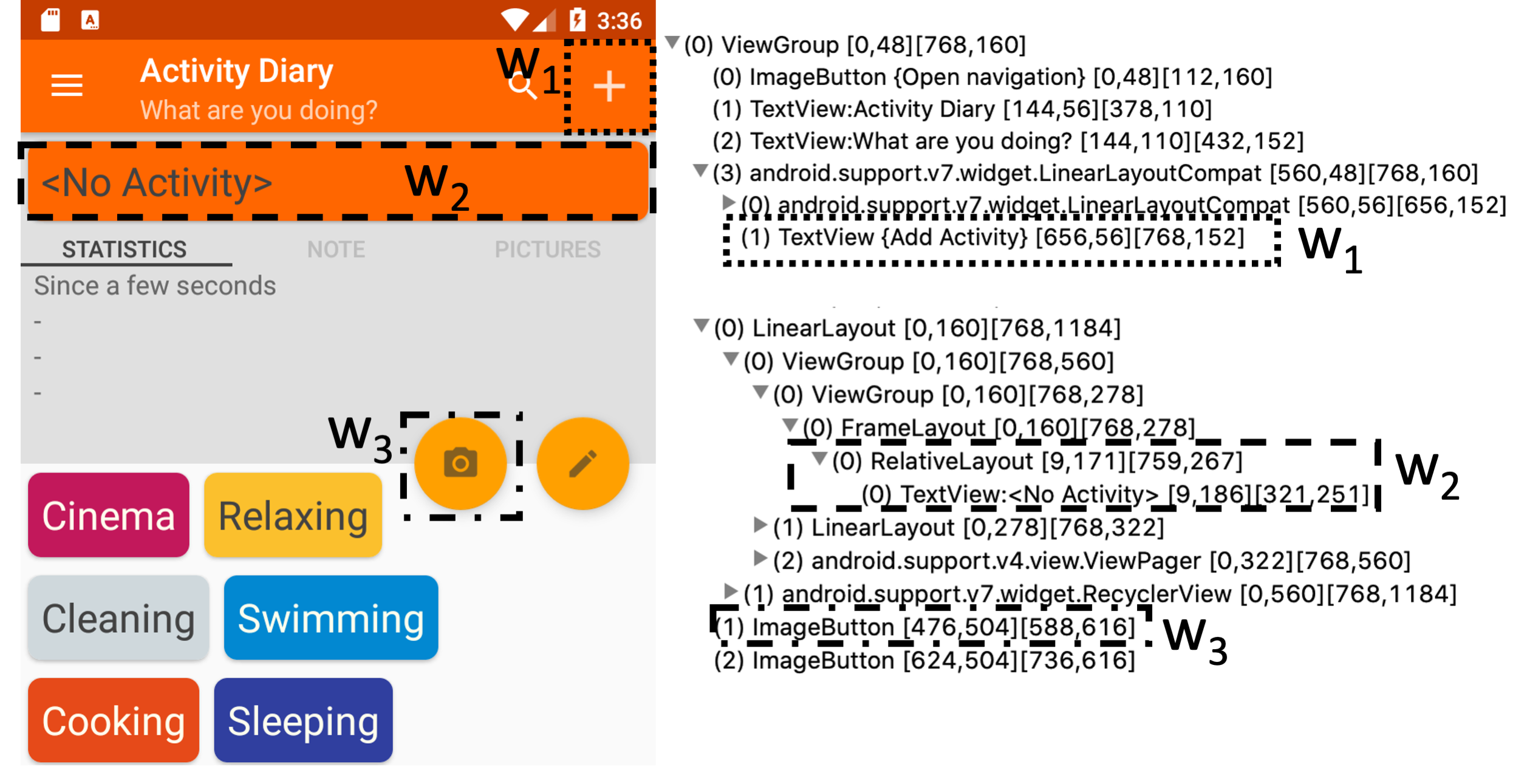} 
      \caption{Example of a GUITree. We use different dotted boxes to match widgets in the tree to the pixels in the screen.}
      \label{fig:android:tree}
\end{minipage}
\end{figure*}

\subsection{App Design and Architecture}
In this paper, we target Android Apps since Android is the most adopted platform and is widely investigated in research ~\cite{Alvarez:WAMA:2019}. However, most of the solutions proposed here should be easily tailorable to other platforms.

When an App is running, the end-user interacts with the active \emph{Window} of the App (i.e., the Window being rendered on the screen). 
Windows consist of a hierarchy tree of widgets; 
\JMR{R3.3}{in this paper, we use the term \emph{GUITree} to refer to such hierarchy trees~\cite{Yang13,Baek2016,stoat17,Gu:APE:ICSE:2019}\footnote{Other work uses the term GUITree to refer to the sequence of App windows encountered during testing~\cite{Amalfitano2017}}}

A widget extends the class \emph{View}. 
Figure~\ref{fig:android:arch} shows a portion of the hierarchy tree of the class \emph{View}.
Figure~\ref{fig:android:tree} shows a portion of the GUITree for a window of 
\JMRCHANGE{Activity Diary, one of our case study Apps (see Section~\ref{sec:empirical})}.

Each widget has a set of properties associated to it.
Widgets can be associated to \emph{EventHandlers} that are invoked by the OS when specific \emph{InputEvents}  are triggered by the end-user. 
Typical InputEvents include click, long click, swipe, and keypress.

In Android, the application logic is typically implemented by Activity classes that are instantiated by the framework and act as controllers of the Model-View-Controller design pattern~\cite{Phillips:AndroidProgramming}.
Inter process communication, instead, is managed by the \emph{Intent} resolution mechanism~\cite{intent}. More precisely, in Android, a system event (e.g., indicating a battery being low) or a message exchanged between apps (e.g, a URL sent by the browser to a music player App) is referred to as an \emph{Intent}. To handle Intents, an App declares in its XML configuration file an Activity that the OS will instantiate and execute when a specific Intent type should be received by the App. 

\subsection{App Testing Automation}

System-level testing of an App through its GUI (i.e., GUI testing) is performed through sequences of \emph{test inputs} that can be either Events or Intents.
Functional GUI testing aims at exercising (i.e., render active) all declared Windows,  trigger all event handlers, and cover all the code of the App under test.
App testing automation aims to generate input sequences to achieve these objectives at the lowest cost possible.

For a complete overview of App testing automation approaches, the reader is referred to recent surveys~\cite{Linares:ICSME:2017,Amalfitano:AndroidTestingSurvey:SQJ:2018}.
\emph{Model-based} solutions are the most commonly ones reported in the literature~\cite{Amalfitano:AndroidTestingSurvey:SQJ:2018}.
In model-based testing approaches, the model used to drive testing is typically a finite state machine (FSM). It can be formally described as a tuple $(S, A,T, \mathcal{L})$ ~\cite{Gu:APE:ICSE:2019}, where
\begin{itemize}
\item S is a set of states. 
\item A is a set of actions. 
\item T is a set of state transitions. Each transition has a source state and a target state. It is triggered by an action $\alpha \epsilon A$.
\item \RF is an abstraction function, which might be used to: (1) assign a Window to a state and (2) match an InputEvent or an Intent to an action.
\end{itemize}

Model-based approaches differ regarding the type of analysis adopted to identify states and transitions (i.e., dynamic~\cite{stoat17,Borges-Droidmate2-ASE-2018} or static~\cite{yang-ase15,yang-jase18}), the abstraction functions used (i.e., predefined~\cite{stoat17,Borges-Droidmate2-ASE-2018} or  adaptable~\cite{Gu:APE:ICSE:2019}), and the model exploration strategies they rely on (i.e., offline~\cite{stoat17} or online~\cite{Borges-Droidmate2-ASE-2018}). 
Adaptable state abstraction functions have been shown to lead to more effective App testing~\cite{Gu:APE:ICSE:2019}
but they have never been used in testing frameworks that enable the effective combination of static and dynamic analysis. 
Further, existing model-based approaches do not prioritize the testing of updated methods, which is our objective here, and thus we require dedicated input generation algorithms.
To leverage the benefits of static and dynamic program analysis, \APPR integrates (1) Gator, a tool that statically identifies states and transitions, (2) DroidMate2 (DM2), a model-based framework that performs online testing, dynamic identification of states and transitions, and enable the combination of static and dynamic analysis, (3) a dedicated algorithm for test input generation, and (4) a custom and adaptable state abstraction function.

\emph{DM2~\cite{Borges-Droidmate2-ASE-2018}} 
consists of two main engines for exploration and automation, respectively.
The former drives the interaction with the App and derives the model of the App under test.
The exploration is driven by a set of user-defined strategies.
The latter 
translates actions into concrete commands on the device. 
In \emph{DM2}, an App state is univocally identified by the set of UI elements in the user interface and their state-related properties. UI elements are identified by means of their descriptive ID or their image bytes cut from a screenshot. State-related properties are given by the position of the elements and other widget specific characteristics.
\APPR relies on the \emph{DM2} automation engine, which provides an API to send inputs to the App under test and retrieve information about the displayed UI elements. In \APPR, App exploration is driven by a custom algorithm (see Section~\ref{sec:approach:test.automation}) that relies on statically derived models and a custom state abstraction function.

\emph{Gator~\cite{yang-ase15,yang-jase18}} is a static analysis tool that creates models of the App under test in the form of window transition graphs (WTGs). 
WTGs are FSMs representing the possible window sequences and their associated events and callbacks. 
Though WTGs can be used to enable model-based testing, they have been mostly used to detect resources leaks~\cite{Wu:Gator:Leak:2019}.

\section{Proposed Approach: \APPR}
\label{sec:approach}

\JMR{2.3}{Within an updated App, we can distinguish among existing features (i.e, features present in previous versions of the App under test) and new features (i.e., features introduced in the App under test). Existing features can be
unchanged (i.e, their functional requirements did not change), modified (i.e, their functional requirements did change), or repaired (i.e., modified because their implementation did not match its functional requirements).} 

\CHANGED{In our work, we aim to automatically exercise \emph{updated features}, including new, modified, and repaired features. 
\JMR{2.3}{More precisely,} we focus on features that are implemented either by introducing new methods or by modifying existing methods~\footnote{Based on related work, 81\% of the updates concern Java files, while only the remaining 19\% concerns manifest files (e.g. permissions) or layout declarations in XML files~\cite{Sharma:QADroid:ISSTA:2019}.}. In this paper, we use the term \emph{updated methods} to refer to both new and modified methods, which are our test targets.}

\JMR{2.3}{The testing activity performed by \APPR is driven by an App model 
with the objective of exercising a set of test targets (i.e., updated methods).
The App model is initially created by static program analysis procedures and then refined during testing.}

The App model metamodel is shown in Figure~\ref{fig:app:model:metamodel:FULL} and described in Section~\ref{sec:approach:app.model.metamodel}. It consists of three parts: (1) an Extended Window Transition Graph (hereafter, EWTG), (2) a Dynamic State Transition Graph (hereafter, DSTG), and (3) a GUI State Transition Graph (hereafter, GSTG). The three graphs are FSMs capturing how input values trigger changes in the state of the App under test. 
The \emph{EWTG} models the sequences of windows being visualized after 
specific inputs (Events or Intents). 
For every input, the EWTG keeps trace of the name of the handlers associated to the input and the list of test targets that may be invoked during the execution of the input handler.
The \emph{GSTG} is a fine-grained model that captures every visual change in the GUI (e.g., the color of a button) that might be triggered by an action 
on the GUI. 
An action is an instance of an input (e.g., click on a specific Button widget). 
Finally, the \emph{DSTG} models the abstract states of the visualized Windows and the state transitions triggered by events. 
Abstract states are identified by a state abstraction function to eliminate possible non-determinism. The DSTG plays a critical role to optimize the test budget and identify a reduced set of input events; indeed, it helps determine a correct and reduced sequence of events necessary to reach a specific Window from another one. 

\begin{figure}[t]
  \centering
	\includegraphics{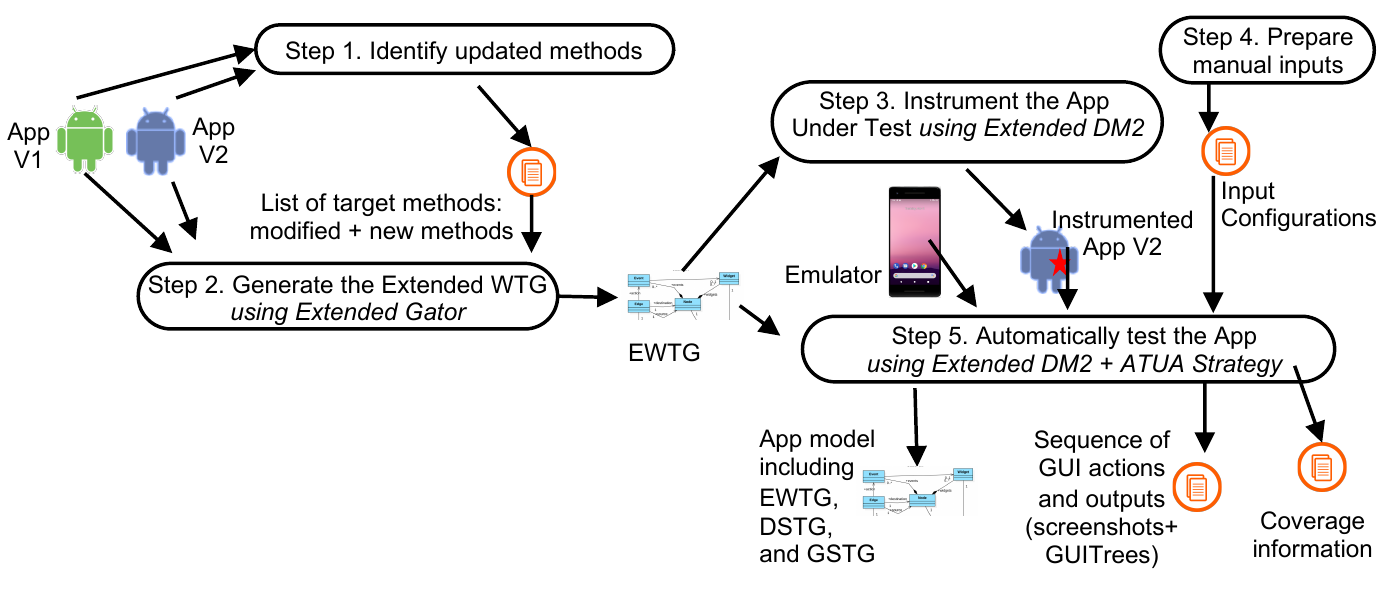}
      \caption{Overview of the \APPR{} process to test App updates.}
      \label{fig:process:updates}
\end{figure}

Figure~\ref{fig:process:updates} provides an overview of the process implemented by \APPR to test App updates.
In Step 1, \APPR compares the previous (App V1 in Figure~\ref{fig:process:updates}) and the updated (App V2) version of the App under test to identify the updated methods.
In Step 2, \APPR relies on Soot~\cite{Rountev:2014}, a static analysis framework, and an extended version of Gator, to generate the EWTG. 
In Step 3, \APPR relies on DM2 to generate a version of the App that is instrumented to trace code coverage.
In Step 4, engineers manually specify test inputs that are unlikely to be generated automatically (e.g., the login credentials for Apps that require a user to be registered on a remote platform).
\JMR{2.3}{In Step 5, \APPR exercises the App under test by relying on an extended version of DM2 that integrates the \APPR test algorithm.}
During testing, \APPR refines the App model and relies on it to identify the actions to perform on the GUI. For example, \APPR uses the App model to identify the action that, in the current window, may lead to the execution of a test target.

The main output of Step 5 is the sequence of GUI actions performed during testing and the outputs (i.e., the screenshot of the active Window and the corresponding GUITree) generated by the App under test after every action. This sequence is used by engineers to verify if the behavior of the App is as expected (test oracle). As mentioned earlier, to verify App results, engineers can rely on two complementary state-of-the-art approaches, not addressed by \APPR, that respectively target regression failures in unchanged features and failures in newly implemented, repaired, and modified features. 
To discover regression failures, engineers can replicate, on a previous App version, the test input sequences generated  for the updated App and automatically compare the generated outputs. Differences in the outputs generated by the two versions should indicate the presence of a regression fault.
To discover failures in new and repaired features, engineers can visualize 
the GUITrees or the screenshots of the active Window rendered after each Action.
The visual inspection of the App outputs enables an engineer to determine the presence of functional failures, based on expected behavior, whether specifications are implicit or documented.
\JMR{3.4}{For example, the engineer shall determine if the Window rendered after each Action includes the expected content and is well positioned.
Also, the engineer shall inspect GUITrees to determine if the widgets within a Window have the expected properties.}
In Section~\ref{sec:empirical}, we discuss to what extent \APPR reduces the cost associated to the manual activities entailed by the test oracle strategies above, with respect to other state-of-the-art test automation solutions.

In addition, \APPR provides, as output of Step 5, an App model including the EWTG, the DSTG, and the GSTG. The App model is generated and continuously refined during testing. 
Further, it reports coverage information, i.e., 
the sets of updated methods and instructions belonging to updated methods that have been exercised during testing.

In the following, we provide additional details about the App model metamodel (Section~\ref{sec:approach:app.model.metamodel}) and
describe Steps 1 to 5 (Sections~\ref{sec:approach:identify.updated.methods}~to~\ref{sec:approach:test.automation}), except for Step 3 which is already automated by DM2.

\subsection{App Model Metamodel}
\label{sec:approach:app.model.metamodel}

\begin{figure*}[tb]
  \centering
	\includegraphics[width=14cm]{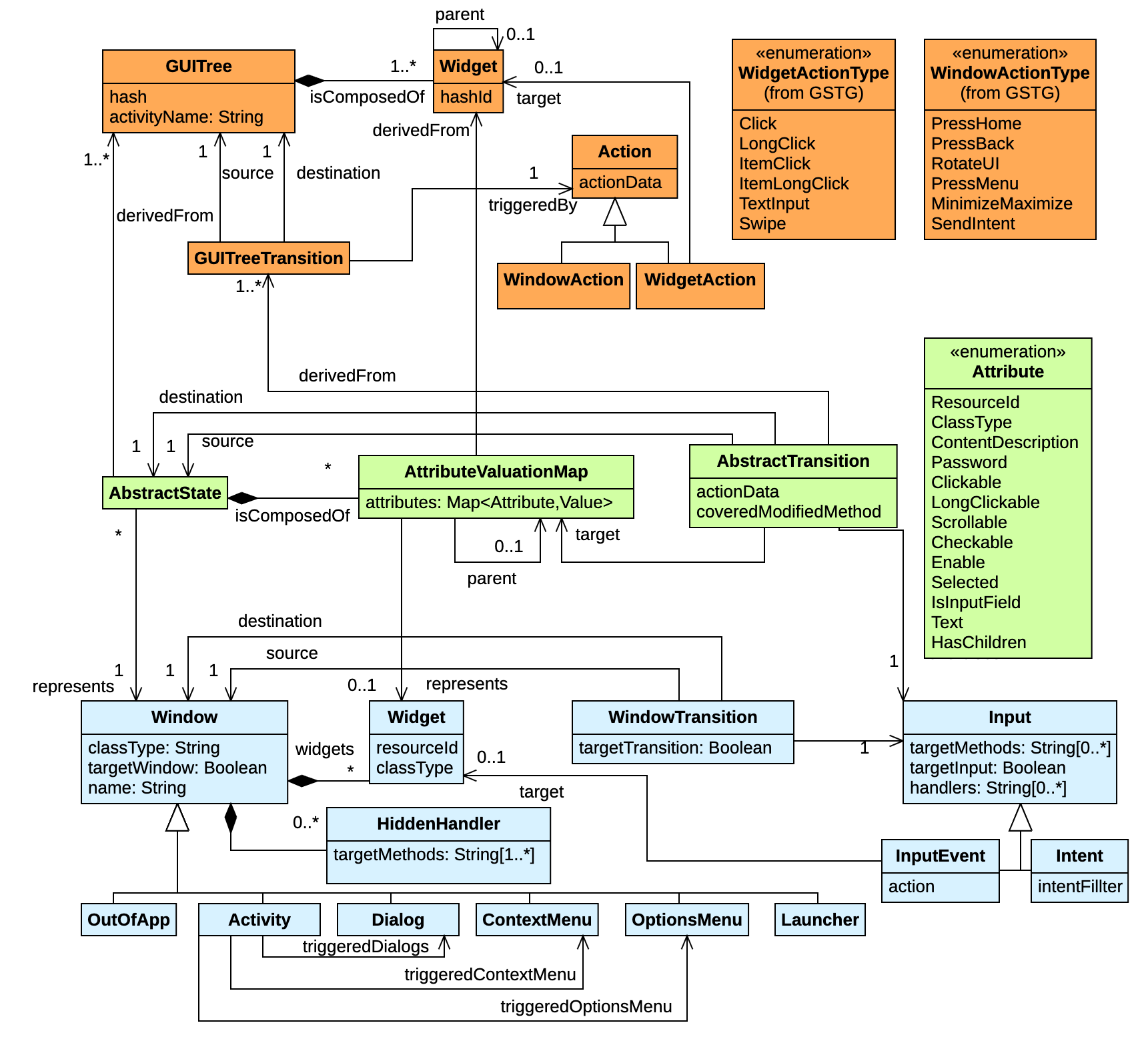}
      \caption{\APPR Metamodel. Colors are used to identify classes belonging to a specific metamodel component: light blue for EWTG (bottom), light green for DSTG (middle), orange for GSTG (top).}
      \label{fig:app:model:metamodel:FULL}
\end{figure*}

\begin{figure*}[t]
  \centering
	\includegraphics[width=14cm]{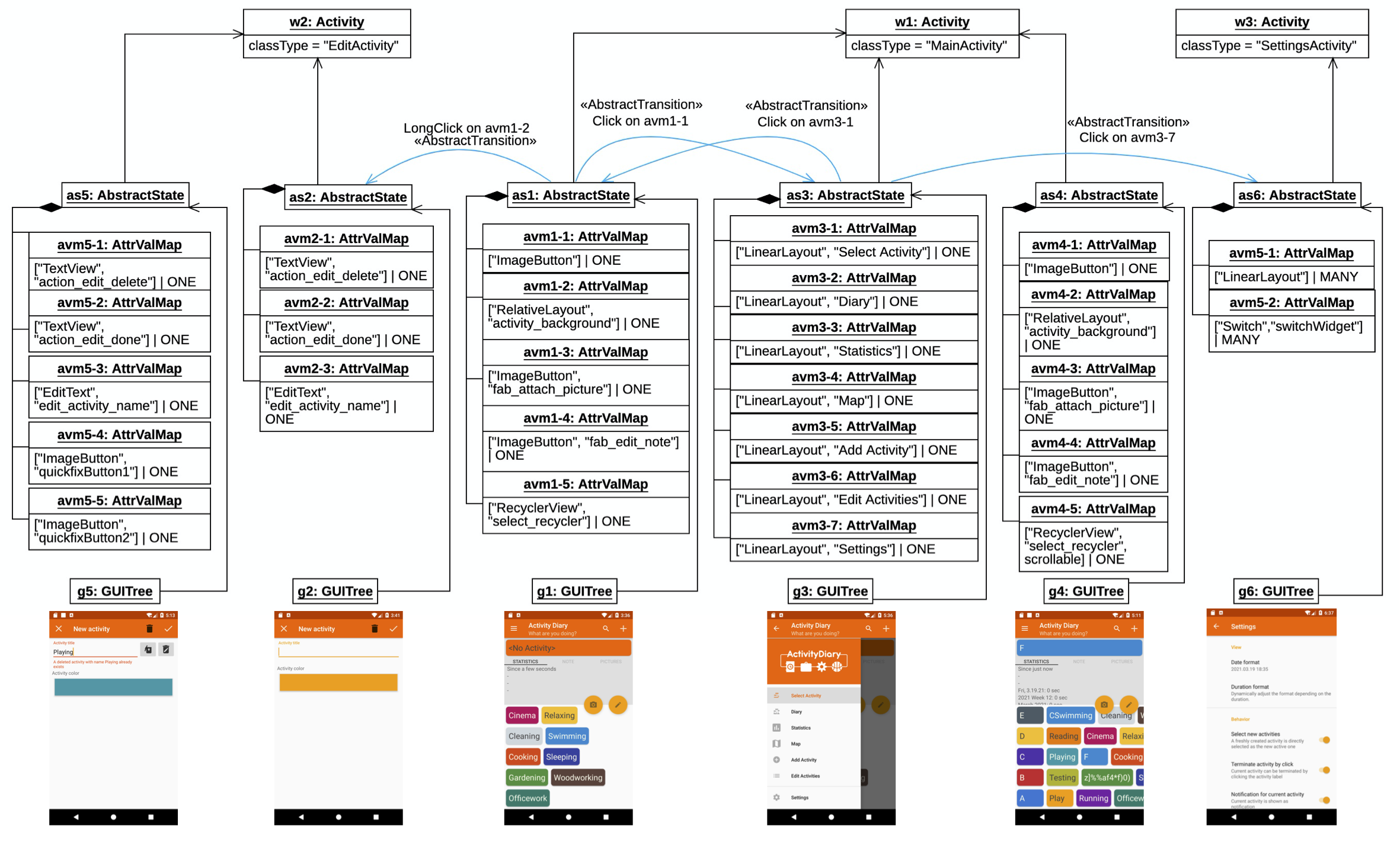}
	\footnotesize
	\textbf{Legend}: Straight, black arrows capture associations. 
	To avoid cluttering, we rely on curved, blue arrows to model AbstractTransitions.
The action and target of abstract transitions are reported in textual form. 
For illustration purposes, we show screenshots instead of GUITree hashes. Cardinality of AttrbuteValuationMaps are reported after the "|" symbol.
      \caption{Example of an App model, represented as a UML object diagram. }
      \label{fig:android:AppModelExample}
\end{figure*}

Figure~\ref{fig:app:model:metamodel:FULL} shows the \APPR metamodel as a UML class diagram. 
Figure~\ref{fig:android:AppModelExample} shows an example App model built when testing \JMRCHANGE{Activity Diary}.

The EWTG is consistent with the WTG generated by Gator. 
Each WindowTransition is triggered by an \emph{Input}, either an \emph{InputEvent} or an \emph{Intent}. 
An InputEvent is associated to the Widget that declares its EventHandler.
If the EventHandler is not declared by a Widget (e.g., for the event \emph{PressHome}), the InputEvent is not associated to any target Widget.
Each Widget belongs to one Window.

In addition to the concepts captured by the WTG, the EWTG generated by \APPR also captures the list of modified methods that can be triggered by the Input 
(i.e., the attribute \emph{targetMethods} of class \emph{Input}), which are used to drive testing.
A WindowTransition triggered by Inputs with associated \emph{targetMethods} is a \emph{target Transition}. 
Similarly, a Window that is the source for at least one target Transition is a \emph{target Window}.
\JMR{3.3}{\emph{TargetMethods} are identified by our Gator extension (see Section~\ref{sec:approach:EWTG}).}
The EWTG also captures the dialogs and menus that can be opened by an Activity (e.g., association \emph{triggeredDialogs}).
Finally, it also models the HiddenHandlers of a Window, which are introduced in Section~\ref{sec:approach:EWTG}.

The GSTG captures the same information provided by the models generated by DM2. 
The state of a Window is captured by its GUITree, which is a composition of Widgets. 
For each Widget, we record the values associated to its properties and derive the Widget hash (to associate an ID to the current state of the Widget) according to the DM2 strategy. 
The hash of the GUITree is then derived from the hash of its Widgets. 
In addition, the GSTG also captures the name of the Activity running when the GUITree is visualized (see attribute \emph{activityName}), which we derive, at runtime, from logcat~\cite{logcat}.
The transition between GUITrees is triggered by an Action, which can be handled either by the Widget (\emph{WidgetAction}) or by the visualized Window (\emph{WindowAction}). The enumerations \emph{WidgetActionType} and \emph{WindowActionType} list the type of actions that can be performed to trigger GUITreeTransitions. Actions might have additional information (i.e., actionData) associated to them; for example, the text provided to the App under test by a TextInput action or the start and end coordinates of a Swipe action.

\CHANGED{The DSTG provides abstract states that group together GUITrees in which a same Action triggers a same App behaviour (e.g., leads to a same abstract state).} 
\CHANGED{The DSTG enables \APPR to efficiently test the App under test by determining the shortest sequence of Actions that reaches a target Window. Also, abstract states capture the conditions under which a specific Action can trigger a modified method, for a certain Window. Abstract states are thus a mean to minimize the number of Actions generated by the test automation approach.}

Each AbstractState consists of a number of AttributeValuationMaps, each one abstracting the state of the widgets belonging to the GUITree that have the same set of attribute valuations.
An \emph{AttributeValuationMap} is a map of pairs $\langle$attribute,value$\rangle$. 
The enumeration \emph{Attribute} in Figure~\ref{fig:app:model:metamodel:FULL} provides the list of attributes appearing in the AttributeValuationMap.
The AttributeValuationMap has a cardinality attribute, which indicates how many widgets have the same attribute valuations. 
For example, in Figure~\ref{fig:android:AppModelExample}, the AbstractState \emph{as6} includes many LinearLayout widgets (\emph{LL} in the Figure, cardinality~\emph{MANY}), one for each item in the displayed list. 

Since the DSTG is used to drive testing, i.e., to select the Actions to be triggered at runtime,
the AbstractState captures only the attributes of widgets that are \emph{interactive}.
A widget is \emph{interactive} when it is enabled, visible, and is an instance of a class that can be the target of any action of type WidgetActionType (See Figure~\ref{fig:app:model:metamodel:FULL}). 

At runtime, during testing, \APPR identifies AbstractStates through a dedicated \emph{abstraction function} (\RF).
\APPR automatically defines a distinct \RF for each Window of the App under test.
\RF relies on a predefined set of \emph{reducers}, i.e., functions that
extract the value of a property of a widget~\cite{Gu:APE:ICSE:2019}.
Table~\ref{table:reducers} shows the list of reducers implemented by \APPR. 
Two AbstractStates differ when at least one value differs across their respective AttributeValuationMaps, or when they have a different cardinality.
For example, in Figure~\ref{fig:android:AppModelExample}, the AbstractStates \emph{as1} and \emph{as3} differ because \emph{as1} does not contain the AttributeValuationMaps for the LinearLayouts belonging to the drawer menu (to save space, we do not report all the  AttributeValuationMaps for  \emph{as3}). The AbstractStates \emph{as1} and \emph{as4} are different because in  \emph{as4} the RecyclerView (avm4-5) becomes scrollable.

In the DSTG, state transitions are captured by AbstractTransitions. 
An AbstractTransition is univocally identified by the \emph{actionType}, its \emph{source}, its \emph{target}, its \emph{destination}, and its \emph{actionData}.
The \emph{actionType} matches one of the items belonging to the enumerations WidgetActionType or WindowActionType.
The \emph{actionData} is captured only for two actionTypes (i.e., Swipe and Intent) that usually lead to distinct AbstractStates depending on their action data. 
In the case of Swipe, the actionData indicates the direction of the Swipe action (i.e., Up, Down, Left, Right). 
\JMRCHANGE{For Intents, the actionData matches the Intent input text because we expect engineers to provide one manual input for each possible Intent type (e.g., one different URL for each of the file types supported by an App)}.
We leave to future work the definition of
functions that provide an abstract representation for the data associated to other types of actions
(e.g., to distinguish between numeric, alphabetic, or non-alphanumeric data provided to TextInputs).

\JMR{2.3}{A DSTG may include non-deterministic AbstractTransitions, that is, transitions with the same actionType, outgoing from the same AbstractState, but reaching different AbstractStates. 
Non-deterministic AbstractTransitions may prevent us from finding the correct sequence of Inputs necessary to reach the states in which target methods could be triggered.}
\APPR detects non-determinism at runtime, during testing, when an Action does not bring the App into the expected AbstractState. 
When this happens, \APPR refines \RF for the AbstractState in which the action had been triggered.
It does so according to five levels of granularity, which are captured in Table~\ref{table:refinement}.
With level L1, \RF distinguishes states based on static information about the widgets (i.e., resource ID and class) and information about how they can be interacted with (i.e., reducers appearing in rows 3 to 12 of Table~\ref{table:reducers}). 
With level L2, in addition to the information accounted for in L1, \RF includes the text associated to the widget, which often affects the behaviour of an app (e.g., invalid characters in a textbox may prevent a state transition).
With level L3, \RF also reports the number of children of a widget (i.e., $R_{HC}$). 
L3 is useful because the interactive widgets captured by \RF may include non-interactive children whose state is not captured by \RF but may characterize the current state (e.g, through descriptive labels).
With levels L4 and L5, \RF captures, for every interactive widget, the same information as L2 and, in addition, for every child, the information captured by levels L1 and L2, respectively.

\begin{table}[tb]
\begin{minipage}{.45\textwidth}

\footnotesize
\begin{tabular}{|p{3mm}|p{1cm}|p{3.8cm}|}
\hline

&\textbf{Reducer}&\textbf{Description}\\
\hline
1& $R_{RID}$&Resource ID.\\

2& $R_{CN}$&Class name.\\

3&$R_{CD}$&Value of \emph{Content description}.\\

4& $R_{P}$&Value of \emph{Password}.\\

5& $R_{C}$&Value of \emph{Clickable}.\\

6& $R_{LC}$&Value of \emph{Long Clickable}.\\

7&$R_{S}$&Value of \emph{Scrollable}.\\

8&$R_{Ch}$&Value of \emph{Checked}.\\

9& $R_{E}$&Value of \emph{Enabled}.\\

10& $R_{S}$&Value of \emph{Selected}.\\

11& $R_{I}$&True if it is an input field.\\

12& $R_{T}$&Value of \emph{Text}.\\

13&$R_{HC}$&True if the widget contains one or more children.\\

\hline
\end{tabular}
\caption{\APPR reducers. We indicate the value of the \emph{property} reported by each.}
\label{table:reducers}

\end{minipage}\hspace{1mm}
\begin{minipage}{.53\textwidth}

\footnotesize
\begin{tabular}{|@{}p{7mm}@{}|@{\hspace{1mm}}p{3.1cm}@{\hspace{1mm}}|@{\hspace{1mm}}p{3.1cm}@{\hspace{1mm}}|}
\hline

\textbf{Level}&\textbf{Reducers applied to interactable Widget}&\textbf{Reducers applied to interactable Widget Children}\\
\hline
L1& $R_{RID}, R_{CN}, R_{CD}, R_{Ch}, R_{E},$ $ R_{P}, R_{S}, R_{I}, R_{C}, R_{LC}, R_{S}$&\\
L2& $R_{RID}, R_{CN}, R_{CD}, R_{Ch}, R_{E},$ $ R_{P}, R_{S}, R_{I}, R_{C}, R_{LC}, R_{S}$, \textcolor{blue}{$R_{T}$}&\\
L3& $R_{RID}, R_{CN}, R_{CD}, R_{Ch}, R_{E},$ $ R_{P}, R_{S}, R_{I}, R_{C}, R_{LC}, R_{S}, R_{T}$, \textcolor{blue}{$R_{HC}$}&\\

L4& $R_{RID}, R_{CN}, R_{CD}, R_{Ch}, R_{E}, $ $R_{P}, R_{S}, R_{I}, R_{C}, R_{LC}, R_{S}, R_{T}$ & 
\textcolor{blue}{$R_{RID}, R_{CN}, R_{CD}, R_{Ch}, R_{E}, $} \textcolor{blue}{$R_{P}, R_{S}, R_{I}, R_{C}, R_{LC}, R_{S}$}\\

L5& $R_{RID}, R_{CN}, R_{CD}, R_{Ch}, R_{E},$  $R_{P},  R_{S}, R_{I}, R_{C}, R_{LC}, R_{S}, R_{T}$ & 
$R_{RID}, R_{CN}, R_{CD}, R_{Ch}, R_{E}, $ $R_{P}, R_{S}, R_{I}, R_{C}, R_{LC}, R_{S}$, \textcolor{blue}{$R_{T}$}\\

\hline

\end{tabular}
\caption{Refinement of \APPR state abstraction function. In blue we show the reducers introduced in finer granularity level.}
\label{table:refinement}

\end{minipage}
\end{table}

\begin{figure}[tb]
  \centering
	\includegraphics[width=14cm]{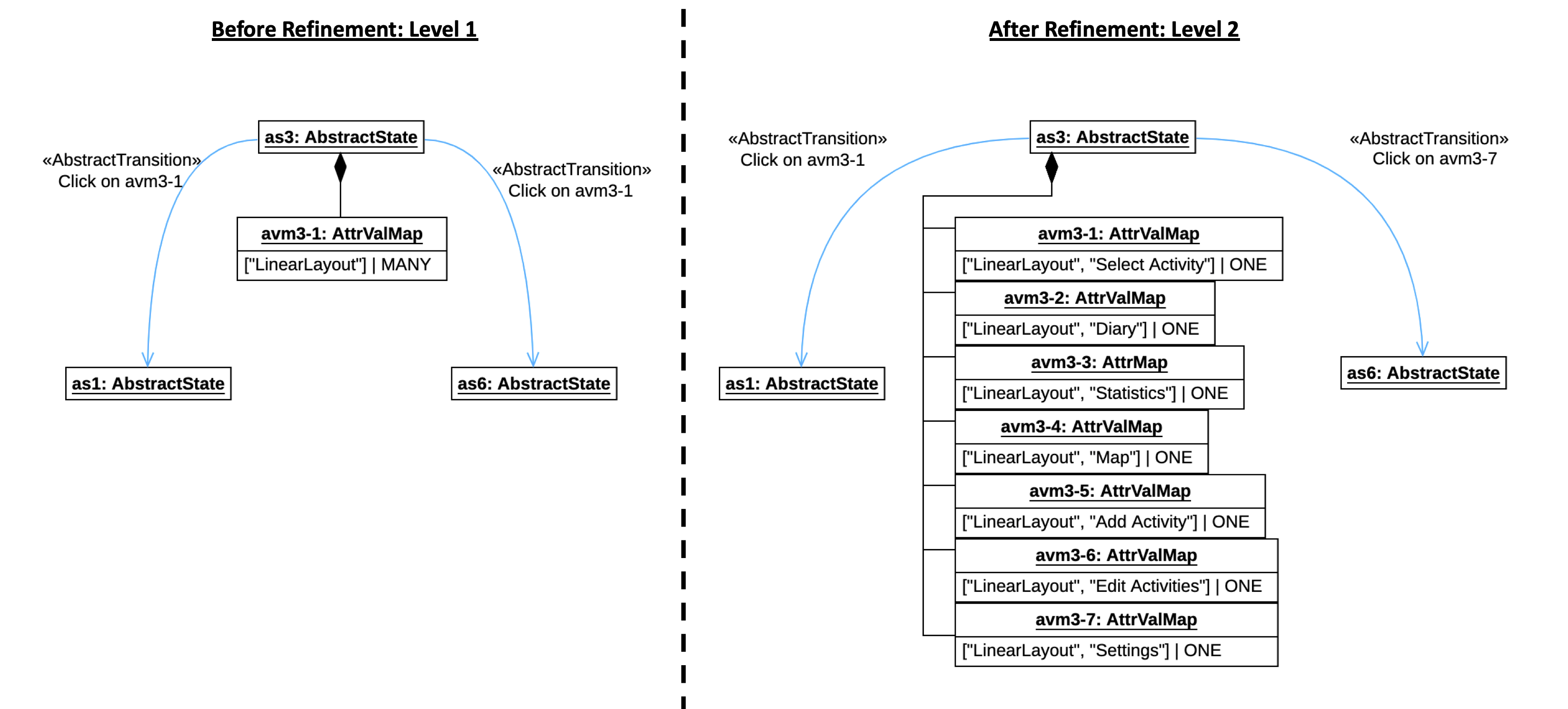}
      \caption{Example of refinement of \RF. For the notation used, see the Legend of Figure~\ref{fig:android:AppModelExample}.}
      \label{fig:app:model:refinement}
\end{figure}

Figure~\ref{fig:app:model:refinement} shows the result of the refinement of \RF for the abstractState \emph{as3}. 
By applying \RF with level L1, all the clickable elements on the Window, which are LinearLayouts with the same properties, except for the text, resulted into a same AttributeValuationMap with cardinality \emph{MANY}. At runtime, \APPR detects non-determinism; indeed, a click on this AttributeValuationMap may lead to two different AbstractStates: \emph{as1} and \emph{as6}. The refinement of \RF, which leads to level L2, allows \APPR to distinguish all the different clickable elements since the text property is included into the AttributeValuationMap, thus eliminating non-determinism.

\DELETE{Concerning the relationships among the classes of the \APPR metamodel that are used to drive testing (see Section~\ref{sec:approach:test.automation}), we note that
a Window of the EWTG can have one or more AbstractStates. Each AbstractState is associated to one or more concrete GUITrees.
Multiple Actions performed during testing are associated to a same AbstractTransition. 
Each AbstractTransition in the DSTG uses one Input. Similarly, each Action in the GSTG uses one Input.}

\subsection{Step 1. Identify Updated Methods}
\label{sec:approach:identify.updated.methods}

In the first step, \APPR identifies methods that have been modified or introduced by the new version of the App (i.e., V2 in Figure~\ref{fig:process:updates}).
This task might be accomplished through source code comparison across versions~\cite{Pastore:ISSRE:12}.
However, 
to enable experiments with commercial Apps,
we developed a toolset (hereafter, \appdiff) that compares compiled Android Apps.

\appdiff is an extension of \emph{LibScout}~\cite{derr:ccs17}, a light-weight static analysis tool for Android.  
It first generates a hashtree over the bytecode for each App. 
The hashtree is a
three-layered Merkle tree in which parent hashes are generated from their child nodes. 
The three layers model the flattened package structure that is preserved in
the compiled code, i.e., packages, classes, and methods. 

The tree is built bottom up starting with the method hashes. 
A method hash is computed over the method signature and the opcodes in bytecode instructions. 
To identify code-level changes across App versions, we additionally store package, class, and method names along with the hashes. 
To efficiently check for differences, two hash trees are matched top-down starting with the package hashes.
Methods that share the same name but have a different hash have been modified. New methods appear only in the most recent version.

\subsection{Step 2. Generate the Extended WTG}
\label{sec:approach:EWTG}

\AutAut generates the Extended WTG by means of static program analysis; more precisely, by performing, on the updated App, the analysis implemented by an extended version of Gator and Soot.

The original version of Gator works by processing Android bytecode and XML layout files.
For the analysis of bytecode, Gator relies on Soot.
Bytecode analysis is used to identify the types of Window (i.e., Activity, Dialog, OptionsMenu, and ContextMenu) that are programmatically specified in the App.
Bytecode analysis is also used to identify the widgets that compose a window and the associated event handlers.
Gator identifies widgets that extend the class \emph{android.view.View} and its handlers.
Event handlers' code is processed to determine window transitions.
XML layout files are processed to identify additional event handlers.

Our extensions to Gator address some known limitations
 ~\cite{Kuznetsov:2018}.
More precisely, we support the identification of window transitions triggered by Fragments and RecyclerView, which are widget containers that are not identified by Gator as such.
Our extensions associate the contained widgets to the window that declares either the Fragment or the RecyclerView.
\JMR{3.3}{Also, in the EWTG, we associate each WindowTransition with the Input triggering the transition (the information is provided by Gator).}

We rely on Soot to traverse the backward call graph of every updated method $m$. During the traversal, when we encounter a method that has been identified by Gator as an event handler $e$, 
we update the EWTG to trace the fact that 
\JMR{3.3}{the inputs associated to the WindowTransition triggered by the event handler $e$ can lead to the execution of the updated method $m$
(i.e., we add the updated method $m$ to the list of target methods for the Input instance).} 
Also, we rely on Soot to extract string literals to be used for testing (see Section~\ref{sec:approach:test.automation}).

Finally, we determine if Gator does not identify some of the event handlers of the App, which is a common problem of static analysis tools for Apps. 
Indeed, these static analysis tools rely on hardcoded procedures for the identification of event handlers (e.g., they look for specific method names~\cite{Rountev:2014}); since OS APIs are under continuous evolution, it is unlikely that static analysis tools will ever be able to identify all the event handlers of an App.
To address this problem, we introduced into \APPR  three solutions, one based on static analysis (described in the next paragraph) and two based on dynamic program analysis (described in Sections~\ref{sec:active} and~\ref{sec:algo:randomExploration}).
 
To identify missing event handlers using static analysis, we rely on the observation that if an event handler is not detected by Gator, the backward traversal of the call graph performed by \APPR will not reach any event handler but will terminate in a method that (i) belongs to a Window class and (ii) is 
not invoked by any other method of the App under test. Such methods are likely event handlers invoked at runtime by the Android APIs. We refer to them as \emph{hidden-handlers}. We keep track of all the hidden-handlers encountered during the analysis along with the list of updated methods reachable from them. We rely on this information during Step 5.


\begin{figure}
\begin{lstlisting}
{
  "BookInsertionAndSearch" : {         //input pattern
    "Windows" : [	  "ACT[bookcatalogue.EditAuthorList]1741", "ACT[bookcatalogue.BookEdit]1802" "ACT[bookcatalogue.BookISBNSearch]1843" ],
    "DataFields" : {  
      "isbn" : {
        "resourceIdPatterns" : [ "isbn_txt" ]
      },
      "title" : {
        "resourceIdPatterns" : [ "title_txt" ]
      },
...
    },
    "Instances" : [
      {
        "isbn" : "0387284540",
        "title" : "Applied probability and statistics",
        "publisher" : "Springer",
        "pages" : "350",
        "list_prices" : "69",
        "format" : "Hard Cover",
        "genre" : "Unfiction",
        "language" : "English"
      }
\end{lstlisting}
\caption{Manual definition of inputs}
\label{fig:manual:input:example}
\end{figure}


\subsection{Step 4. Prepare manual inputs}
\label{sec:approach:manual.inputs}

Certain input values are unlikely to be automatically generated, consequently certain Apps' features might not be automatically exercised without an appropriate solution to handle such cases. 
\JMR{3.5}{In related work, these inputs are referred to as \emph{Unlocking GUI Input Event Sequences} (hereafter, \emph{unlocking inputs}, for brevity)~\cite{AMALFITANO201995}.
Examples include login credentials, 
files of a specific type, and data to be received by the App under test through the Android Intent mechanism. 
To handle these cases, in \APPR, engineers specify unlocking inputs in a JSON file capturing a set of input values and the patterns identifying the Windows requiring such input values.}
An example of our input definition format is provided in Figure~\ref{fig:manual:input:example}.

According to our format, engineers can specify one or more input insertion patterns (e.g., \emph{BookInsertionAndSearch} in Figure~\ref{fig:manual:input:example}, Line 2).
For each pattern, they specify the Windows in which the pattern should be used (Line 3), and the widgets which should be used to provide the input data specified (i.e., \emph{DataFields} field in Line 4). Each widget is identified by a name (e.g., \emph{isbn} in Line 5) and a regular expression that enables its selection in the GUITree, based on its name (e.g., \emph{isbn\_txt} in Line 6). Finally, multiple input instances (e.g., book names in this case) can be specified (see field \emph{Instances} in line 13). 

\JMR{3.5}{Since \APPR relies on software engineers to identify unlocking inputs, its effectiveness might be affected by engineers' mistakes. For example, in our experiments, we specify manual inputs only for login operations and key features on the App under test (see Section~\ref{sec:empirical:setup}), thus potentially omitting unlocking inputs concerning other App features.
The integration of state-of-the-art solutions to automatically discover unlocking inputs is part of our future work~\cite{AMALFITANO201995}.}

\subsection{Step 5. Automatically test the App}
\label{sec:approach:test.automation}

\begin{figure}[tb]
  \centering
	\includegraphics[width=14cm]{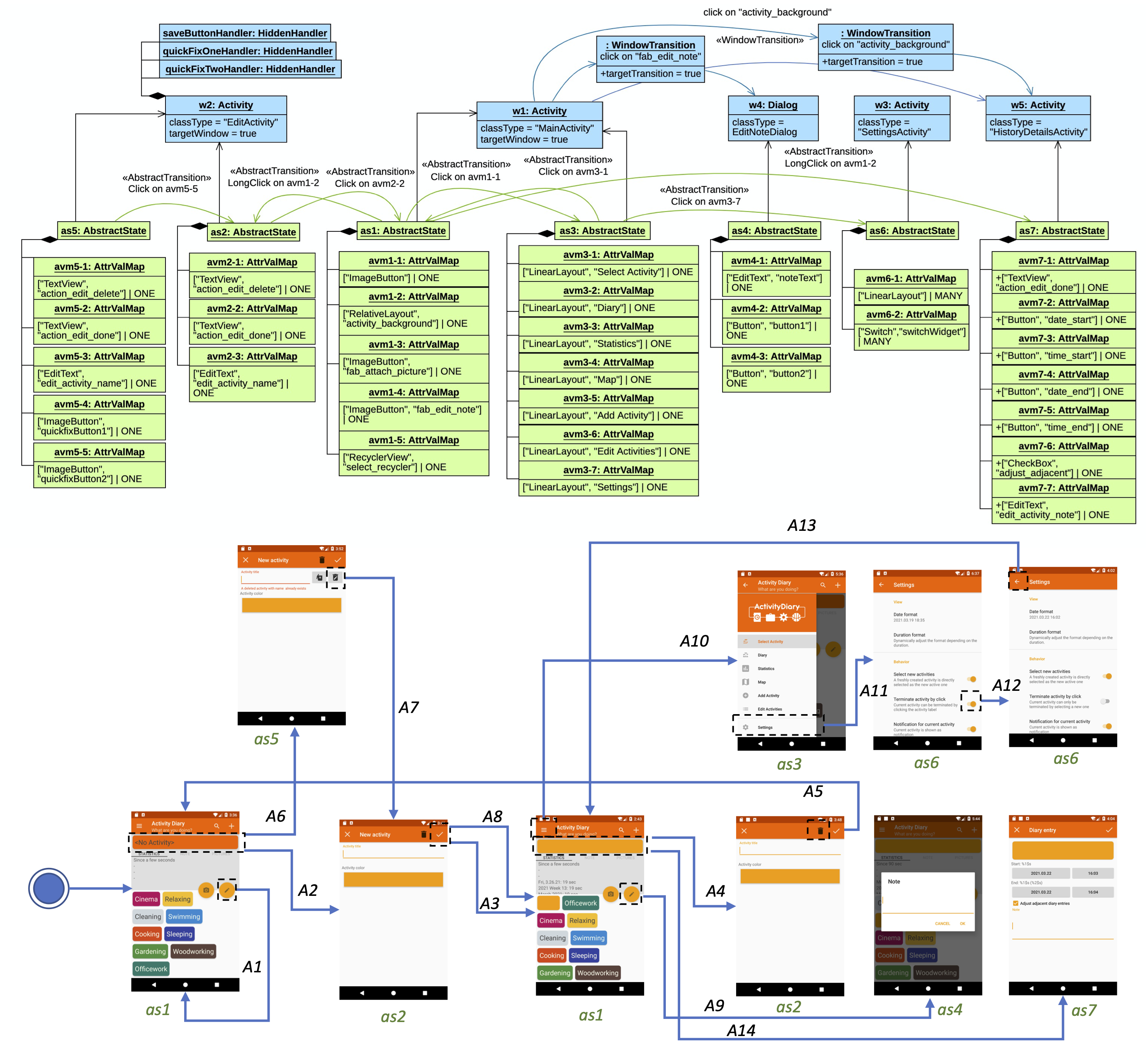}
\footnotesize
\textbf{Legend:}
The top part shows the EWTG, the DSTG is in the middle, while, to simplify reading, the GSTG is represented by means of screenshots corresponding to each GUITree. Labels below screenshots are used to associate GSTG states to AbstractStates.
{The Actions appearing in the GSTG are}:  
\emph{A1}, click on \emph{EditNote} button.
\emph{A2}, long click on \emph{Current activity} widget.
\emph{A3}, click on \emph{Save} button.
\emph{A4}, long click on \emph{Current activity} widget.
\emph{A5}, click on \emph{Delete activity} button.
\emph{A6}, long click on \emph{Current activity} widget.
\emph{A7}, click on the button to automatically rename the deleted activity.
\emph{A8}, click on \emph{Save} button.
\emph{A9}, click on \emph{EditNote} button.
\emph{A10}, click on \emph{Open navigation} button.
\emph{A11}, click on \emph{Settings} button.
\emph{A12}, click on \emph{Terminate activity by click} button.
\emph{A13}, click on \emph{Back} button.
\emph{A14}, click on \emph{Current activity} widget.
      \caption{App Model for the Activity Diary running example}
      \label{fig:app:model:runningexample}
\end{figure}

\AutAut automatically tests the updated App by triggering the Actions required to exercise target Transitions.
When testing starts, the App model consists of an instance of the EWTG for the App under test. GSTG and DSTG are dynamically constructed and extended at runtime by \APPR.

The test execution process includes three distinct phases, each one relying on a different strategy for the generation of Actions. 
In \emph{Phase 1}, \APPR triggers one Action for every target Input.
The goal of Phase 1 is to handle the simplest scenario, i.e., exercise instructions that are executed every time data provided through a target Input is processed.
In \emph{Phase 2}, \APPR exercises target Windows with multiple, diverse sets of Inputs. The goal of Phase 2 is to exercise those instructions that are executed only when specific constraints on input values provided in a Window are satisfied.
In \emph{Phase 3}, \APPR exercises both target Windows and Windows they depend on. The goal of Phase 3 is to exercise those instructions that can be executed only when certain constraints on the input values provided in related Windows (e.g., preferences Windows) are satisfied.

\subsubsection{Running Example}{\ }

\JMR{3.1}{To illustrate our approach, we describe part of the actions taken by \APPR when testing the upgrade to version 134 of Activity Diary. 
Activity Diary enables end-users to record a diary for their activities. It includes features to categorize activities, report statistics, remind users about recurrent activities, and attach notes and pictures to activities.}

\JMRCHANGE{We consider a subset of the features updated in version 134 of Activity Diary, which aim to
(1) visualize the details of the current activity by clicking on the activity name,
(2) edit the current activity or create a new activity with a long click on the current activity name, (3) automatically fix a duplicated name for an activity, (4) edit an activity note.
Figure~\ref{fig:app:model:runningexample} shows the App model for our running example.}

\JMRCHANGE{In the EWTG of Figure~\ref{fig:app:model:runningexample}, the transition between MainActivity (\emph{w1})  and EditNoteDialog (\emph{w4}) is a target transition since it triggers one modified method: the handler of the EditNote button (i.e., \emph{editNoteHandler}, not shown in Figure~\ref{fig:app:model:runningexample}).
The EditActivity Window (\emph{w2}) is a target Window since it contains three hidden-handlers:  \emph{saveButtonHandler}, \emph{quickFixHandlerOne}, and \emph{quickFixHandlerTwo}. The hidden-handler \emph{saveButtonHandler} reaches the modified method \emph{checkConstraints} while the other two hidden-handlers are the event handlers for the quick fix buttons appearing in the UI. \APPR classifies these three methods (i.e., \emph{saveButtonHandler}, \emph{quickFixHandlerOne}, and \emph{quickFixHandlerTwo}) as hidden-handlers because they are not associated by GATOR to any WindowTransition in the EWTG (see Section~\ref{sec:approach:EWTG}). Indeed, GATOR cannot correctly process the control flow that reaches function \emph{setListener}, which is the function used to assign the three handlers to their corresponding buttons\footnote{GATOR does not correctly process control flows starting within event handlers, likely because of their recursive nature; in our running example, the control flow starts within event handlers triggered by changes in color selectors and text boxes.}.}

\JMRCHANGE{The GSTG in Figure~\ref{fig:app:model:runningexample} captures the sequence of Actions triggered by \APPR during testing\footnote{A demo video for the running example is available online~\cite{replicability}}. They are described in the following sections, where appropriate.}

\subsubsection{Detection of the active Window}{\ }
\label{sec:active}

\JMR{2.3}{A building block of our test automation strategy is the detection of the active Window.
More precisely, \APPR should determine if a pop-up (i.e., a Dialog, an OptionsMenu, or a ContextMenu) is open on top of the active Window.} 
Neither Android nor DM2 provides such information.
To determine if a pop-up is open, \APPR
relies on the dimensions of the active Window on the screen. Indeed, if an Activity is displayed, its dimensions should match the dimensions of the screen. Otherwise, the currently displayed Window is either a Dialog, an OptionsMenu, or a ContextMenu. 
To determine which Dialog or Menu is open, \APPR identifies the Window of the EWTG with the highest portion of Widgets visualized on the screen. 
\APPR computes the ratio, $R_w$, of Widgets belonging to Window $w$ that appear in the displayed GUITree.
More precisely, \APPR computes $R_w$ for all the Dialogs and Menus that can be triggered by the current Window. The active Window is the one with the highest values for $R_w$.

\JMRCHANGE{If, due to the limitations described in Section~\ref{sec:approach:EWTG}, static analysis does not detect that, for a certain Window $w$, there is an event handler that will pop-up a certain Dialog or Menu $p$, }
\APPR may not be able to find a Dialog or Menu to be matched with the displayed GUITree. More precisely, \APPR may observe that (1) the current Window has no Dialogs and Menus associated to it in the EWTG or (2) the score computed for every Dialog and Menu of the current Window is zero. To overcome such a problem, in these scenarios, \APPR updates the EWTG to include a new pop-up Window.

\subsubsection{\APPR testing algorithm}{\ }
\label{sec:algo}

At runtime, after detecting the active Window, \APPR automatically derives the current AbstractState according to the procedure described in Section~\ref{sec:approach:app.model.metamodel}. The subsequent activities depend on the current testing phase.

The activities performed in the three phases follow the same algorithm, which is presented in Algorithm~\ref{alg:testing}. 
What differentiates the three phases are the strategies adopted to 
\JMRCHANGE{exercise the App}
and the test budget allocated
.
Line~\ref{algo:test:loop} in Algorithm~\ref{alg:testing} shows that the algorithm iterates till the test budget for the current phase is consumed (function \emph{phaseBudgetConsumed}), all the targets for the current phase have been covered (function \emph{coverageTargetsExercised} ), or it cannot further improve coverage (function \emph{stagnation}).

The iteration starts by identifying a test target (function \emph{selectTarget}, Line~\ref{algo:test:select}).
\CHANGED{The test target is either a Window or a WindowTransition.}
A new test target is identified when no target has been selected yet or the current target has already been fully exercised (Line~\ref{algo:test:checkSelectedTarget}). 
After identifying the test target, \APPR relies on the App model to identify the test target path  (Line~\ref{algo:test:path}), i.e., a sequence of Actions that makes the App render the target Window or reach the target AbstractState. 

The test target path is derived with a breadth-first traversal of the App model. The traversal starts from the current AbstractState. The traversal proceeds through both AbstractTransitions and WindowTransitions. A WindowTransition is taken only if an AbstractTransition 
is not available. The visit of the model stops when we reach the test target or all the reachable nodes are explored. 

So long as a test target is not reached (Line~\ref{algo:test:notTarget}), \APPR executes function \emph{reachTargetNode} (Line~\ref{algo:test:reachTarget}), which triggers the next Action in the test target path. 
For each Action in the test target path, we know the Window or the AbstractState to expect. 
After executing an Action, function \emph{reachTargetNode}  checks if the App is in the expected Window or AbstractState. If not, 
function \emph{reachTargetNode} flags the target as not reached, and returns to the main execution loop to look for a different path to reach the test target (Line~\ref{algo:test:target:unreachable}). 
When a target is reached (Line~\ref{algo:test:TargetReached}), \APPR exercises the target according to the Action generation strategy for the current phase (function \emph{exerciseTarget}).

\CHANGED{Finally, random exploration of the active Window might be triggered by functions 
\emph{reachTargetNode} and \emph{identifyPathToTarget} to improve the EWTG (Line~\ref{algo:test:execRandom}). This is  described in Section~\ref{sec:algo:randomExploration}.}

\CHANGEDOCT{
To regulate the allocation of the phase budget (i.e., how many Actions each function invoked by the algorithm is allowed to generate), the \APPR algorithm makes use of three budget variables: (1) \emph{reachabilityBudget},
which specifies the maximum number of Actions to be used to reach a target node,  (2) \emph{targetBudget}, 
which specifies the number of Actions to be used to exercise the target node,
(3) \emph{randomBudget}, which specifies the number of Actions to be used for random exploration.
At runtime, when counting the number of Actions performed,
we ignore Actions of type TextInput and Click on checkboxes since they generally do not trigger WindowTransitions.
The budget variables are initialized with different values depending on the current test phase. Table~\ref{table:budgets} provides an overview of the criteria adopted, which are described in details in the following paragraphs. To define budgets, a \emph{scale factor} is used to optimally distribute the test budget across phases and test targets. For example, with a test budget of five hours, we can invest more time in Phase 2 than with a test budget of one hour.}

\begin{algorithm}[tb]

\begin{algorithmic}[1]

\scriptsize

\While { ( NOT stagnation() ) AND ( NOT phaseBudgetConsumed( \emph{phaseBudget}) ) AND (NOT coverageTargetsExercised() ) } \label{algo:test:loop}
	\If { target not selected OR target already exercised OR \emph{visitBudget} exhausted}	\label{algo:test:checkSelectedTarget}
		\State selectTarget() \label{algo:test:select}
		\State identifyPathToTarget() \label{algo:test:path}
	\ElsIf { target unreachable}	\label{algo:test:target:unreachable}
		\State identifyPathToTarget() \label{algo:test:unreachable:path}
	\EndIf

\If { NOT\ targetReached() }   \label{algo:test:notTarget}

		\State reachTargetNode( \emph{reachabilityBudget} ) \label{algo:test:reachTarget}
\Else	\label{algo:test:TargetReached}
	\State exerciseTarget( \emph{targetBudget} )	\label{algo:test:execTarget}
\EndIf

	\If {additional random exploration required }   \label{algo:test:randomRequired}
		\State performRandomExploration( \emph{randomExplorationBudget} )	\label{algo:test:execRandom}
	\EndIf
	
\EndWhile

\end{algorithmic}

\caption{\APPR testing algorithm.}
\label{alg:testing}
\end{algorithm}

\begin{table}[tb]
\footnotesize
\begin{tabular}{|p{7mm}|p{1.2cm}|p{10.5cm}|}
\hline

\textbf{Phase}&\textbf{Budget}&\textbf{Strategy}\\
\hline
\multirow{2}{*}{Phase1}&Phase&Infinite, i.e., all the target windows are exercised till stagnation is detected or all the targets are covered.\\
\cline{2-3}
&Reachability&Infinite, i.e., all the paths are traversed in this phase.\\
\cline{2-3}
 &Target&Infinite, i.e., all the target Inputs are tried in this phase.\\
\cline{2-3} 
  &Random Exploration&Set to $\mathit{scaleFactor} \cdot \mathit{NumberOfActionsForActiveWindow}$. 
\JMRCHANGE{$\mathit{NumberOfActionsForActiveWindow}$ is the number of distinct Actions that can be performed in the active window; it is based on the interactive information associated to a widget (e.g., we perform a Click Action if the widget is clickable, or four Swipe Actions - one for each swipe direction - if it is scrollable).}
  In our experiments, we set $\mathit{scaleFactor}$ to $1$ for an overall test budget of one hour, to $2$ for a test budget of five hours. Random exploration is triggered by either \emph{reachTargetNode} or \emph{identifyPathToTarget}.\\
 \hline 
 \multirow{3}{*}{Phase2}&Phase&Set to $\mathit{scaleFactor} \cdot \mathit{NumberOfTargetWindows}$.\\
  \cline{2-3}
& Reachability&Set to $\mathit{scaleFactor} \cdot \mathit{actionsThreshold}$. The value is reset every time a new TargetWindow is identified.  We set $\mathit{actionsThreshold}$ to $25$.\\
 \cline{2-3}
 & Target&Set to be equal to what remains of the {ReachabilityBudget} after the target window is reached. In other words, $\mathit{ReachabilityBudget} + \mathit{TargetBudget} = \mathit{scaleFactor}*\mathit{actionsThreshold}$ \\
  \cline{2-3}
 &Random Exploration&Set to $\mathit{scaleFactor} \cdot \mathit{randomThreshold}$. Random exploration is triggered by either \emph{reachTargetNode} or \emph{identifyPathToTarget}. We set $\mathit{randomThreshold}$ equal to $\mathit{actionsThreshold}$.\\
\hline
 \multirow{3}{*}{Phase3}&Reachability&Not used in this phase.\\
 \cline{2-3}
 & Target&Set to $\mathit{scaleFactor} \cdot \mathit{actionsThreshold}$. It is reset every time a new TargetEvent is identified.\\
 \cline{2-3}
&Random Exploration&Set to $\mathit{scaleFactor} \cdot \mathit{actionsThreshold}$. Random exploration is triggered (1) to explore the related Window, (2) when the related Window cannot be reached through the identified path, (3) 
when the target Window cannot be reached through the identified path (see Section~\ref{sec:phase3}). We set $\mathit{randomThreshold}$ to $5$.\\
\hline
\end{tabular}
\caption{Strategies adopted, in different \APPR phases, to define the budget allocated to distinct test activities.}
\label{table:budgets}
\end{table}%

\subsubsection{Random exploration}{\ }
\label{sec:algo:randomExploration}

Functions 
\emph{reachTargetNode} and \emph{identifyPathToTarget} in Algorithm~\ref{alg:testing} may trigger the random exploration of the App under test to improve the EWTG. This is done when Inputs cannot be exercised and a test target cannot be reached. 

Function 
\emph{reachTargetNode} determines that an Input cannot be exercised when the associated Widget is not visible or enabled in the GUITree. It happens, for example, when the content of a \emph{NavigationDrawer} varies based on the buttons pressed in the active Window.
To make the required Widget visible, function
\emph{reachTargetNode} randomly exercises the active Window. 
This is done by iteratively and randomly selecting one widget among the ones that have been exercised less frequently in the active Window. The selected widget is then exercised according to the strategies listed in Table~\ref{table:exerciseWidgets}. 
The exploration of the active Window terminates when the desired widget is found or when the test budget for random exploration is exhausted.

Function \emph{identifyPathToTarget} may determine that it is not possible to find a path to a test target. This happens when
the EWTG does not include all the WindowTransitions,
which is due to the limitations of static analysis tools mentioned in Section~\ref{sec:approach:EWTG}.
For example, Gator does not detect the Animation design pattern~\cite{Animator:online}, which leads to a WindowTransition.
When a test target path is not found, \APPR performs a random exploration of the active Window and then resumes the execution from the beginning of the main execution loop. \APPR records of all the unreachable targets identified.

\begin{table}[tb]
\footnotesize
\begin{tabular}{|p{2cm}|p{11cm}|}
\hline

\textbf{Widget type}&\textbf{Input generation procedure}\\
\hline
Any widget& Trigger an InputEvent among the ones for which an event handler has been declared.\\
\hline
Textarea & Randomly apply one of the following: 
(1) leave it empty,
(2) reuse a string already used in the past,
(3) reuse a string already used for the same widget,
(4) reuse a string literal extracted with static analysis,
(5) use a randomly generated alphabetic string~\cite{Borges-Droidmate2-ASE-2018},
(6) use a randomly generated non alphabetic char.\\
\hline
Radio buttons and check boxes& Randomly select one of the possible options (e.g., checked/not checked, for check boxes). \\
\hline
Widgets with manual input& Randomly select one of the available InputInstances, if more than one is available, and then assign the specified value.\\
\hline
Intent& If the current activity declares an Intent, it triggers the Intent specified by the engineer.\\
\hline
\end{tabular}
\caption{\APPR Input generation procedures.}
\label{table:exerciseWidgets}
\end{table}%

\subsubsection{Phase 1}{\ }

In Phase 1, a test target is any Window with target Inputs that have not been triggered yet.
Function \emph{selectTarget} randomly selects a Window with such characteristics (Line~\ref{algo:test:select} in Algorithm~\ref{alg:testing}).

After selecting the target Window, \APPR follows the path to reach the test target.\\
\DELETE{However, to optimize test execution, \APPR exercises any test target accidentally reached before the target Window. In other words, function \emph{targetReached} returns true whenever the active Window is a test target not yet fully exercised.}

Function \emph{exerciseTarget}, first produces \emph{user-like inputs} (i.e., input values for text areas, radio buttons, and check boxes), as specified in Table~\ref{table:exerciseWidgets}.  Then it triggers an Action that exercises a randomly selected target Input.
Function \emph{exerciseTarget} keeps triggering Actions that exercise target Inputs until all the target Inputs have been exercised or another Window has been visualized. \APPR then resumes the execution of the main loop (i.e., Line~\ref{algo:test:loop} in Algorithm~\ref{alg:testing}). When the target Window is associated to hidden-handlers that can reach target methods, \APPR also performs a random exploration of the Window. If an Action triggers the execution of an hidden-handler, \APPR  introduces a corresponding WindowTransition into the EWTG.

To maximize the chances of exercising every target Input, which is the objective of Phase 1, the phase, reachability, and target budgets are infinite. 
More precisely, we try to reach every TargetWindow (infinite \emph{phaseBudget}) by trying every possible path (infinite \emph{reachabilityBudget}); furthermore, we exercise every TargetWindow with all the target Inputs (infinite \emph{targetBudget}). 

In Phase 1, we observe \emph{stagnation} when all the remaining targets 
either cannot be reached or all their target Inputs cannot be exercised. 

\JMR{3.1}{\emph{Running Example.} In Phase 1, \APPR triggers one Action for every target input. By default, Activity Diary starts by rendering the MainActivity with a predefined set of activities and no current activity being selected. Since MainActivity is a target Window, \APPR exercises it. First \APPR clicks on the edit note button (Action A1 in Figure~\ref{fig:app:model:runningexample}) and, since there is no current activity selected, Action A1 partially covers the target methods (indeed, Activity Diary does not open the EditNote Window, which will be achieved in Phase 2). Then, \APPR triggers a long click on the current activity widget (A2), which leads to an instance of the EditActivity Window. Since EditActivity is a target Window, \APPR aims to exercise it. However, EditActivity contains only hidden-handlers, not target Transitions. For this reason, \APPR performs a random exploration of the window; during the random exploration, after filling the window with random inputs,
\APPR clicks on the save button (A3), which triggers the execution of one of the updated methods (i.e., \emph{checkConstraints}) and thus \APPR introduces into the DSTG a new AbstractTransition associated to the updated method (i.e., the abstract transition between \emph{avm2-2} and \emph{as1} in Figure~\ref{fig:app:model:runningexample}, which does not have a corresponding WindowTransition in the EWTG).
After these three actions, \APPR has exercised both the MainActivity and the EditActivity and can thus move to Phase 2. In Phase 1, \APPR has thus exercised all the easy-to-reach target methods in a few steps.}

\subsubsection{Phase 2}{\ }

\JMR{2.3}{In Phase 2, we aim to maximize the coverage of those target methods that have not been fully covered. 
To this end, the target Window shall be the one that can trigger the execution of the highest number of uncovered instructions.
Also, since the AbstractState of an App might affect the reachability of a target method, 
we should give higher priority to Windows with target methods exercised 
in fewer AbstractTransitions. 
}

To achieve the above-mentioned objectives, we select as target Window the one that maximizes the score $\mathit{WS}_w$,
$$\mathit{WS}_w=\sum\limits_{m \epsilon \mathit{MT}_w}^{}{c_w \cdot u_m}$$
where $MT_w$ is the set of target methods associated to the target Inputs of Window $w$. Term $u_m$ is the number of uncovered instructions belonging to method $m$. A target Window x can thus be any Window with $\mathit{WS}_w>0$. 
\JMRCHANGE{A Window $w$ is selected as target with a probability proportional to $\mathit{WS}_w$.}

In the formula above, $c_w$ is a weight introduced to focus first on those methods that have been covered less. 
 It is the complement of the proportion of AbstractTransitions that exercise the method:
  $$c_m = 1 - \frac{\mathit{AA}_m}{\mathit{AA}}$$ where $\mathit{AA}_m$ is the number of AbstractTransitions that covered method $m$ and $\mathit{AA}$ is the number of AbstractTransitions in the App model. 

To select the test target path, \APPR identifies the AbstractState 
with the highest number of uncovered instructions belonging to interactive widgets, 
which is captured by the $\mathit{AS}_{as_w}$ score:
$$\mathit{AS}_{as_w} = \sum\limits_{m \epsilon \mathit{MT}_{as_w}}^{}{c_w\cdot u_m}$$
 where $as_w$ is an AbstractState for the Window $w$, 
 and $\mathit{MT}_{as_w}$ is the set of target methods that might be covered through $as_w$. 
The set $\mathit{MT}_{as_w}$ consists of all target methods triggered by either  (1) an Intent or (2) an InputEvent for an interactive widget in $\mathit{as_w}$.

\CHANGEDOCT{Function \emph{exerciseTarget} works in the same way as in Phase 1. However, Phase 2 differs from Phase 1 with respect to the target, phase, and reachability budgets. Indeed, to uniformly distribute the phase budget across the selected target Windows, in Phase 2, the budget for reaching a test target and exercising it is set to ``$\mathit{scaleFactor}\cdot \mathit{actionsThreshold}$'', with  $\mathit{actionsThreshold}$ representing a number of Actions that, based on preliminary experiments, is sufficient to reach a target and exercise it (in our experiments, we set its value to $25$).
In Phase 2, the \emph{phase budget} is exhausted when \APPR has exercised a number of windows that is equal to
``$\mathit{scaleFactor}\cdot $ \emph{overall number of TargetWindows}''.}
Also, a same Window can be selected as a target multiple times.
By repeatedly generating different sets of Actions for a same Window, 
\APPR covers different combinations of user-like inputs, which may include combinations that lead to the coverage of different sets of instructions.

In Phase 2, we observe \emph{stagnation} when, after exercising all the available targets, the coverage of target methods has not increased.

\JMR{3.1}{\emph{Running Example.} Phase 2 is necessary to maximize the coverage of updated methods reached through the MainActivity and the EditActivity Windows.
The target Window with the highest $\mathit{WS}_w$ score is EditActivity because some of the instructions of method \emph{checkConstraints} and all the instructions implementing the quick fix feature have not been exercised in Phase 1. 
MainActivity has a lower $\mathit{WS}_w$ score since only a few instructions of the EditNote button handler are not covered.}

\JMRCHANGE{\APPR selects the EditActivity Window as first target; at this stage, it has only one AbstractState (i.e., \emph{as2} in  in Figure~\ref{fig:app:model:runningexample}). \APPR reaches the EditActivity Window with a long click (Action A4) on the the current activity (an activity with an empty name). EditActivity includes one target AbstractTransition, the one exercising \emph{checkConstraints}.
While generating inputs to maximize the coverage of method \emph{checkConstraints}, \APPR clicks on the button that deletes the activity and brings the App back to the MainActivity (A5). From the main Activity, \APPR performs again a long click on the currentActivity widget (A6), which leads to an instance of EditActivity for the definition of a new activity where the quick fix buttons are visualized. In this case, the quick fix buttons are visualized because an activity with an empty name (the default for new activities) had ben selected and Activity Diary already contains an activity with an empty name (i.e., the one deleted by Action A5). 
Because of the presence of the two quick fix buttons, \APPR introduces a new abstract state into the DSTG (i.e., \emph{as5}).}

\JMRCHANGE{To cover the hidden-handlers of EditActivity, 
\APPR exercises the quick fix button that automatically renames the activity having a conflicting name (A7).
Finally, \APPR selects MainActivity as target 
and exercises the EditNote button (A9), which pops-up the EditNote Dialog thus covering the missing lines.
In Phase 2, \APPR successfully covered all the target methods triggered within a target Window (i.e., EditActivity), by repeatedly exercising it.}

\subsubsection{Phase 3}{\ }
\label{sec:phase3}
In Phase 3, we aim to cover those target method instructions that exhibit data dependencies from state variables defined by Windows different than the one reaching a target method. Examples include instructions that can be executed only after enabling specific options in the preferences Window of the App. 
For this reason, in Phase 3, the test target is a WindowTransition presenting associated targetMethods that remain to be fully covered. 
Also, for each WindowTransition to be tested, we need to identify a set of related Windows that should be exercised before executing it.

Function \emph{selectTarget} returns a WindowTransition belonging to a target Window selected according to the same criteria as for Phase 2, i.e., with a probability proportional to its $\mathit{WS}$ score. To minimize the effort spent in reaching target Windows, once a target Window has been selected, function \emph{selectTarget} iteratively returns each target WindowTransition belonging to it.

When a test target has been selected, in function \emph{exerciseTarget}, \APPR (1) identifies the related Window that should be exercised first, (2) identifies a path to this related Window, (3) reaches the related Window and randomly exercises it, (4) identifies a path to the closer AbstractState for the target Window in which the target WindowTransition is enabled, and (5)  reaches the identified AbstractState and triggers 
a target Input.
In Phase 3, function \emph{identifyPathToTarget} is not invoked because testing starts from the related Window.
Consistent with Phase 2, the \emph{targetBudget} is set to $\mathit{scaleFactor} \cdot \mathit{actionsThreshold}$.

Function \emph{exerciseTarget} relies on random exploration (1) to explore the related Window, (2) when the related Window cannot be reached through the identified path, (3) when the target Window cannot be reached through the identified path. 
The random exploration budget is set to $\mathit{scaleFactor} \cdot \mathit{randomThreshold}$. Since random exploration has been largely used in previous phases and to limit the time spent in related windows, in Phase 3,
we set $\mathit{randomThreshold}$ to a value lower than the one used in Phase 2 (e.g., we used $5$ in our experiments).

To identify related Windows,  we rely on
 information retrieval techniques.
We do not rely on traditional data-flow analysis~\cite{FlowDroid}
because data dependencies might be implemented in many different forms (e.g., setting a state variable in a shared object or saving a property in a key-value registry) that are not fully identified by such analysis.

Related Windows are retrieved through the computation of the term frequency (TF) and inverse document frequency (IDF) metrics, 
which are standard information retrieval metrics~\cite{IRbook}.
In the following, we discuss how we compute these metrics.

Since dependencies between Windows are due to either state variables defined in shared objects or property values in key-value registries, the executable code of Windows presenting such dependencies should share a subset of \emph{class attributes} and \emph{literals}.
For this reason, the terms used to identify dependencies are \emph{class attributes} and \emph{literals} appearing in the implementation of the methods of the App (extracted with Soot). 

We compute $\mathit{TF}(t,h)$, the frequency of the term $t$ for an Input handler $h$, as the number of methods in which the term appears, considering the handler itself and any of the methods invoked by the handler. To include only terms that characterize the functionality triggered by the WindowTransition, we consider only methods declared in the same class of the handler or in its inner or outer classes.

The frequency of the term $t$ for a WindowTransition $wt$ is computed as the sum of the term frequency for all the handlers of the Input ($HI_{wt}$) that triggers the transition, 

$$TF(t,\mathit{wt})= \sum\limits_{}^{h \in HI_{wt}} \mathit{TF}(t,h)$$

The frequency of the term $t$ for a Window $w$, $\mathit{TF}(t,\mathit{w})$, instead, is computed as the number of methods in which the term appears, considering the methods that are either declared in the class that implements the Window or in its parent class.

The inverse document frequency of a term is computed as 
$$\mathit{IDF}(t)=\mathit{log}(\frac{\mathit{total\ number\ of\ Windows}}{\mathit{number\ of\ Windows\ in\ which\ t\ appears}})$$

The related Windows for a WindowTransition $\mathit{wt}$ can be identified by computing a dependency score for every Window $\mathit{w}$ of the App, as follows

$$DS(\mathit{w},\mathit{wt})= \sum\limits_{}^{t \in T}{\mathit{NW}(t,\mathit{w})  \cdot  \mathit{NW}(t,\mathit{wt})}$$

where $T$ is the set of terms for the App, and $\mathit{NW}(t,\mathit{d})$ is the normalized term weight, which captures the extent to which a term is representative for either a Window or a WindowTransition. $\mathit{NW}(t,\mathit{d})$ is computed according to a standard formula~\cite{IRbook}:

$$\mathit{NW}(t,\mathit{d}) = \frac{TF(t,d) \cdot  IDF(t)}{EL(d)}$$

where EL(d) is the Euclidean Length of the element $d$ (i.e., a Window or a WindowTransition).
It is computed as the square root of the sum of the terms' weight squared~\cite{IRbook}.

\APPR randomly selects related Windows using the dependency score as probability distribution.

Phase 3 terminates when the overall test budget is exhausted or all the instructions of the target methods have been covered.

\JMR{3.1}{\emph{Running Example.} Phase 3 enables \APPR to test the Activity Diary feature that 
visualize the details of the current activity after a click on the activity name. This feature requires a specific configuration to be enabled in the settings page (by default, a click on the activity name terminates the activity). After selecting the MainActivity as target Window (EditActivity had been fully exercised in Phase 2), \APPR selects the SettingsActivity as related Window to be exercised first (it is reached with Actions A10 and A11 in Figure~\ref{fig:app:model:runningexample}). While exercising the SettingsActivity, it deselects the option \emph{Terminate activity by click} (Action A12). After exercising the related Window, \APPR reaches the MainActivity (A13) and triggers the Action that exercises the modified feature (i.e., click on current activity widget - A14). Phase 3 thus enabled \APPR to test the updated feature (i.e., visualize the current activity's details after a click) in a few steps, which is unlikely with random-based, state-of-the-art approaches (see Section~\ref{appendix:complementarities} for additional examples).}

\section{Empirical Evaluation}
\label{sec:empirical}

The objective of the empirical evaluation is to compare \APPR with state-of-the-art approaches in terms of cost-effectiveness. It is motivated by our need to achieve high test coverage (effectiveness) while enabling the verification of test results within an acceptable budget (cost) in a CI context.

When a new release is ready for testing, a test automation technique is \emph{effective} when it enables engineers to  
verify updated features in the App under test automatically;
more precisely, when it extensively exercises updated methods and their instructions. Measuring the effectiveness of App testing techniques in terms of method and instruction coverage is common practice~\cite{Choudhary-AutomatedTestInputGeneration-ASE-2015}. 
Although engineers may aim to exercise all the methods that could be impacted by the changes (e.g., the ones identified by means of change impact analysis as mentioned in Section~\ref{sec:approach}), in our empirical evaluation, we focus on updated methods since exercising them is the minimum requirement of any testing criterion targeting software updates.
 
What we refer to as \emph{cost} comes in two forms, (1) test execution time and (2) human effort \JMR{3.8}{, which we define as the time spent by a trained engineer to execute the manual tasks required by a test automation technique}.
In general, \emph{test execution time} does not necessarily need to be minimized but it should be practical. For example, we expect a test budget of one hour to be practical in a continuous integration context, while a budget of five hours might be acceptable when testing overnight.

\emph{Human effort}, in our context, is mostly driven from the absence 
of a solution to automate test oracles, even in the presence of automated test input generation.
\JMRTWO{2.1}{More precisely, it is not possible to define automated oracles working with any feasible test input; consequently, the evaluation of the correctness of App outputs should, in general, be done manually.}
As described in Section~\ref{sec:approach}, to address the oracle problem in the context of App updates, engineers can rely on two complementary state-of-the-art approaches that respectively address regression failures and failures in newly implemented, repaired, or modified features. 
To discover regression failures, engineers can replicate, on a previous App version, the test input sequences generated  for the updated App. With \APPR, input sequences correspond to the sequences of Actions generated by \APPR to test an App.
\JMRTWO{2.1}{In such context they address the oracle definition problem by verifying that the outputs generated by the two App versions match.}
To discover failures in new, repaired, and modified features engineers should visualize the screenshots of the Windows or the GUITrees generated for the provided inputs. 
The effort in doing so can potentially be reduced through crowdsourcing. 
\JMRTWO{2.1}{In this context, the oracle definition problem remains unaddressed; a manual oracle can be defined (i.e., a human can decide if an output is correct based on the App specifications) but oracle evaluation remains manual and, therefore, expensive.}
Regardless of the situation and context, human effort, given the scarcity of qualified human resources, should in general be minimized.

We assume that human effort is proportional to the number of inputs generated by test automation.
Indeed, 
to overcome execution errors due to changes in the GUI and be able to replicate test input sequences in a different App version, engineers may need to manually repair the sequence of inputs (e.g., by changing the ID of a widget to be clicked).
A large number of inputs may thus lead to numerous repair operations and make test oracle automation infeasible\footnote{Related work reports that repairing a single test input takes 15 minutes, on average~\cite{Hammoudi:2016}.}. 
\JMR{3.4}{Also, the number of outputs that shall be visually inspected by an engineer  is proportional to  the number of inputs; indeed, an engineer shall visualize the GUITree or the screenshots of the active Window rendered after each Action (see Section~\ref{sec:approach}).
The effort required to inspect a Window or a GUITree depends on the specific output generated by the App and the specification of the App.
For example, in our running example (Figure~\ref{fig:app:model:runningexample}), it is reasonable to believe that it is simpler to inspect the Window rendered after Action A4, which contains only a text box and a color selector, rather than inspecting the output of Action A10, which leads to a menu dialog with many textual items. 
However, the specifications of the App may also impact the required human effort. 
For example, the color visualized when editing an activity (i.e., A4) depends on the color previously selected for the same activity, which requires the engineer to verify such consistency based on execution history. The menu dialog visualized with Action 10, instead, does not depend on execution history and thus may require less human effort for verification.
For these reasons, the cost for the visual inspection of results can be estimated only through an experiment involving human subjects. 
However, since our objective is to \emph{compare the human effort required by different test generation techniques},
not to estimate the cost of failure detection approaches,
we can rely on the number of inputs.
Indeed, under the assumption that the distribution of the effort for visually inspecting an output is similar for different test automation techniques, we can conclude that 
techniques leading to a larger number of inputs may lead 
to proportionally increased costs for output verification.
For our experiments, 
such assumption holds
because all the considered approaches exercise a common subset of target methods and we expect a same method to lead to similar outputs across different executions. 
Indeed, we have observed that more than 50\% of the target methods exercised in our experiments are covered by all the testing techniques and more than 80\% of them are covered by at least two techniques (see Section~\ref{appendix:complementarities}).
}

Our research questions are organised according to the two cost measures above. They evaluate the extent to which we have achieved the objectives mentioned in the introduction; RQ1 addresses O2, while RQ2 addresses O1. Further, beyond comparisons, RQ3 aims to characterize how \APPR and state-of-the-art approaches complement each other.

\begin{itemize}
\item[RQ1] \emph{Can \APPR reduce the human effort required for testing Apps, compared to state-of-the-art approaches?} 
We aim to determine if the number of inputs generated by \APPR is significantly lower than the number of inputs generated by  state-of-the-art approaches, for a same execution time budget. A lower number of  inputs makes test automation more widely applicable in practice since it reduces the effort related to the definition of test oracles and repair of input sequences. Also, to determine if the effort of using \APPR is justified by practical benefits, we aim to verify if  \APPR provides higher effectiveness per unit of effort than state-of-the-art approaches.

\item[RQ2] \emph{Can \APPR effectively test Apps within practical time budgets, compared to state-of-the-art approaches?} 
This research question aims to determine if \APPR performs significantly better than state-of-the-art approaches in terms of coverage of updated methods and their instructions, for a same execution time budget.

\item[RQ3] \JMR{3.11}{\emph{Is there any difference in the functionalities that are automatically exercised across test automation approaches?} We aim to determine if the testing approaches considered in our empirical evaluation are complementary and to which extent. Specifically, we aim to determine if there are differences in the inputs triggered by the different approaches (e.g., input sequence length, widgets being exercised, or program states being reached) that lead to a diverse and complementary set of functionalities being exercised.}

\end{itemize}

A replicability package is made available online~\cite{replicability}.

\subsection{Study subjects}

To perform our experiments, we considered as experimental subjects a number of Apps available on the Android Play Store that are highly popular (i.e., more than 100,000 downloads, on average) and that were used for validation in recent papers reporting on related techniques, i.e., APE~\cite{Gu:APE:ICSE:2019} and DM2~\cite{Borges-Droidmate2-ASE-2018}.
We considered only the Apps that can be executed on the recent Android simulator version supported by our toolset (i.e., Android API Level above 23). 
For each App, we considered the latest 10 released versions at the time of running our experiments (hereafter referred to as V0, ..., V9), when available\footnote{Preliminary experiments to setup \APPR had been conducted with Jamendo, a music streaming App~\cite{Jamendo}.
}. In Table~\ref{table:caseStudies}, for each version of each subject, we report the overall number of methods, the number of updated methods, the overall number of bytecode instructions, and the number of bytecode instructions belonging to the updated methods. 
In total, we downloaded and processed 83 App versions, 74 being updated versions.

\begin{table}[tb]
\caption{Overview of subject systems.}
\label{table:caseStudies}
\footnotesize
\begin{tabular}{|p{2mm}|p{18mm}|p{3mm}|
>{\raggedleft\arraybackslash}p{10mm}@{\hspace{1pt}}|
@{\hspace{1pt}}>{\raggedleft\arraybackslash}p{8mm}@{\hspace{1pt}}|
@{\hspace{1pt}}>{\raggedleft\arraybackslash}p{8mm}@{\hspace{1pt}}|
@{\hspace{1pt}}>{\raggedleft\arraybackslash}p{8mm}@{\hspace{1pt}}|
@{\hspace{1pt}}>{\raggedleft\arraybackslash}p{8mm}@{\hspace{1pt}}|
@{\hspace{1pt}}>{\raggedleft\arraybackslash}p{8mm}@{\hspace{1pt}}|
@{\hspace{1pt}}>{\raggedleft\arraybackslash}p{8mm}@{\hspace{1pt}}|
@{\hspace{1pt}}>{\raggedleft\arraybackslash}p{8mm}@{\hspace{1pt}}|
@{\hspace{1pt}}>{\raggedleft\arraybackslash}p{8mm}@{\hspace{1pt}}|
@{\hspace{1pt}}>{\raggedleft\arraybackslash}p{8mm}@{\hspace{1pt}}|
@{\hspace{1pt}}>{\raggedleft\arraybackslash}p{8mm}@{\hspace{1pt}}|
}
\hline
\multicolumn{2}{|c|}{\textbf{Subject App}}&\multicolumn{2}{c|}{\textbf{Details: (V0)}}
&\textbf{V1}
&\textbf{V2}
&\textbf{V3}
&\textbf{V4}
&\textbf{V5}
&\textbf{V6}
&\textbf{V7}
&\textbf{V8}
&\textbf{V9}
&\textbf{V10}
\\

\hline

\hline
1&Wikipedia								
&\textbf{V}&110& 144& 146& 159& 190& 198& 10239& 10263& 10264& 10269 &\\
&&\textbf{AM}&3767& 	5009& 	5646& 	6435& 	6943& 	7477& 	8814& 	8751& 	8759& 	8793&\\
&&\textbf{UM}&3767& 	446& 	195& 	108& 	370& 	292& 	1430& 	535& 	13& 	94&\\
&&\textbf{AI}&32208& 	38913& 	43753& 	48761& 	51147& 	54759& 	68207& 	69471& 	69533& 	69856&\\
&&\textbf{UI}&32208& 	11000& 	4157& 	2441& 	6606	& 6345& 	24536& 	12724& 	281& 	2698&\\

\hline
2&Activity Diary
&\textbf{V}&105& 	111& 	115& 	117*& 	118& 	122& 	125& 	130& 	131& 	134 &\\
&&\textbf{AM}&260& 	333& 	333& 	333& 	333& 	450& 	479& 	540& 	540& 	659&\\
&&\textbf{UM}&260& 	18& 	3& 	7& 	12& 	117& 	39& 	28& 	1& 	49&\\
&&\textbf{AI}&3667& 	4832& 	4831& 	4834& 	4880& 	6613& 	7052& 	8247& 	8251& 	10622&\\
&&\textbf{UI}&3667& 	558& 	21& 	295& 	599& 	3393& 	2027& 	1535& 	15& 	2459&\\
\hline
3&File Manager
&\textbf{V}&44& 	53& 	77& 	79& 	82& 	84 &\multicolumn{5}{c|}{}\\
&&\textbf{AM}&2042& 	2132& 	3422& 	3430& 	3648& 	3648&\multicolumn{5}{c|}{}\\
&&\textbf{UM}&2042& 	306& 	415& 	11& 	644& 	2&\multicolumn{5}{c|}{}\\
&&\textbf{AI}&34389& 	34931& 	48241& 	48294& 	51755& 	51789&\multicolumn{5}{c|}{}\\
&&\textbf{UI}&34389& 	14510& 	13744& 	703& 	24960& 	143&\multicolumn{5}{c|}{}\\
\hline
4&Nuzzel
&\textbf{V}&302& 	303&    318*& 	323& 	325& 	328& 	329& 	330& 	331& 	333&     334\\
&&\textbf{AM}&4223& 	4220& 	4524& 	4498& 	4527& 	4650& 	4771& 	4832& 	4833& 	4833&    4834\\
&&\textbf{UM}&4223& 	8& 	717& 	75& 	33& 	41& 	21& 	21& 	1& 	1&   1\\
&&\textbf{AI}&40522& 	40449& 	43309& 	43083& 	43403& 	44234& 	45331& 	45908& 	45913& 	45916&   45940\\
&&\textbf{UI}&40522& 	151& 	18335& 	2952& 	1593& 	1990& 	647& 	1378& 	35& 	69&  45\\
\hline
5&Yahoo weather
&\textbf{V}&1.16.0& 	1.16.1& 	1.16.2& 	1.17.3& 	1.18.1& 	1.19.1& 	1.20.1& 	1.20.3& 	1.20.5& 	1.20.7 &\\
&&\textbf{AM}&2932&  2904&  2904& 	2630& 	3105& 	3109& 	3178& 	3255& 	3255& 	3303 &\\
&&\textbf{UM}&2932&  5&   4& 	243& 	10& 	16& 	118& 	101& 	12& 	9 &\\
&&\textbf{AI}&38015&     37867&     37857& 	34220& 	38219& 	38211& 	39086& 	39439& 	39439& 	39462&\\
&&\textbf{UI}&38015&     417&     272& 	10198& 	857& 	588& 	4295& 	3842& 	689& 	961&\\
\hline
6&Wikihow
&\textbf{V}&2.7.3& 		2.8.0& 	2.8.1& 	2.8.3& 	2.9.1& 	2.9.2& 	2.9.3&\multicolumn{4}{c|}{}\\
&&\textbf{AM}&333& 	333& 	333& 	333& 	325& 	322& 	319&\multicolumn{4}{c|}{}\\
&&\textbf{UM}&333& 	111& 	1& 	1& 	65& 	4& 	18&\multicolumn{4}{c|}{}\\
&&\textbf{AI}&3704& 	3992& 	3941& 	3944& 	3808& 	3761& 	3657&\multicolumn{4}{c|}{}\\
&&\textbf{UI}&3704& 	2279& 	39& 	42& 	1370& 	93& 	543&\multicolumn{4}{c|}{}\\

\hline
7&BBC Mobile
&\textbf{V}&5.1.0& 		5.10.0& 	5.11.0& 	5.12.0& 	5.13.0& 	5.4.0& 	5.5.0& 	5.6.0& 	5.8.1& 	5.9.0&\\
&&\textbf{AM}&10706& 	8724& 	8792& 	8902& 	8926& 	9945& 	10380& 	10696& 	10200& 	8939&\\
&&\textbf{UM}&10706& 	649& 	27& 	44& 	25& 	603& 	242& 	553& 	77& 	95&\\
&&\textbf{AI}&76649& 	61604& 	62078& 	62937& 	63232& 	71053& 	73082& 	73950& 	72439& 	61618&\\
&&\textbf{UI}&76649& 	11182& 	1557& 	2288& 	2101& 	10637& 	6324& 	9484& 	1638& 	3274&\\

\hline
8&VLC player
&\textbf{V}&3.1.4& 		3.1.5& 	3.1.7& 	3.2.12& 	3.2.2& 	3.2.3& 	3.2.6& 	3.2.7& 	3.2.9&\multicolumn{2}{c|}{}\\
&&\textbf{AM}&6796& 	6843& 	6854& 	8681& 	8544& 	8551& 	8621& 	8641& 	8676&\multicolumn{2}{c|}{}\\
&&\textbf{UM}&6796& 	672& 	26& 	3& 	149& 	13& 	51& 	33& 	42&\multicolumn{2}{c|}{}\\
&&\textbf{AI}&86266& 	87560& 	87886& 	117207& 	115344& 	115412& 	116409& 	116647& 	117071&\multicolumn{2}{c|}{}\\
&&\textbf{UI}&86266& 	34086& 	1522& 	150& 	9611& 	1163& 	3527& 	1961& 	3010&\multicolumn{2}{c|}{}\\

\hline
9&City-mapper
&\textbf{V}&9.1& 		9.2& 	9.3& 	9.4& 	9.5& 	9.6& 	9.7& 	9.8& 	9.9& 	10.0 &\\
&&\textbf{AM}&9629& 	9499& 	9599& 	9491& 	9602& 	9761& 	9868& 	9929& 	9884& 	10050 &\\
&&\textbf{UM}&9629& 	51& 	37& 	55& 	73& 	119& 	76& 	73& 	12& 	69 &\\
&&\textbf{AI}&155117& 	154086& 	157036& 	153200& 	155950& 	161914& 	164267& 	165747& 	165747& 	163303 &\\
&&\textbf{UI}&155117& 	2726& 	2286& 	2075& 	3160& 	6262& 	2756& 	2340& 	1372& 	3775 &\\

\hline
\end{tabular}
\footnotesize{\textbf{Legend:} V: ID of the version under test. AM: number of methods implemented in V (All Methods). UM: number of Updated Methods in V. AI: number of instructions in V (All Instructions). UI: number of instructions belonging to updated methods. An asterisk (*) is used to indicate not tested versions.}
\end{table}%

For all subjects, we treat version V0 as the first released version. 
The number of updated methods ranges from one (e.g., version V8 of subject App 2) to 1430 (e.g., version V6 of subject App 1), thus being representative of a wide variety of release scenarios (i.e., from simple bug fixes to major releases). The number of bytecode instructions, ranging from 3667 to 165747, shows that the considered Apps have varied degrees of complexity. Further, the growing number of instructions belonging to the different versions of the Apps (e.g., from 32208 to 69856 in the case of Wikipedia) suggests that they are representative of typical Apps where features are incrementally introduced at every release, thus further motivating the adoption of \APPR.

\subsection{Experimental setup}
\label{sec:empirical:setup}

\begin{table}[tb]
\footnotesize
\caption{Case studies with manual inputs.}
\label{table:manualInputs}
\begin{tabular}{
|@{\hspace{1pt}}p{1.5cm}@{\hspace{1pt}}|
p{6cm}|
p{1cm}|
p{1cm}|
p{1cm}|
}
\hline
\textbf{Case study}&
\textbf{Feature tested}&
\multicolumn{1}{c|}{\textbf{\# Windows}}&
\multicolumn{1}{c|}{\textbf{\# Data fields}}&
\multicolumn{1}{c|}{\textbf{\# Instances}}\\
\hline
\multirow{2}{*}{Wikipedia}&
Log-in functionality&1&2&1\\
\cline{2-5}
&Creation of a new account&
1&
4&
14								
\\
\hline
\multirow{2}{*}{VLC}
&
Play a video stream using a URL
&
1
&
1
&
2
\\
\cline{2-5}
&
Populate the library with all the videos on the device
&
1
&
1
&
1
\\
\hline
Nuzzel
&
Request an e-mail newsletter
&
1
&
1
&
1
\\
\hline
\end{tabular}
\end{table}%

In our experiments, we compare \APPR with three state-of-the art test automation tools, i.e., DM2, Monkey, and APE. These tools do not specifically target updated methods, but simply aim to maximize the coverage of the whole App.

DM2 has been selected because it is the framework on top of which we implemented \APPR. We configured it to use the biased-random testing strategy, which matches the random input selection strategy of \APPR. The comparison with DM2 enables us to determine if the additional analyses performed by \APPR (i.e., static program analysis, adaptable state abstraction function, and inputs generation based on information retrieval) contribute to generating better results than a simpler solution based on dynamic program analysis only.

Monkey is a program that runs on the Android emulator and generates pseudo-random streams of events. 
It is used as baseline for App testing approaches and surprisingly fares better in many benchmarks~\cite{Wang:EmpStudy:2018}. The reason is that the time saved by not processing the App GUITree can be used to further explore the App state space. 

APE is a state-of-the-art App testing toolset that overcomes existing approaches thanks to an adaptable state abstraction function (see Section~\ref{sec:related}).

In our experiments, we considered two possible execution scenarios, 
with respectively test budgets of one hour (a practical choice in a continuous integration context) and five hours (a reasonable choice for overnight execution).

To use \APPR, for three subjects, we specified a set of manual inputs necessary to exercise the primary features of the Apps (e.g., to login and use the App). 
Support for manual inputs is a necessary feature of test automation tools because Apps often require domain-specific information that cannot be derived automatically (e.g., login data).
Table~\ref{table:manualInputs} provides a summary of the manual inputs defined; for each, we provide a description, and the number of windows in which the manual input might be triggered.
For Wikipedia, we configured two manual inputs, one with the information for creating a new account, another one with information to log-in. In the case of VLC, we provide the URL of a stream to be reproduced and the indication of a checkbox to be checked in order to populate the library with device data (otherwise, no content can be played and testing is limited). Regarding Nuzzel, we provide the e-mail address to receive a newsletter.
The effort required to define manual inputs is limited; indeed, for each input, we have specified a single Window where it is applied, between one and four data fields, and a very limited number of input instances, \CHANGEDOCT{that is, one for every tested feature except for the creation of a new Wikipedia account and the playback of a video stream with VLC. When creating a new account, it is necessary to specify a larger set of inputs to exercise the feature under test multiple times; indeed, a same e-mail address cannot be shared by distinct Wikipedia accounts. As for VLC, since one of its main features is to play video streams, it makes sense to test it with both a working and a corrupted video stream. This example illustrates that the effort required to specify manual inputs is negligible.}

To account for randomness, we executed each tool against each updated version 10 times. 
\CHANGEDOCT{We report results for 72 of the 74 versions available since, for two App versions of Nuzzel and Activity Diary (indicated with an asterisk in Table~\ref{table:caseStudies}), it was not possible to execute all the testing tools. More precisely, 
for version 318 of Nuzzel, 
the App starts but gets stuck in the first Activity,
while version 117 of ActivityDiary can be tested only with \APPR and DM2, but not with Monkey and APE.}
 In total, we executed 5760 test sessions (4 tools $\times$ 72 versions $\times$ 10 runs $\times$ 2 test budgets) for a total of 17280 test execution hours. To perform our experiments, we relied on the Grid 5000 infrastructure~\cite{grid5000,grid5000Web},
 which provides access to 800 compute-nodes grouped into homogeneous clusters. We rely on nodes with 
16x2.1 GHz and 18x2.2 GHz CPU cores.

In the following sections, we analyze differences in results using a non-parametric Mann Whitney U-test (with $\alpha$ = 0.05).
\JMR{3.10}{Particularly, we discuss the p-values computed by the Mann Whitney U-test to reject null hypotheses stating that there is no difference between \APPR and each of the state-of-the-art solutions, a common practice in software testing research~\cite{Briand:Statistical:2011,Briand:Hitchhiker:2014}.} We discuss effect size based on Vargha and Delaney’s $A_{12}$ statistics~\cite{VDA}, a non-parametric effect size measure. The $A_{12}$ statistic, given observations (e.g., code coverage, in our context) obtained with two treatments X and Y (testing tools, in our context), indicates the probability that treatment X leads to higher values than treatment Y. Based on $A_{12}$, effect size is considered small when $0.56 \le A_{12} < 0.64$, medium when $0.64 \le A_{12} < 0.71$, large when $A_{12} \ge 0.71$. Otherwise the two populations are considered equivalent~\cite{VDA}.
\CHANGEDOCT{In contrast, when $A_{12}$ is below $0.50$, it is more likely that treatment X leads to lower values than treatment Y. Symmetrically to the case above, effect size is small when $0.36 < A_{12} \le 0.44$, medium when $0.29 < A_{12} \le 0.36$, large when $A_{12} \le 0.29$.}

\subsection{RQ1: Human Effort}

\subsubsection{Experimental setup}

\JMRCHANGE{In line with the discussion concerning human effort reported above,} to address RQ1, we count the number of inputs generated by each testing tool, for each test execution run. For DM2 and \APPR, we rely on the CSV file generated by the ActionTrace component of DM2, which reports all the inputs triggered during testing. For Monkey and APE, we record the number of test inputs reported by the tool at the end of execution.  

\paragraph{Metrics} For each subject App, we compare \emph{distributions of the number of inputs generated across tools}.  \JMR{3.6}{We also analyze the \emph{target instructions/input ratio}, that is, the ratio between the number of target instructions (i.e., instructions belonging to updated methods) that are automatically exercised and the number of inputs triggered by the test automation tool.}
This ratio captures how useful it is for a software engineer, on average, to invest time in repairing a single input of the test sequence or verifying the output produced by an input. For example, a target instructions/input ratio of five indicates that, for every input, the test automation approach exercises, on average, five instructions belonging to updated methods.

To answer positively this research question, \APPR, compared to other tools, should generate less test inputs and have the highest  \emph{target instructions/input ratio}.

\subsubsection{Results}

\begin{figure*}
\begin{minipage}{.45\textwidth}
	\includegraphics[width=7cm]{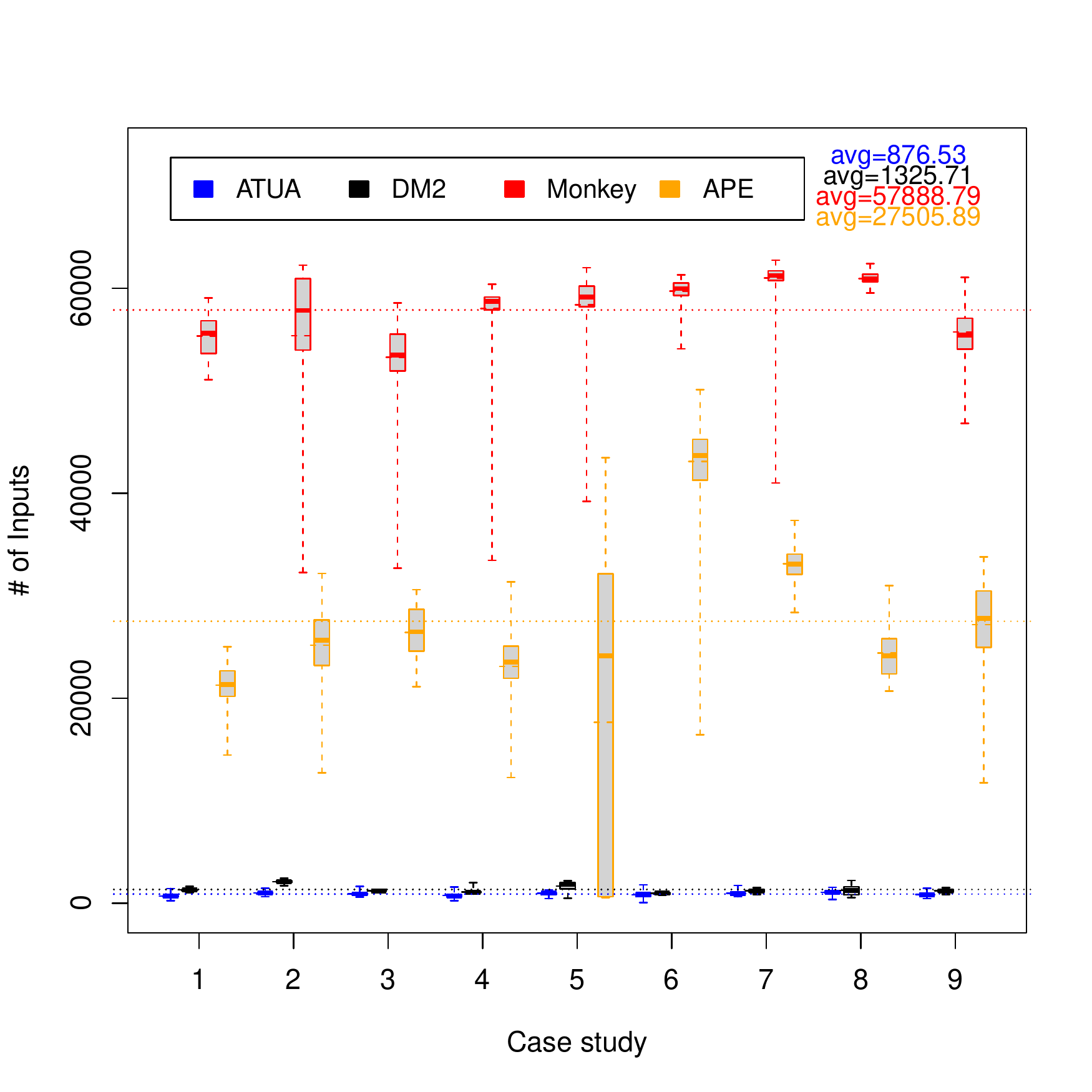}
      \caption{Number of inputs generated (budget=1 hour).}
      \label{fig:RQ3:Inputs:1}
\end{minipage}\hspace{5mm}
\begin{minipage}{.45\textwidth}
	\includegraphics[width=7cm]{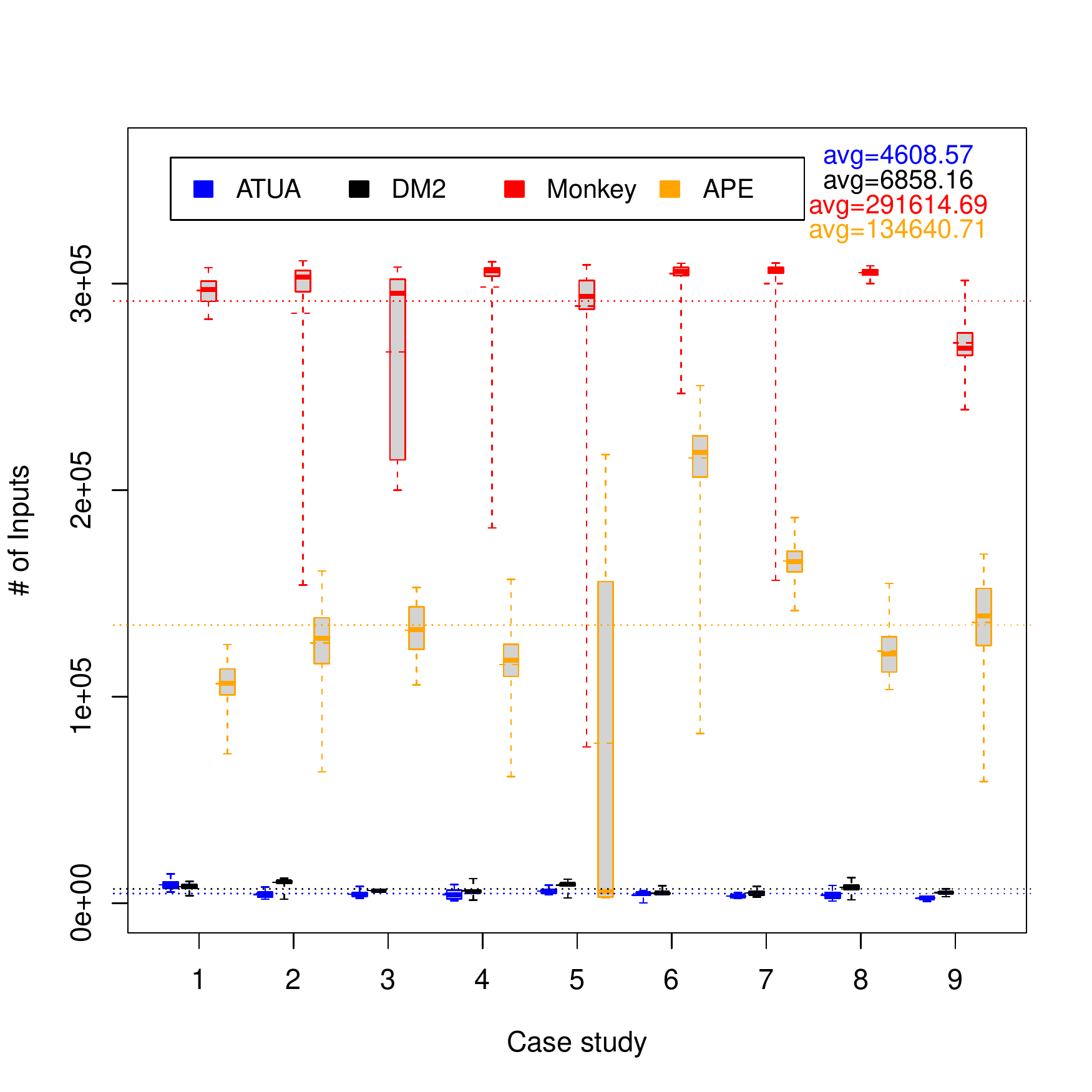}
      \caption{Number of inputs generated (budget=5 hours).}
      \label{fig:RQ3:Inputs:5}
\end{minipage}
\end{figure*}

Figures~\ref{fig:RQ3:Inputs:1} and~\ref{fig:RQ3:Inputs:5} show boxplots capturing the number of inputs generated by each approach in every run, for every subject App, for the two distinct test budgets considered (i.e., one hour and five hours). 
Please note that, in all the boxplots presented in this paper: (1) horizontal dashed lines show the average across the data points of the boxplot (i.e., average for the subject App),
(2) horizontal dotted lines traversing the whole chart show the average across all the runs, (3)
whiskers are used to report min and max values across runs for all versions.

Figures~\ref{fig:RQ3:Inputs:1} and~\ref{fig:RQ3:Inputs:5} shows that \APPR generates, on average, the lowest number of inputs: \FIXME{876.53} for one hour, \FIXME{4608.57} for five hours. 
\APPR is thus the most suitable approach to minimize test automation effort. Monkey generates the largest number of inputs (i.e., \FIXME{57888.79} for  one hour, \FIXME{291614.69} for five hours) because it does not invest any of the time budget into analyzing execution data but simply generates purely random inputs. APE relies on Monkey to generate inputs; however, APE generates less inputs than Monkey (i.e., \FIXME{27505.89} for one hour, \FIXME{134640.71} for five hours) because it spends time refining the state abstraction function (see Section~\ref{sec:related}). 
Finally, DM2 generates a number of inputs (i.e., \FIXME{1325.71} for one hour, \FIXME{6858.16} for five hours) that is closer to those generated by \APPR. This is mostly due to the fact that both approaches are model-based and share the same dynamic analysis infrastructure; however, on average, \APPR generates less inputs because it invests more of the time budget into the analysis of run-time data. 

\begin{figure*}
\begin{minipage}{.45\textwidth}
	\includegraphics[width=7cm]{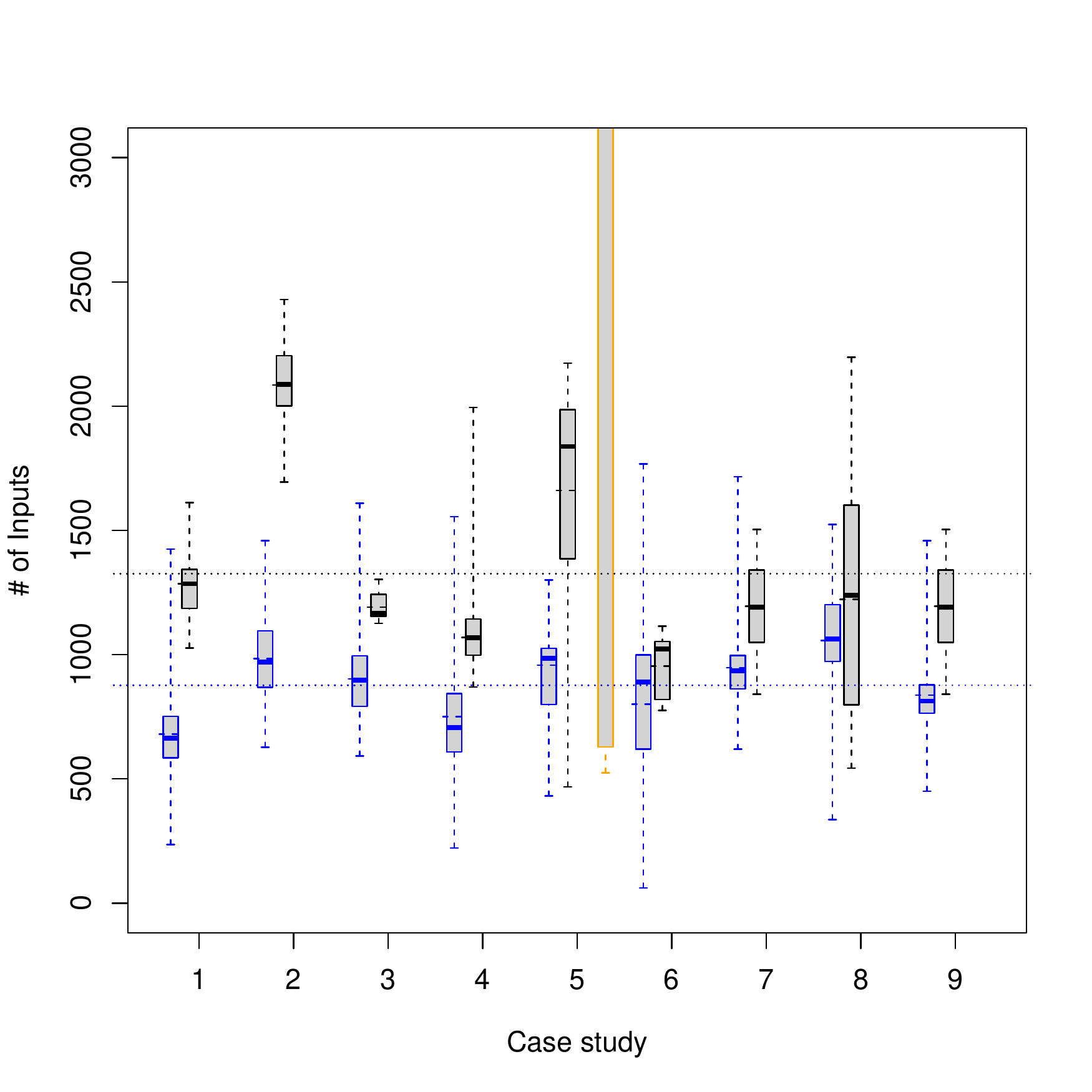}
      \caption{Number of inputs generated (budget=1 hour). Zoom on \APPR and DM2 data.}
      \label{fig:RQ3:Inputs:1:Zoom}
\end{minipage}\hspace{5mm}
\begin{minipage}{.45\textwidth}
	\includegraphics[width=7cm]{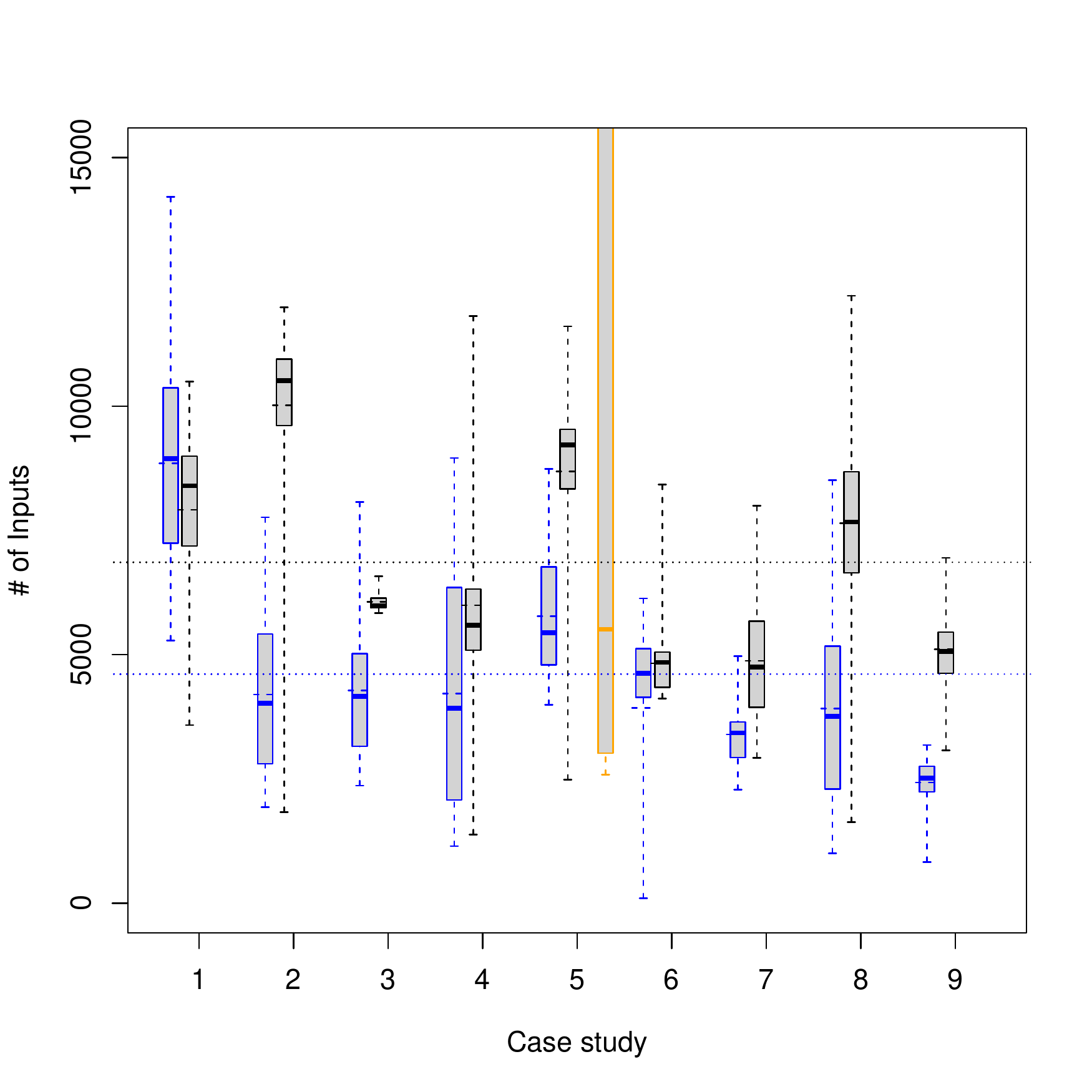}
      \caption{Number of inputs generated (budget=5 hours). Zoom on \APPR and DM2 data.}
      \label{fig:RQ3:Inputs:5:Zoom}
\end{minipage}
\end{figure*}

Figures~\ref{fig:RQ3:Inputs:1:Zoom} and~\ref{fig:RQ3:Inputs:5:Zoom} present the same boxplots as Figures~\ref{fig:RQ3:Inputs:1} and~\ref{fig:RQ3:Inputs:5} but zoom in on \APPR and DM2 data to highlight their differences. 
Table~\ref{table:effectSize:inputs} reports the p-value and $A_{12}$ statistics obtained with the Mann Whitney U-test and the Vargha and Delaney’s  method, respectively. Recall that we aim to minimize the number of inputs and are thus interested in effect sizes below 0.46, i.e., the probability that \APPR generates a number of inputs higher than another approach should be below 0.46, ideally close to zero.

For a budget of one hour, for all the subject Apps, \APPR generates, on average, less inputs than DM2. Differences are statistically significant 
\JMR{3.10}{(i.e., we reject the null hypothesis that \emph{there is no difference in the number of inputs generated by \APPR and the state-of-the-art approach})} 
and effect size is always in favor of \APPR  and large for seven of the nine subjects.
Differences with Monkey and APE are always significant and effect size is always large except for subject 5 (YahooWeather), in which APE does not interact properly with one App version, apparently because of a bug in APE.

With a budget of five hours, \APPR also generates, on average, less inputs than DM2 across subjects. But in the case of Wikipedia, effect size is not in favor of \APPR (i.e., it is likely to generate more inputs than DM2). 
However, this is due to a bug in DM2 rather than a feature; indeed, in the presence of WebViews, the communication between the DM2 device daemon and the DM2 client are delayed, thus reducing the number of inputs being generated. In the case of \APPR, which is built on top of DM2, this problem is less evident because such communication is triggered less frequently.
The differences between the number of inputs generated by \APPR and the ones generated by APE and Monkey are always statistically significant with a large effect size in favor of \APPR (except for YahooWeather, as already discussed above). 

\begin{table}[tb]
\footnotesize
\caption{Statistical significance and effect size for Figure~\ref{fig:RQ3:Inputs:1} and ~\ref{fig:RQ3:Inputs:5}.}
\label{table:effectSize:inputs}
\begin{tabular}{
|>{\raggedleft\arraybackslash}p{2mm}@{\hspace{1pt}}|
>{\raggedleft\arraybackslash}p{6mm}@{\hspace{1pt}}|
>{\raggedleft\arraybackslash}p{6mm}@{\hspace{1pt}}|
>{\raggedleft\arraybackslash}p{6mm}@{\hspace{1pt}}|
>{\raggedleft\arraybackslash}p{5mm}@{\hspace{1pt}}|
>{\raggedleft\arraybackslash}p{5mm}@{\hspace{1pt}}|
>{\raggedleft\arraybackslash}p{5mm}@{\hspace{1pt}}|
>{\raggedleft\arraybackslash}p{6mm}@{\hspace{1pt}}|
>{\raggedleft\arraybackslash}p{6mm}@{\hspace{1pt}}|
>{\raggedleft\arraybackslash}p{6mm}@{\hspace{1pt}}|
>{\raggedleft\arraybackslash}p{5mm}@{\hspace{1pt}}|
>{\raggedleft\arraybackslash}p{5mm}@{\hspace{1pt}}|
>{\raggedleft\arraybackslash}p{5mm}@{\hspace{1pt}}|
}
\hline
&
\multicolumn{6}{c|}{\textbf{1 hour budget}}&
\multicolumn{6}{c|}{\textbf{5 hours budget}}\\
&
\multicolumn{3}{c|}{\textbf{p-value}}&
\multicolumn{3}{c|}{\textbf{$A_{12}$}}&
\multicolumn{3}{c|}{\textbf{p-value}}&
\multicolumn{3}{c|}{\textbf{$A_{12}$}}\\
\cline{2-13}
\textbf{S}&
D&M&A&
D&M&A&
D&M&A&
D&M&A\\
\hline
1&<0.05&<0.05&<0.05&.018&.000&.000&<0.05&<0.05&<0.05&\R{.630}&.000&.000\\
2&<0.05&<0.05&<0.05&.000&.000&.000&<0.05&<0.05&<0.05&.020&.000&.000\\
3&<0.05&<0.05&<0.05&.039&.000&.000&<0.05&<0.05&<0.05&.093&.000&.000\\
4&<0.05&<0.05&<0.05&.099&.000&.000&<0.05&<0.05&<0.05&.278&.000&.000\\
5&<0.05&<0.05&<0.05&.074&.000&.397&<0.05&<0.05&\R{0.81}&.090&.000&\R{.490}\\
6&<0.05&<0.05&<0.05&.360&.000&.000&\R{0.15}&<0.05&<0.05&.425&.000&.000\\
7&<0.05&<0.05&<0.05&.117&.000&.000&<0.05&<0.05&<0.05&.123&.000&.000\\
8&<0.05&<0.05&<0.05&.405&.000&.000&<0.05&<0.05&<0.05&.075&.000&.000\\
9&<0.05&<0.05&<0.05&\R{.048}&.000&.000&<0.05&<0.05&<0.05&.001&.000&.000\\
\hline
\end{tabular}

\textbf{Legend}: \emph{S}, subject. \emph{D}, comparison with DM2, \emph{M}, Monkey, \emph{A}, APE. We underline the few cases in which statistics indicate that \APPR shows no significant difference (i.e., \emph{p-value} $\ge 0.05$) or no higher chances of generating less instructions ($A_{12} > 0.44$) than state-of-the-art approaches.
\end{table}%

\begin{figure*}
\begin{minipage}{.45\textwidth}
	\includegraphics[width=7cm]{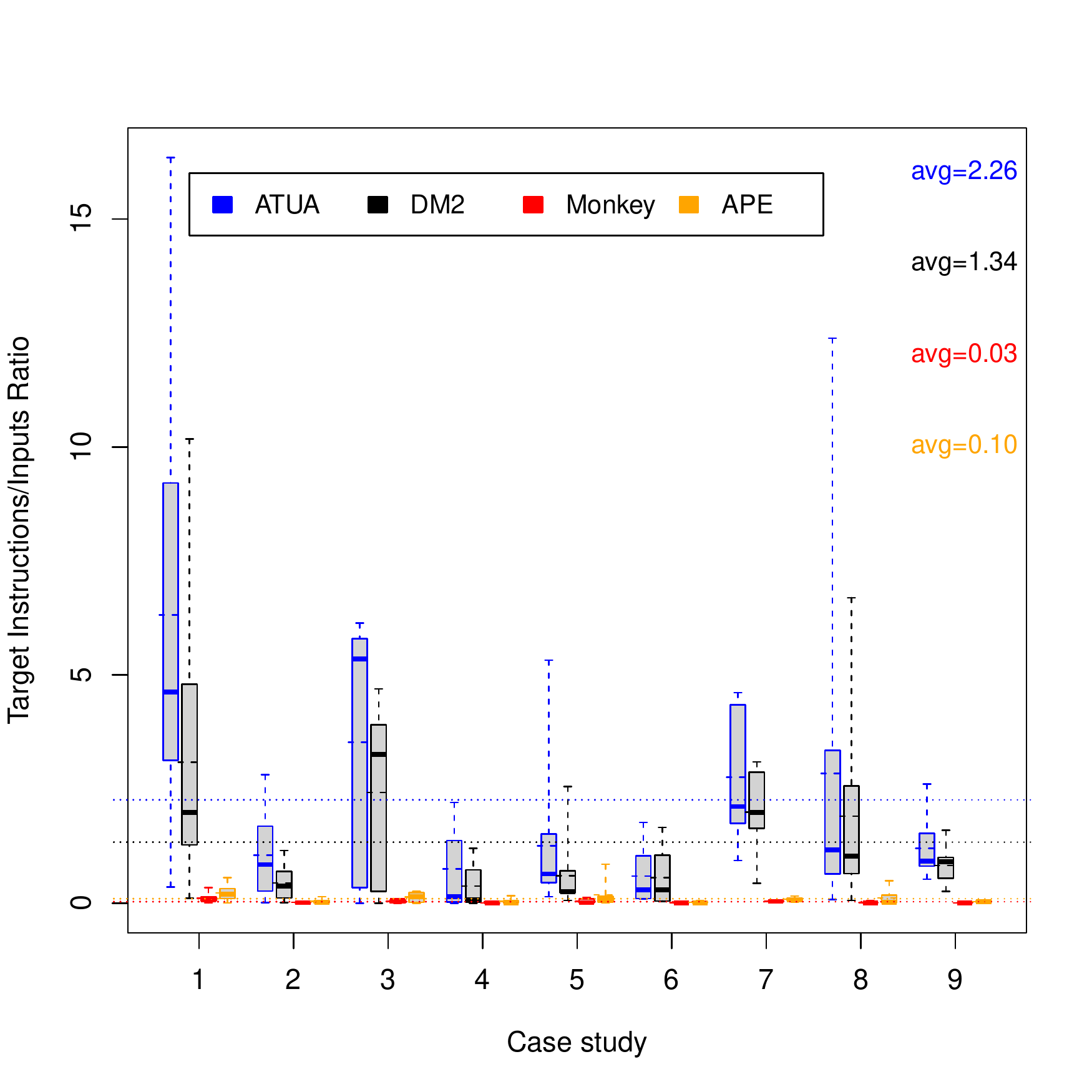}
      \caption{Target instructions/inputs ratio (budget=1 hour).}
      \label{fig:RQ3:CostEffectivenessRatio:1}
\end{minipage}\hspace{5mm}
\begin{minipage}{.45\textwidth}
	\includegraphics[width=7cm]{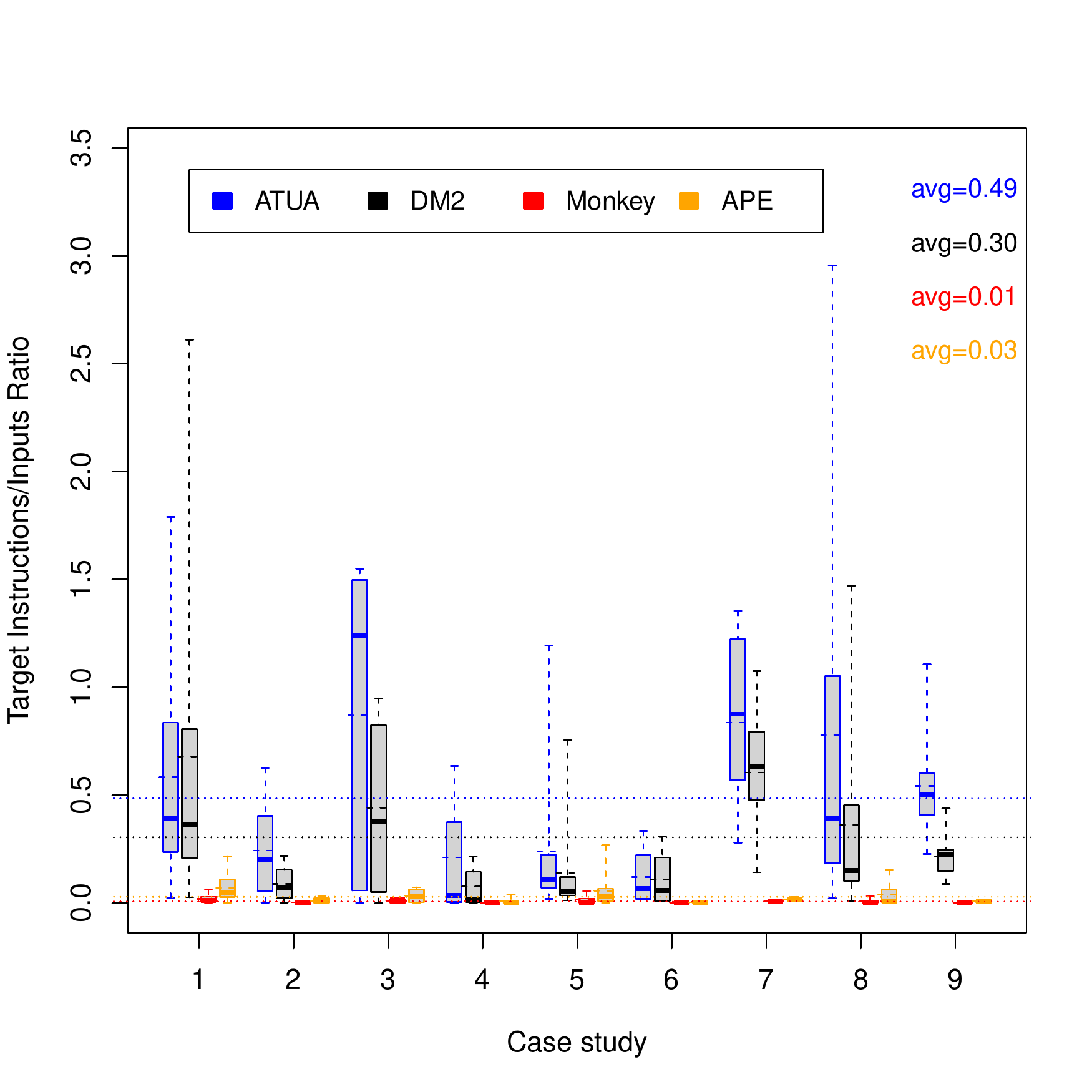}
      \caption{Target instructions/inputs ratio (budget=5 hours).}
      \label{fig:RQ3:CostEffectivenessRatio:5}
\end{minipage}
\end{figure*}

Figures~\ref{fig:RQ3:CostEffectivenessRatio:1} and ~\ref{fig:RQ3:CostEffectivenessRatio:5} show, for each subject, the distribution of the target instructions/inputs ratio. \APPR has the highest ratio: \FIXME{2.26} for one hour, \FIXME{0.49} for five hours.  
For the one-hour budget, \APPR's test automation effort (i.e., manual repair of a GUI input, visual inspection of outputs) is thus more beneficial because each input enables the verification of 2.26 additional target instructions. 
As a comparison, other state-of-the-art approaches yield lower ratios: 1.34 (DM2), 0.03 (Monkey), and 0.10 (APE). These results show that, though Monkey and APE are known for effectively triggering crashes, they are unlikely to be applicable in a testing context where the number of generated inputs should be minimized. For a time budget of five hours, average differences are less pronounced but the same trends hold.

Table~\ref{table:effectSize:ratio} provides p-values and $A_{12}$ statistics. Since we aim to determine if \APPR is likely to generate a higher target instructions/inputs ratio, we look for $A_{12}$ values above $0.50$.   
For a one-hour budget, effect size is always in favor of \APPR (i.e., is more likely to generate a higher instructions/inputs ratio); effect size is always large with respect to Monkey and APE. Even if in a few cases differences 
are not statistically significant \JMR{3.10}{(i.e., we cannot reject the null hypothesis that \emph{there is no difference between \APPR and the state-of-the-art approach concerning the ratio between instructions covered and inputs being triggered})}, effect size trends provides a clear picture of the benefits: \textbf{\APPR  is likely to yield a higher instructions/inputs ratio}.
The same conclusions can be drawn for a five-hour budget, though for two subjects (Wikipedia and VLC) \APPR performs similarly to DM2.

To summarize, regarding the human effort required for practical execution time budgets, \APPR performs better than the other approaches since it saves around 
\FIXME{33.8\%} (1 hour budget) and \FIXME{32.8\%} (5 hours budget) of the effort compared with DM2, while it shows huge savings compared to the other two approaches.
As for the effectiveness per unit of effort, \APPR provides tangible gains of \FIXME{68.7\%} (1h) and \FIXME{63.3\%} (5h) compared with DM2, and huge differences with the others. \textbf{\APPR therefore significantly decreases the human effort required for repairing inputs and defining oracles when compared to state-of-the-art approaches.}

\begin{table}[tb]
\footnotesize
\caption{Statistical significance and effect size for target instructions/inputs ratios.}
\label{table:effectSize:ratio}
\begin{tabular}{
|>{\raggedleft\arraybackslash}p{2mm}@{\hspace{1pt}}|
>{\raggedleft\arraybackslash}p{6mm}@{\hspace{1pt}}|
>{\raggedleft\arraybackslash}p{6mm}@{\hspace{1pt}}|
>{\raggedleft\arraybackslash}p{6mm}@{\hspace{1pt}}|
>{\raggedleft\arraybackslash}p{5mm}@{\hspace{1pt}}|
>{\raggedleft\arraybackslash}p{5mm}@{\hspace{1pt}}|
>{\raggedleft\arraybackslash}p{5mm}@{\hspace{1pt}}|
>{\raggedleft\arraybackslash}p{6mm}@{\hspace{1pt}}|
>{\raggedleft\arraybackslash}p{6mm}@{\hspace{1pt}}|
>{\raggedleft\arraybackslash}p{6mm}@{\hspace{1pt}}|
>{\raggedleft\arraybackslash}p{5mm}@{\hspace{1pt}}|
>{\raggedleft\arraybackslash}p{5mm}@{\hspace{1pt}}|
>{\raggedleft\arraybackslash}p{5mm}@{\hspace{1pt}}|
}
\hline
&
\multicolumn{6}{c|}{\textbf{1 hour budget}}&
\multicolumn{6}{c|}{\textbf{5 hours budget}}\\
&
\multicolumn{3}{c|}{\textbf{p-value}}&
\multicolumn{3}{c|}{\textbf{$A_{12}$}}&
\multicolumn{3}{c|}{\textbf{p-value}}&
\multicolumn{3}{c|}{\textbf{$A_{12}$}}\\
\cline{2-13}
\textbf{S}&
D&M&A&
D&M&A&
D&M&A&
D&M&A\\
\hline
1&\R{0.10}&<0.05&<0.05&.728&1.00&.975&\R{0.69}&<0.05&<0.05&\R{.555}&.963&.901\\
2&\R{0.17}&<0.05&<0.05&.703&.906&.875&\R{0.14}&<0.05&<0.05&.719&.938&.875\\
3&\R{0.29}&\R{0.09}&\R{0.09}&.700&.820&.820&\R{0.25}&\R{0.08}&\R{0.17}&.720&.840&.760\\
4&\R{0.33}&<0.05&\R{0.05}&.636&.895&.772&\R{0.38}&<0.05&<0.05&.623&.895&.747\\
5&\R{0.17}&<0.05&<0.05&.691&1.00&.901&\R{0.17}&<0.05&<0.05&.691&.950&.827\\
6&\R{0.63}&<0.05&<0.05&.583&1.00&1.00&\R{0.52}&<0.05&<0.05&.611&1.00&1.00\\
7&\R{0.27}&<0.05&<0.05&.654&1.00&1.00&\R{0.27}&<0.05&<0.05&.654&1.00&1.00\\
8&\R{0.60}&<0.05&<0.05&.578&1.00&.953&\R{0.92}&<0.05&<0.05&\R{.516}&.968&.906\\
9&\R{0.39}&<0.05&<0.05&.642&1.00&1.00&<0.05&<0.05&<0.05&.914&1.00&1.00
\\
\hline
\end{tabular}

\textbf{Legend}: \emph{S}, subject. \emph{D}, comparison with DM2, \emph{M}, Monkey, \emph{A}, APE. We underline the few cases in which statistics indicate that \APPR shows no significant difference (i.e., \emph{p-value} $\ge 0.05$) or no higher likelihood of achieving a higher instructions/inputs ratio (i.e., $A_{12} < 0.56$) than state-of-the-art approaches.
\end{table}%

\subsection{RQ2: Effectiveness within Time Budget}
\label{sec:rq2}
\subsubsection{Experiment design}

To address RQ2, we focus on code coverage results obtained with the updated versions of our subject Apps, i.e., versions V1 to V9. 
More precisely, we keep trace of the updated methods (hereafter, \emph{target methods}) and instructions belonging to updated methods (hereafter, \emph{target instructions}) that are exercised by the test automation tools considered in our study. 
To collect data for \APPR and DM2, we rely on the Soot-based code coverage extension integrated into DM2; for Monkey and APE, we rely on MiniTracing, a toolset developed to measure code coverage with APE~\cite{MiniTracing}. Since all these code coverage tools measure the coverage of the whole App under test, to determine the coverage of target methods and instructions, we filter results based on the list of updated methods generated by AppDiff. 

Since \APPR and state-of-the-art approaches require a different degree of human effort (see RQ1) and human effort is measured in terms of inputs generated, we shall set an identical limit to the number of inputs that might be generated by the test generation techniques. The rationale is that we try to emulate, in our experiments, realistic conditions where testers are limited by both execution time and human resources. This is thus expected to yield unbiased comparisons of practical value. It is also consistent with our objective, stated earlier, of minimizing human effort while keeping execution time within acceptable bounds. 
More precisely, for each software version $v$, we define an inputs budget equal to the maximum number of inputs generated, over ten runs, by \APPR, which is the approach generating the fewest test inputs for a given time budget, based on RQ1 results.

\JMR{2.2}{Though the fault detection rate (i.e., the proportion of faults being detected by a test automation technique) would be a useful, complementary metric to evaluate test automation effectiveness, it is inapplicable in our context since an important subset of our subject Apps (BBC, YahooWeather, Wikihow, Nuzzel) are not open source, a choice made to include representative Apps.
Indeed, the unavailability of source code and bug repositories prevented us from determining if a failure was due to a fault introduced by an App upgrade. Further, supporting the use of coverage for our experiments, it has been recently shown that there is moderate to high correlation between code coverage and the detection of real faults~\cite{Kochhar2015}. Finally, without automated functional oracles, in existing studies, effectiveness is typically measured by looking at runtime failures (i.e., due to uncaught exceptions or crashes), which represent only a small proportion of the failures that are typically observed in the field~\cite{Gazzola}.}

\paragraph{Metrics} 
\JMR{3.6}{Since the number of target methods and instructions varies across App versions, \APPR and state-of-the-art approaches shall be compared in terms of percentage of target methods and percentage of target instructions} covered.
In addition, such coverage metrics shall be
 obtained when testing a subject App for a maximum and practical execution time budget (i.e., one hour and five hours, as discussed in Section~\ref{sec:empirical:setup}), while not exceeding a maximum input budget determining human effort.

To positively answer this research question, \APPR should, in statistical terms, \JMRCHANGE{exercise a larger percentage of} target methods and instructions than the other approaches.

\subsubsection{Results}

\begin{figure*}
\begin{minipage}{.45\textwidth}
	\includegraphics[width=7cm]{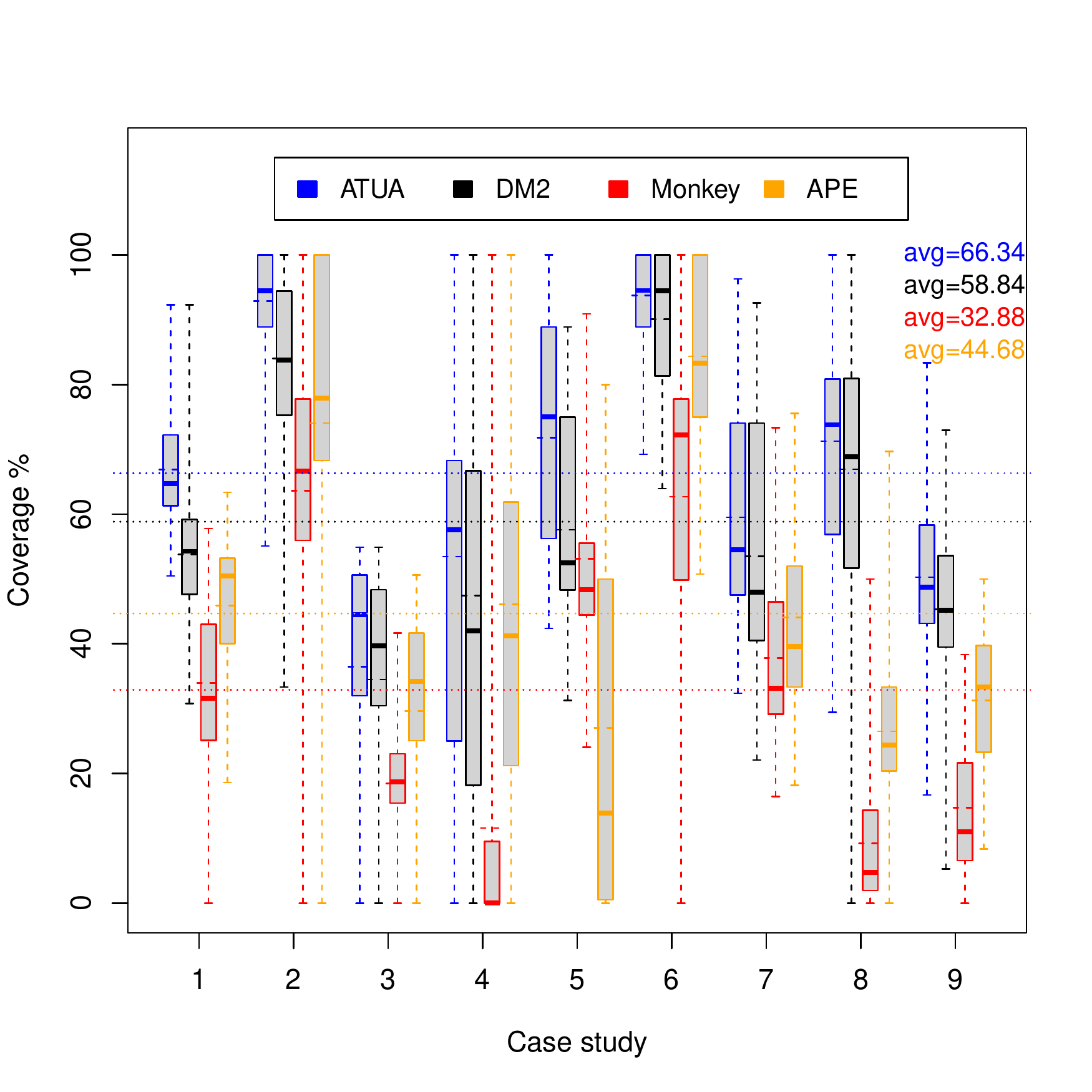}
	\vspace{-1cm}
      \caption{Percentage of updated methods covered for each version of the case studies (budget=1 hour).}
      \label{fig:RQ1:Method:1}
\end{minipage}\hspace{5mm}
\begin{minipage}{.45\textwidth}
	\includegraphics[width=7cm]{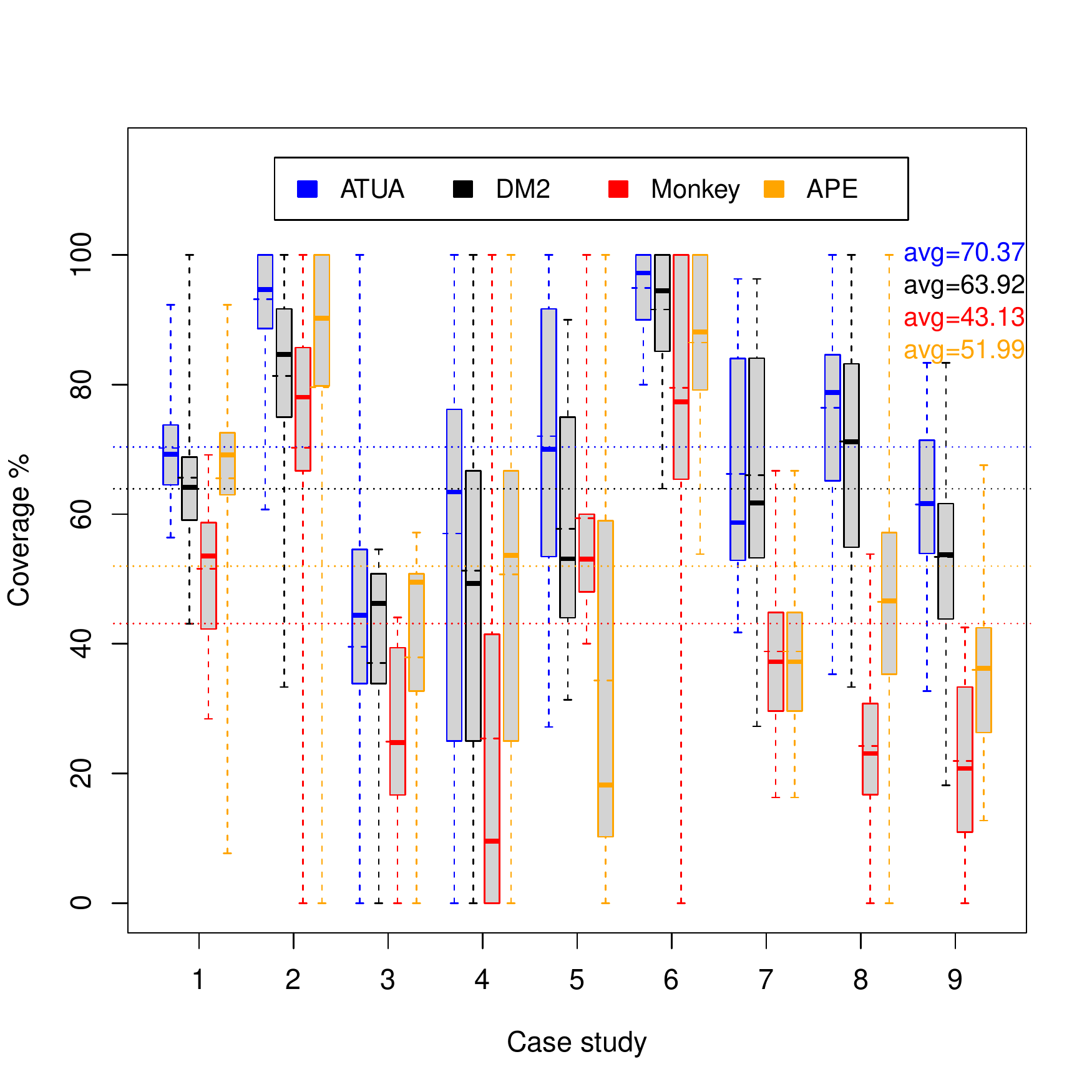}
		\vspace{-1cm}
      \caption{Percentage of updated methods covered for each version of the case studies (budget=5 hours).}
      \label{fig:RQ1:Method:5}
\end{minipage}
\begin{minipage}{.45\textwidth}
	\vspace{-7mm}
	\includegraphics[width=7cm]{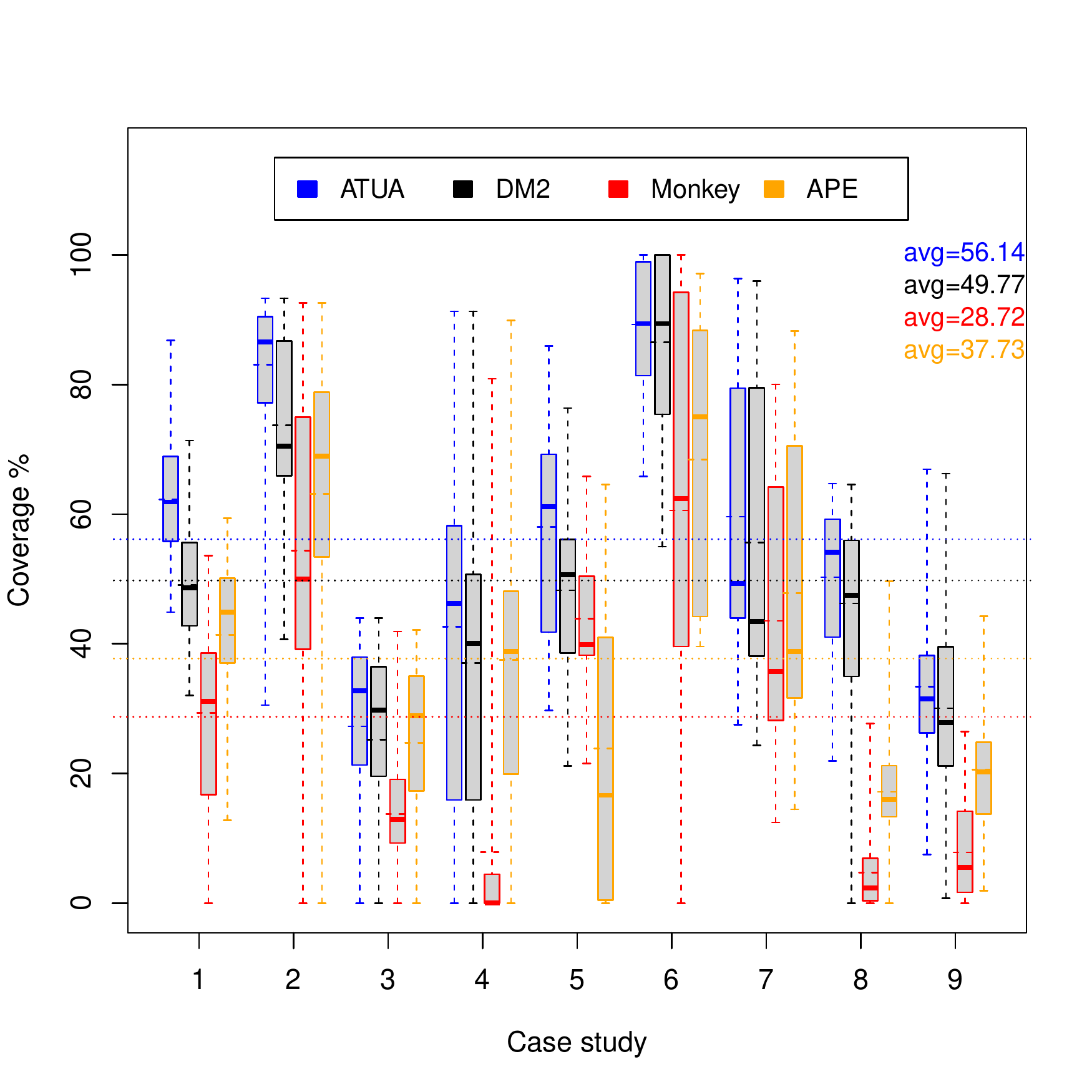}
		\vspace{-1cm}
      \caption{Percentage of instructions belonging to updated methods that are covered for each version of the case studies (budget=1 hour).}
      \label{fig:RQ1:Instr:1}
\end{minipage}\hspace{5mm}
\begin{minipage}{.45\textwidth}
	\vspace{-7mm}
	\includegraphics[width=7cm]{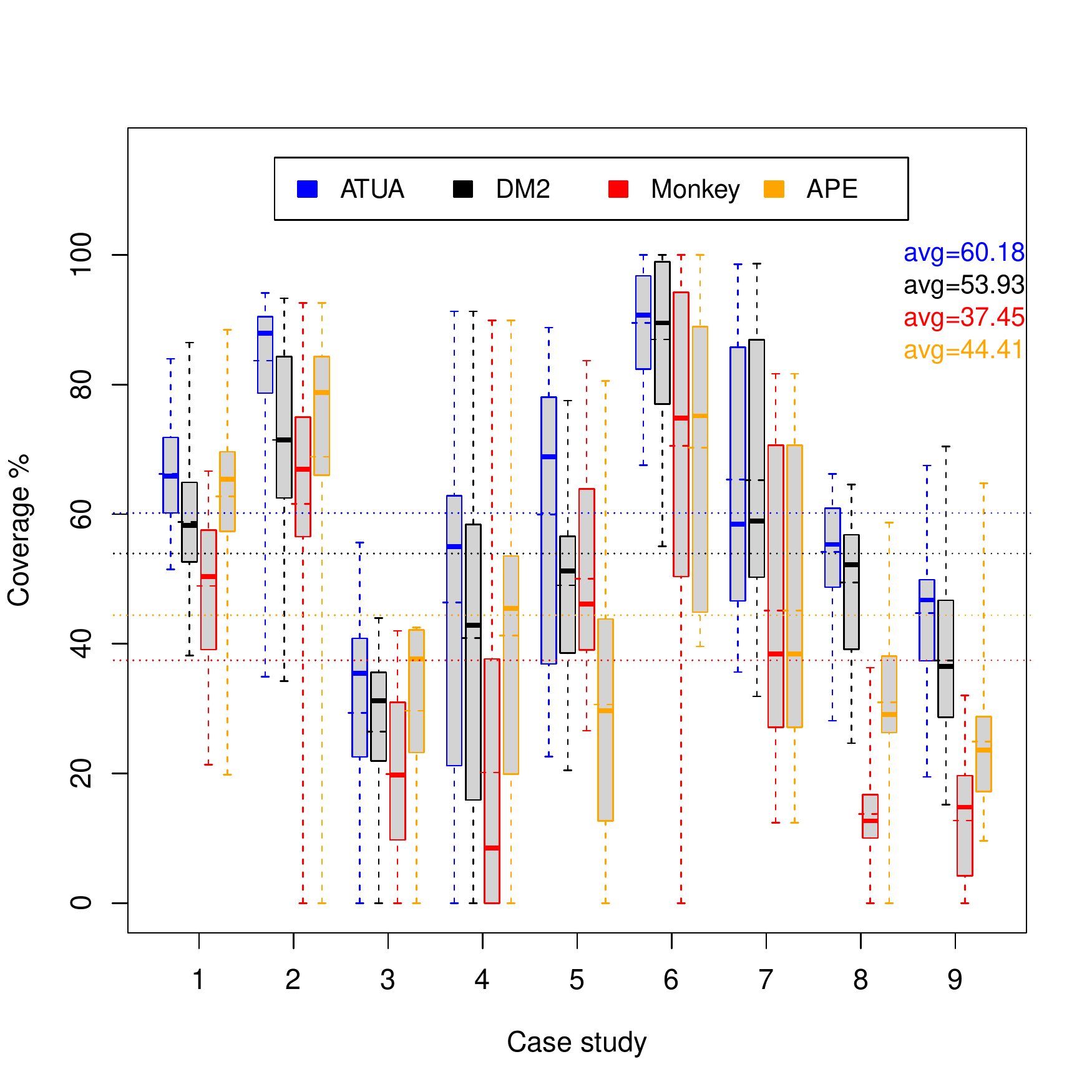}
		\vspace{-1cm}
      \caption{Percentage of instructions belonging to updated methods that are covered for each version of the case studies (budget=5 hours).}
      \label{fig:RQ1:Instr:5}
\end{minipage}
\end{figure*}

Figures~\ref{fig:RQ1:Method:1} and ~\ref{fig:RQ1:Instr:1} show the distribution of the percentage of target methods and instructions that have been covered by the selected testing tools for the subject Apps, with a test budget of one hour. Figures~\ref{fig:RQ1:Method:5}  and ~\ref{fig:RQ1:Instr:5} report the same measurements for a budget of five hours.

With a test execution budget of one hour, \APPR is the approach with the highest percentage of target methods and instructions being exercised on average, with \FIXME{66.34\%} and \FIXME{56.14\%}, respectively.
The largest differences are observed when \APPR is compared to Monkey; indeed, on average, \APPR exercises \FIXME{33.46\%} and \FIXME{27.42\%} more methods and instructions than Monkey, respectively. 
Since Monkey implements a pure random exploration strategy, our results show that a limit on the number of inputs generated by Monkey highly affects its performance.
In contrast, the APE state abstraction function enables a more effective generation of test inputs, thus leading to, on average, higher coverage than Monkey.
However, \APPR outperforms APE; indeed, on average, \APPR exercises \FIXME{21.66\%} and \FIXME{18.41\%} more target methods and instructions than APE, respectively.
Though DM2 fares better than Monkey and APE, as it relies on a model-based approach leveraging dynamic analysis, \APPR exercises \FIXME{7.50\%} and \FIXME{6.37\%} more target methods and instructions. 
\JMR{2.2}{Note that, 
the increase achieved by \APPR is particularly significant, +12.74\% (i.e., +7.50\%/58.84\%) and +12.79\% (i.e., +6.37\%/49.77\%) for methods and instructions coverage, respectively}. 
\CHANGEDNOV{This is explained by the
transition-driven exploration based on
static analysis (\APPR Phase 1 and 2) and information retrieval (Phase 3), which are not part of DM2.}

When executed with a test budget of one hour, for all the subject Apps, both the median and the average obtained with \APPR are higher than those obtained with other approaches.

\begin{table}[tb]
\footnotesize
\caption{Statistical significance and effect size for RQ2.}
\label{table:effectSize:rq21}
\begin{tabular}{
|>{\raggedleft\arraybackslash}p{2mm}@{\hspace{1pt}}|
>{\raggedleft\arraybackslash}p{6mm}@{\hspace{1pt}}|
>{\raggedleft\arraybackslash}p{6mm}@{\hspace{1pt}}|
>{\raggedleft\arraybackslash}p{6mm}@{\hspace{1pt}}|
>{\raggedleft\arraybackslash}p{5mm}@{\hspace{1pt}}|
>{\raggedleft\arraybackslash}p{5mm}@{\hspace{1pt}}|
>{\raggedleft\arraybackslash}p{5mm}@{\hspace{1pt}}|
>{\raggedleft\arraybackslash}p{6mm}@{\hspace{1pt}}|
>{\raggedleft\arraybackslash}p{6mm}@{\hspace{1pt}}|
>{\raggedleft\arraybackslash}p{6mm}@{\hspace{1pt}}|
>{\raggedleft\arraybackslash}p{5mm}@{\hspace{1pt}}|
>{\raggedleft\arraybackslash}p{5mm}@{\hspace{1pt}}|
>{\raggedleft\arraybackslash}p{5mm}@{\hspace{1pt}}|
}
\hline
&
\multicolumn{6}{c|}{\textbf{1 hour budget}}&
\multicolumn{6}{c|}{\textbf{5 hours budget}}\\
&
\multicolumn{3}{c|}{\textbf{p-value}}&
\multicolumn{3}{c|}{\textbf{$A_{12}$}}&
\multicolumn{3}{c|}{\textbf{p-value}}&
\multicolumn{3}{c|}{\textbf{$A_{12}$}}\\
\cline{2-13}
\textbf{C}&
D&M&A&
D&M&A&
D&M&A&
D&M&A\\
\hline
&
\multicolumn{12}{c|}{\textbf{Coverage of target methods}}\\
\cline{2-13}
1&<0.05&<0.05&<0.05&.857&.994&.956&<0.05&<0.05&\R{0.14}&.703&.935&.564\\
2&<0.05&<0.05&<0.05&.733&.868&.703&<0.05&<0.05&<0.05&.774&.832&.623\\
3&\R{0.41}&<0.05&<0.05&\R{.548}&.795&.684&\R{0.61}&<0.05&0.99&\R{.529}&.767&\R{.501}\\
4&\R{0.16}&<0.05&\R{0.07}&.560&.852&.577&\R{0.16}&<0.05&\R{0.08}&.560&.760&.575\\
5&<0.05&<0.05&<0.05&.735&.814&.920&<0.05&<0.05&<0.05&.713&.702&.841\\
6&\R{0.08}&<0.05&<0.05&.587&.838&.689&\R{0.2}&<0.05&<0.05&.564&.719&.681\\
7&<0.05&<0.05&<0.05&.614&.836&.762&\R{0.78}&<0.05&<0.05&\R{.488}&.879&.879\\
8&\R{0.23}&<0.05&<0.05&\R{.554}&.997&.972&\R{0.05}&<0.05&<0.05&.588&.995&.892\\
9&<0.05&<0.05&<0.05&.587&.981&.878&<0.05&<0.05&<0.05&.685&.992&.918\\
\hline
&\multicolumn{12}{c|}{\textbf{Coverage of target instructions}}\\
\cline{2-13}
1&<0.05&<0.05&<0.05&.853&.991&.936&<0.05&<0.05&\R{0.15}&.735&.884&.562\\
2&<0.05&<0.05&<0.05&.707&.843&.783&<0.05&<0.05&<0.05&.756&.829&.721\\
3&\R{0.31}&<0.05&\R{0.14}&\R{.559}&.774&.585&\R{0.18}&<0.05&\R{0.70}&.577&.705&\R{.478}\\
4&<0.05&<0.05&\R{0.07}&.600&.861&.579&\R{0.05}&<0.05&<0.05&.584&.764&.594\\
5&<0.05&<0.05&<0.05&.684&.748&.886&<0.05&<0.05&<0.05&.659&.623&.822\\
6&\R{0.42}&<0.05&<0.05&\R{.542}&.788&.827&\R{0.47}&<0.05&<0.05&\R{.538}&.727&.786\\
7&<0.05&<0.05&<0.05&.598&.735&.695&\R{0.65}&<0.05&<0.05&\R{.480}&.780&.780\\
8&\R{0.06}&<0.05&<0.05&.586&.999&.984&\R{0.33}&<0.05&<0.05&.633&.999&.936\\
9&0.05&<0.05&<0.05&.584&.980&.822&<0.05&<0.05&<0.05&.671&.988&.896\\
\hline
\end{tabular}

\textbf{Legend}: \emph{C}, case study. \emph{D}, comparison with DM2, \emph{M}, Monkey, \emph{A}, APE. We underline the few cases in which statistics indicate that \APPR shows no significant difference (i.e., \emph{p-value} $\ge 0.05$ ) or no higher likelihood (i.e., $A_{12} < 0.56$) of covering more targets  than state-of-the-art approaches.
\end{table}%

To discuss differences across subjects, we report in Table~\ref{table:effectSize:rq21} the p-value and $A_{12}$ statistics obtained with the Mann Whitney U-test and Vargha and Delaney’s  method, respectively. Overall, differences are statistically significant\JMR{3.10}{\footnote{We can reject the null hypothesis that \emph{there is no difference in the number of target methods and instructions exercised by \APPR and the i-th state-of-the-art approach}.}} but there are exceptions:  when \APPR is compared to APE for Nuzzel  (subject 4), and when  \APPR is compared to DM2 for Nuzzel, File Manager (subject 3), Wikihow (subject 6), and VLC (subject 8). However, for most of the subjects, \APPR is likely to exercise more target methods and instructions than other approaches; this is shown by the $A_{12}$ statistics being always above 0.56, except for File Manager, Wikihow, and VLC in the case of DM2\footnote{Average $A_{12}$ is 0.77 for target methods and 0.76 for target instructions coverage; median is 0.80 for target methods and 0.77 for target instructions coverage.}. \CHANGEDNOV{Regarding VLC, the effectiveness of \APPR is limited by the need for setup operations that require some human effort. Indeed, since certain features can be tested only on specific devices (e.g., an Android TV), identifying target methods through static analysis is of limited usefulness and \APPR performs similarly to DM2.}
\CHANGEDNOV{However, such limitations could be surmounted after investing some effort to carefully setup \APPR. For example, by configuring \APPR to be executed on an Android TV in addition to a mobile emulator (i.e., what we used in our experiments).}
\CHANGEDNOV{Concerning File Manager and Wikihow, \APPR is affected by some limitations of static analysis, which cannot determine that certain WindowTransitions are associated to specific data types provided as input.
More precisely, in the case of File Manager, a number of updated features can be exercised only through specific files (e.g., the decompress operation can be executed only with files having ZIP or RAR filename extension). The static analysis currently implemented in \APPR cannot determine that certain features are enabled only in the presence of specific runtime data (e.g., file names) and thus \APPR, similar to DM2, exercises such features only if it accidentally triggers them thanks to random exploration. A similar but more evident problem occurs also in the case of Wikihow, where static analysis does not
identify the WindowTransitions triggered by the inputs sent to WebViews. Indeed, the input handlers executed after sending an input to a WebView (e.g., a click on an anchor) depend on the content of the page (e.g., the file type appearing in the URL of the anchor) and thus cannot be identified by static analysis, which does not process the content of the HTML pages displayed at runtime. For this reason, \APPR cannot fully take advantage of static analysis results in the presence of WebViews. In such cases, similar to DM2, \APPR exercises App features thanks to random exploration. However, current \APPR results with Apps using WebViews largely depend on the proportion of features implemented through WebViews. For example, in the case of WikiHow, which mainly relies on WebViews (five out of eight content types are displayed through a WebView), \APPR performs similarly to DM2; instead, in the case of Wikipedia, which implements only one out of 35 Windows using a WebView\footnote{In Wikipedia, WebViews are used to  display Wikipedia pages while other Views are used for other features such as displaying news, image galleries, or editing the content of a page.}, \APPR outperforms all the other approaches ($A_{12} \ge 0.56$). To overcome the limitations of static analysis and thus improve \APPR results, it might be necessary to develop dedicated strategies relying on dynamic analysis; for example, by extending the state abstraction function of \APPR to use reducers dedicated to HTML anchors or file objects.}

With a test budget of five hours, all the approaches achieve better coverage results; however, the ranking observed for a one-hour budget remains unchanged.
\APPR is the approach with the highest percentage of exercised target methods and instructions, on average, with \FIXME{70.37\%} and \FIXME{60.18\%}, respectively.
The largest differences are still observed when \APPR is compared to Monkey; indeed, on average, \APPR exercises \FIXME{27.24\%} and \FIXME{22.73\%} more target methods and instructions than Monkey, respectively. 
\APPR exercises \FIXME{18.38\%} and \FIXME{15.77\%} more target methods and instructions than APE.
The second best approach remains DM2, as \APPR exercises \FIXME{6.45\%} and \FIXME{6.25\%} more target methods and instructions than DM2\JMR{2.2}{, with a gain of +10.09\% (+6.45\%/63.92) and +11.59\% (+6.25\%/53.93\%) in the number of target methods and instructions covered with respect to DM2, respectively.}

A larger time budget enables \APPR to achieve higher coverage. 
This is set with the \emph{scaleFactor} configuration parameter (see Section~\ref{sec:algo}), which we increase for a five-hour budget, thus augmenting the time spent to perform random exploration, reach the test target, and exercise targets. We leave to future work the study of the effect of different configuration values for the \emph{scaleFactor} parameter of  \APPR.

With a test budget of five hours, the difference between \APPR and other approaches decreases though. Unsurprisingly, with a larger test budget, random-based approaches can more easily reach updated features than with a one-hour budget, for which leveraging static analysis is more important. For example, in subject App 7 (BBC Mobile), in five hours, DM2 achieves the same average coverage as \APPR. 

To discuss differences across subjects, we refer to 
the p-value and $A_{12}$ statistics reported in 
the rightmost columns of Table~\ref{table:effectSize:rq21}.
Because of the larger test budget benefiting random exploration, differences between \APPR and other approaches are not significant in 7 out of 27 cases (i.e., $3 \times 9$, which is the number of pairwise comparisons between \APPR and the other approaches), two cases more than with a one-hour budget. However, \APPR is still likely to exercise more target methods and instructions than other approaches. Indeed, for both method and instruction coverage, the $A_{12}$ statistics is above 0.56 for 24 out of 27 cases. \CHANGEDNOV{In general, effect size is slightly lower than for a one-hour budget, with an average $A_{12}$ of 0.73 and 0.72 for the coverage of target methods and instructions.}
\CHANGEDNOV{In particular, we observe that the larger time budget enables random-driven approaches to achieve the same effectiveness as \APPR when \APPR is negatively affected by static analysis limitations. This happens for File Manager (subject 3), where APE performs similarly to \APPR, Wikihow (subject 6), where DM2 performs similarly to \APPR, and BBC mobile (subject 7), where the additional time budget enables DM2 to exercise the few updated features depending on WebViews (in BBC Mobile, WebViews are used to display BBC Web pages)}.

\textbf{To summarize, \APPR is the approach that, on average, most effectively test updated Apps within practical time budgets and human effort.}
It tends to cover more target methods and instructions than other approaches.
The second best approach is DM2. For a one-hour budget, on average, \APPR automatically exercises \FIXME{7.50\%} and \FIXME{6.37\%} more target methods and instructions than DM2, \JMR{2.2}{with a gain of +12.74\% and +12.79\%, respectively}.
With a five-hour budget, on average, \APPR automatically exercises 
\FIXME{6.45\%} and \FIXME{6.25\%} more target methods and instructions than DM2, \JMR{2.2}{with a gain of +10.09\% and +11.59\%, respectively}. For seven out of nine subjects, for both time budgets, \APPR tends to exercise more target methods and instructions than DM2. \CHANGEDNOV{For the remaining two subjects, DM2 and \APPR are comparable, due mostly to the current limitations of static analysis.}

\subsection{\JMR{3.11}{RQ3 - Complementarity  of Testing Approaches}}
\label{appendix:complementarities}

\subsubsection{Experiment design.}

\JMRCHANGE{A software testing approach is complementary to other approaches if it exercises a set of functionalities not exercised by the others. 
Since we measure effectiveness based on method coverage, to determine complementarity, we look for methods that are univocally covered by each testing approach considered in our experiments.
A method is \emph{univocally covered} by approach $A$ for version $V$ of a subject App $S$ if it is exercised by $A$ in at least one of the ten test execution runs on version $V$ and is not exercised by any other approach in any test execution run of that same version.
We cannot compare testing approaches based on instruction coverage because some of our subjects are commercial Apps released without source code  (e.g., to understand the semantics of the covered instructions).
Since the number of target methods varies for each App version, we compare the percentage of target methods 
that are univocally covered by each approach.
Finally, it shall be possible to identify common characteristics in the inputs triggering the univocally exercised methods.}

\JMRCHANGE{\paragraph{Metrics}To compare testing approaches, we thus report (1) the \emph{overall number of univocally covered methods across all the subject App versions} and (2) \emph{the distribution of the percentage of tested methods that are univocally covered by each approach, across all the subject App versions}. 
Furthermore, we manually inspect the list of univocally covered methods. Based on their signatures\footnote{Since most of our subject Apps are commercial Apps released without source code, the functionality implemented by a method is inferred from its signature.} and, for \APPR, the data collected in the GSTG, we (3) \emph{determine the characteristics of the inputs and upgraded functionalities that are better targeted by each of the testing approaches}. Since the test budget affects the performance of testing approaches, we discuss the results achieved for one-hour and five-hour budgets, separately.
}

\JMRCHANGE{For each testing approach, we analyze if (1) it covers a large number of methods not covered by other approaches across all Apps, (2) its distribution, across Apps, of the percentage of tested methods not covered by other approaches has a significantly larger average than that of other approaches, 
and (3) we can characterize the situations in which the approach univocally exercises some updated App methods.}

\subsubsection{Results.} 

\begin{figure*}
\begin{minipage}{.45\textwidth}
	\includegraphics[width=7cm]{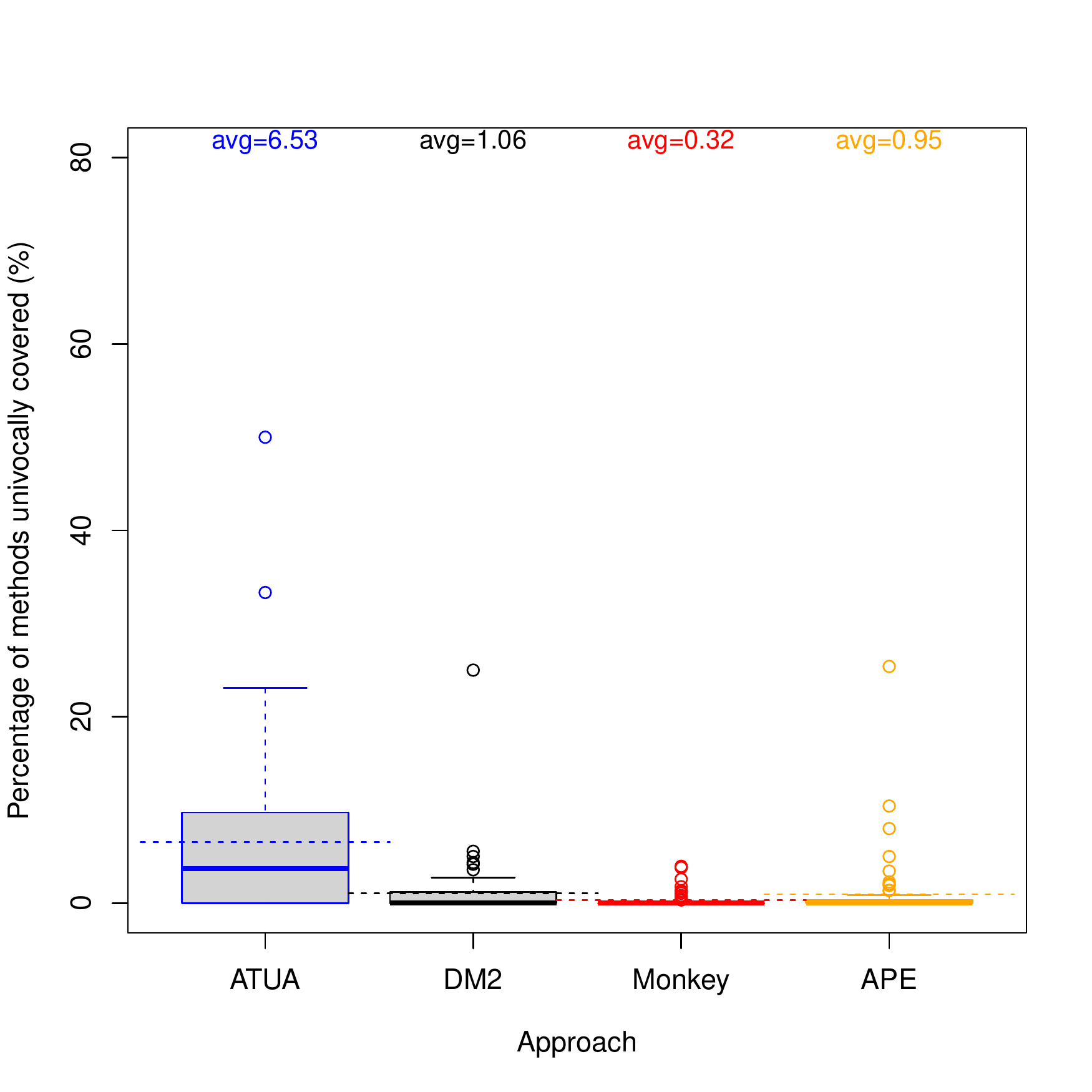}
	\vspace{-1cm}
      \caption{Proportion of methods that are univocally covered by one testing approach, distribution across all the tested subject versions (budget=1 hour).}
      \label{fig:RQ3:MethodPercentage:1}
\end{minipage}\hspace{5mm}
\begin{minipage}{.45\textwidth}
	\includegraphics[width=7cm]{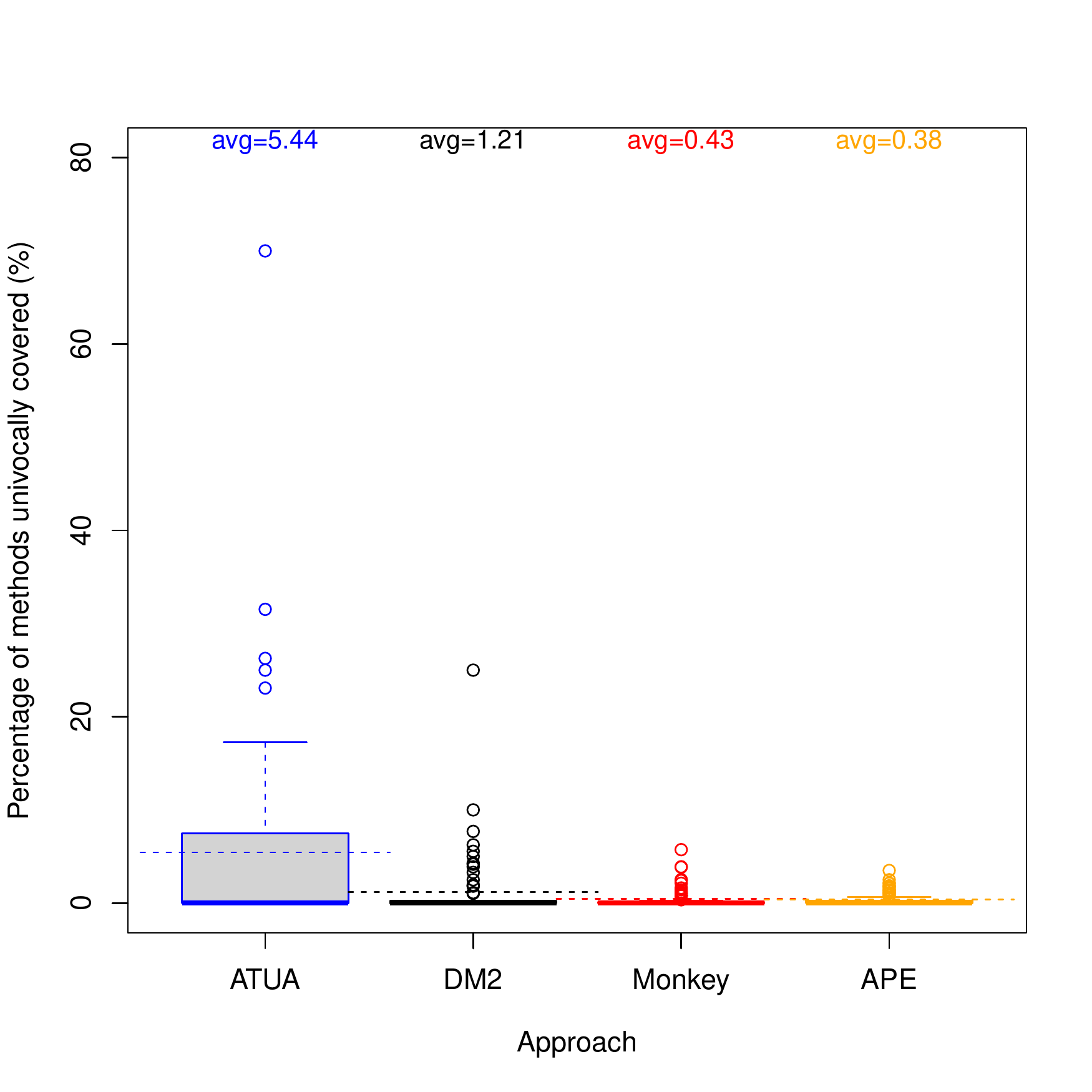}
		\vspace{-1cm}
      \caption{Proportion of methods that are univocally covered by one testing approach, distribution across all the tested subject versions (budget=5 hours).}
      \label{fig:RQ3:MethodPercentage:5}      
\end{minipage}
\end{figure*}

\JMRCHANGE{For the one-hour budget, a total of 6982 methods belonging to the different App versions have been exercised in our experiments. Overall, 784 out of 6982 methods (11\%) are exercised only by one testing approach, 6198 methods (89\%) are covered by at least two approaches, while 3648 methods (52\%) are covered by all the approaches.
\APPR exercises the largest number of methods not exercised by other approaches, 518 (66\%);
it is followed by APE (156, i.e., 20\%), DM2 (78, i.e., 10\%), and Monkey (32, i.e., 4\%).}

\JMRCHANGE{Figure~\ref{fig:RQ3:MethodPercentage:1} shows
the distribution of the percentage of methods exercised in our experiments that are univocally covered by one testing approach, across versions, for a one-hour budget. 
On average, 7\% of the methods exercised in our experiments are covered only by \APPR, while the other approaches univocally cover only 1\% or less.
Differences are statistically significant.
Also, we report that for 52\% of the App versions, \APPR exercises more univocally covered methods than other approaches (92\% if including versions with the same number of univocally covered methods).
These results show that, across a majority of individual versions, \APPR provides coverage capabilities that cannot be obtained with other approaches.}

\JMRCHANGE{The effectiveness of \APPR is  primarily due to its capability of reaching target Windows and target Widgets that are difficult to reach by solely relying on random exploration. We identify three distinct cases.
\emph{First}, \APPR can trigger complex sequences of inputs that enable the visualization of target Widgets. This is the case of SettingsActivity for Nuzzel, which requires opening a drawer, swipe up, and then click on the settings button.
Similarly, in Yahoo Weather, it is necessary to swipe up the weather information fragment and click on the map to trigger methods on the WeatherMapView. Other similar cases concern the renaming of files and the opening of the preferences activity in File Manager.
Such complex sequences of events are unlikely to be triggered by approaches relying on random exploration;
instead, they are selected by \APPR thanks to the use of the App model to  identify both the sequence of events that reaches a target Window and the target events that exercise the updated methods.
The \emph{second} case concerns \APPR being able to bring an App into a specific state required for testing, which is enabled by the fact that, in Phase 2, \APPR exercises the inputs that trigger updated methods multiple times, when the App has likely reached different App states.
For example, \APPR is the only approach exercising the method \emph{undeleteActivity} of class \emph{ActivityHelper} in Activity diary, which requires to first create an activity, then delete it, and finally undelete it.
The \emph{third} case concerns \APPR's capacity to select the Apps' options required to exercise certain Apps' features, which is the objective of Phase 3.
For example, in BBC mobile, to exercise the methods in class \emph{MyNewsByTimeFragment}, it is necessary to reach the settings window, enable the option \emph{My News By Topic}, and then open the tab \emph{My News}.}

\JMRCHANGE{The methods not covered by \APPR but covered by other approaches, instead, are generally the ones whose triggering Actions cannot be identified by \APPR because of the limitations of static analysis. We have identified three different scenarios in which \APPR is less effective than state-of-the-art approaches. 
First, though \APPR  can trigger complex input sequences, it cannot, in certain cases (e.g., in classes extending SettingsActivity), identify the events that trigger specific WindowTransitions, which APE can instead trigger. By relying on an adaptable state abstraction function, APE can direct random exploration towards App states that contain the widgets required to test the updated methods, which Monkey and DM2 do not achieve. 
For example, this happens when testing the updated methods of \emph{AboutActivity} in File Manager. 
Second, in the case of WebViews, instead, DM2, by investing more budget on random exploration, can reach Windows that can't be reached by \APPR because static analysis does not identify the required WindowTransitions in the EWTG. 
Also, based on the observed results, the specific random exploration strategy implemented by DM2 appears to be more effective than the one of APE and Monkey.
Finally, Monkey performs better than \APPR and the other approaches when the execution of updated methods depends on specific environmental conditions; for example, the internet connection being disabled, which is the case for testing methods \emph{onGoOffline} and \emph{onPageLoadError} in Wikipedia.}

\JMRCHANGE{Similar findings can be observed for a test budget of five hours.
Overall, a total of 7326 methods have been exercised in our experiments, which is expectedly higher than for the one-hour budget. Overall, 675 out of 7326 methods (9\%) are exercised only by one testing approach, 6767 methods (92\%) are covered by at least two approaches, while 4308 methods (58\%) are covered by all approaches. Although the larger test budget enables all the approaches to exercise a larger common set of methods, we still observe a high degree of complementarity (i.e., 9\% of the covered methods are univocally covered).
In particular, \APPR remains the approach that exercises the largest number of methods not exercised by other approaches: 478 (71\%); the other approaches, instead, show similar numbers of univocally covered methods: 71 for APE, 64 for DM2,	62 for Monkey.}

\JMRCHANGE{Figure~\ref{fig:RQ3:MethodPercentage:5} depicts
the distribution of the percentage of covered methods that are univocally covered by one testing approach, across versions, for a budget of five hours. 
On average, across versions, 5\% of the methods exercised in our experiments are covered only by \APPR, while the other approaches univocally cover only 1\% or less, thus confirming that, even for a test budget of five hours, \APPR complements all the other approaches.
Differences are statistically significant.
Such complementarity is also stressed by the fact that for 42\% of the App versions, \APPR covers more univocally covered methods than other approaches (81\% if including versions with the same number of univocally covered methods).
}

\JMRCHANGE{Also, for a five-hour budget, \APPR confirms its capacity to trigger complex sequences of inputs not generated by other approaches. This is the case for File Manager, where \APPR successfully starts the FTP client by (1) clicking on the  "add" button, (2) then clicking on "Cloud connection", (3) then clicking on SCP/SFTP connection and (4) finally, within the SCP/SFTP connection dialog, fill all the compulsory fields and (5) click on the "Create" button. Such a complex sequence of inputs (including filling FTP connection information) is unlikely to be generated by random approaches. \APPR, instead, once it finds the sequence of inputs that reaches the SCP/SFTP connection dialog, can, in Phase 2, trigger multiple sequences of inputs until it (randomly) finds the one that successfully starts the FTP client.
For APE, Monkey, and DM2, we can observe the same characteristics observed as for a one-hour test budget.
}

\JMRCHANGE{To summarize, for both one-hour and five-hour budgets, \textbf{\APPR is the approach that exercises the largest number of univocally covered methods}. Across versions, the percentage of exercised methods univocally covered by \APPR is significantly larger than that of other approaches. 
In practice, the results above also suggest that it might be useful to combine approaches since they may complement each other to cover a larger number of methods. However,  \APPR should always be included in the selected combination since it exercises a larger set of upgraded functionalities that cannot be exercised using other approaches.}

\subsection{Discussion}
\label{sec:discussion}

\paragraph{Human effort} RQ1 results have shown that \APPR performs better than the other approaches since it saves around 
\FIXME{33.8\%} (one-hour budget) and \FIXME{32.8\%} (five-hour budget) of the effort compared with DM2, the second-best approach. 
Hereafter, we discuss practical implications concerning testing costs based on related work about the nature of App upgrades~\cite{Mcilroy:FrequentlyUpdatedApps:ESE:2016} and the maintainability of GUI test cases~\cite{Pan2020}.

On average, \APPR generates 450 (one hour) and 2251 (five hours) fewer inputs than DM2 for each App version across all subject Apps. Since related work~\cite{Mcilroy:FrequentlyUpdatedApps:ESE:2016} has shown that roughly 35\% of the updates concern the introduction of new features, under the assumption that inputs are uniformly distributed across updated features, we can estimate that \APPR generates, on average for each App version and across all subjects, 
158 (one hour) and 788 (five hours) fewer inputs than DM2 for testing new features. Consequently, \APPR generates  292 (one hour) and 1463 (five hours) fewer inputs than DM2 for testing bug fixes and improved features (i.e., changes concerning non-functional requirements).

When testing new features, the output generated by each input should be manually verified; for example, by inspecting the screenshots of the GUI trees visualized after triggering an input (they are automatically captured by \APPR) and determining if they match the expected results. Unfortunately, the software engineering literature lacks studies about the cost of manual verification of GUI trees; assuming, for the sake of illustration, that visual inspection of GUI trees takes a few minutes, say ranging from one minute to five minutes, \APPR may lead to savings within the following intervals of [158-790] and [788-3940] minutes, respectively the for one-hour and five-hour test budgets.
In the App development context, where Apps are frequently released (e.g., weekly or bi-weekly) and, additionally, test cases might need to be executed every day following continuous integration practices, such effort savings appear to be particularly beneficial, especially considering that testing should be performed by highly-trained engineers with a deep understanding of the App's features.

When testing updated features, engineers can re-execute the generated test input sequences on previous App versions in order to compare results and, ideally, eliminate oracle costs. However, we have to expect that a number of maintenance operations are required in order to adapt test sequences to a different App version. Pan et al.~\cite{Pan2020}, for example, report that 26.5\% of the test inputs need to be repaired.
\APPR will thus save engineers from manually repairing 77 and 388 inputs, respectively for the one-hour an five-hour test budget. Under time pressure, which is the case when Apps are frequently released, this is a significant advantage.

\paragraph{Effectiveness}

\APPR is the approach that, on average, most effectively tests updated Apps within practical time budgets and human effort. For the one-hour budget, better from competing approaches, it exercises more than 60\% of target methods and 50\% of target instructions. 
With a five-hour budget, it exercises more than 70\% of target methods and 60\% of target instructions. Higher percentages can probably be reached with longer execution budgets, which were not possible in our context given the computational costs of our experiments. Based on these results, we can claim that \APPR can contribute to reducing development costs; indeed, engineers would then be able to focus their manual testing effort on a reduced portion of the developed App.

When comparing with other approaches, we observed that for both one-hour and five-hour budgets, on average, \APPR %
\JMR{2.2}{achieves method and instruction coverage results increased by at least 10\% with respect to the second-best approach (DM2), a practically significant improvement.}
The effectiveness of \APPR is comparable to the effectiveness of DM2 and APE only when \APPR cannot fully leverage static analysis to determine the relation between inputs and WindowTransitions, i.e., when Apps integrate input handlers that are selected at runtime based on the nature of input data, which happens, for example, in the presence of WebViews.
For the six subjects for which static analysis can effectively be exploited, the percentage of improvement rises above 8\%.
Among our subject Apps, in the worst case (i.e., five-hour budget), \APPR is comparable to other approaches for one third of the subjects and otherwise fares better; considering that (1) no single competing approach achieves similar coverage as \APPR for these three subject Apps (e.g., DM2 achieves the same results as \APPR for at most two), (2) competing approaches never outperform \APPR but at best reach the same effectiveness, \APPR remains the best choice.
To further improve \APPR effectiveness, part of our future work concerns the development of an additional set of reducers that will enable the \APPR state abstraction function to distinguish between widgets containing different types of data.

\JMR{3.11}{Finally, \APPR has shown to be complementary to other approaches. Indeed, for both one-hour and five-hour budgets, it can exercise 518 and 478 target methods not covered by other approaches, three and six times the number of the second best approach.
Thus, when combining testing approaches to cover higher target method coverage, \APPR should be included. 
Finally, as an explanation of the above results, we have observed that the three testing phases integrated in \APPR enable the generation of complex input sequences, specific App states, or diverse App settings that are required to test updated features.}

\subsection{Threats to validity}

We discuss internal, conclusion, construct and external validity according to standard practice~\cite{Feldt2010,Wohlin2012,Ralph2018}.

\subsubsection{Internal validity}
\JMR{3.6}{To address threats to \emph{internal validity}, we should ensure that
the observed outcome (inputs and code coverage, in our case) depends on the treatment (i.e., the test automation approaches)
and not external factors (e.g., implementation errors and diverse experimental conditions)~\cite{Feldt2010}.} 

\JMRCHANGE{To minimize \emph{implementation errors}, we have carefully inspected and tested \APPR before running our experiments. Also, for the state-of-the-art approaches, we relied on the software released by their authors, which had been used in several experiments.} 

\JMRCHANGE{To \emph{ensure the same conditions} for all the experiments, we executed each tool on a clean instance of the same Android emulator with newly installed Apps. 
However, the same experimental conditions may not be guaranteed for Apps that depend on external data sources (e.g., to visualize news)~\cite{DiMartino2020}; indeed, in the presence of external data source, test results may depend on the content being visualized at a specific instant (e.g., the presence of a video in the latest news). 
Our case study subjects include six Apps loading external data (i.e., Nuzzel, Wikipedia, Yahoo weather, Wikihow, BBC Mobile, and Citymapper) because they are highly popular and representative.}
 
\JMRCHANGE{To address this threat, taking advantage of our Grid infrastructure, for each test budget (i.e., one hour and five hours), for each subject App version, we executed all the testing tools in parallel in five batches with two sequential executions each. In practice, for each subject App's version, for each tool, we ran ten executions distributed over a time frame of two hours (for a one-hour budget) and ten hours (for a five-hour budget).
Our experimental configuration should minimize the threat for Apps (i.e., Wikipedia, Wikihow, and Citymapper) loading remote content 
that unlikely changes in the time frame of our executions (i.e., ten hours max). In addition, we believe that our configuration also addresses the threat for the remaining three Apps (i.e., Nuzzel, Yahoo weather, BBC Mobile) because, by running all the different testing tools in parallel, we maximize the likelihood of processing the same remote content (i.e., news or weather forecasts) when triggering the same Actions.
}

\subsubsection{Conclusion validity}
\JMR{3.6}{Threats to \emph{conclusion validity} concern the statistical power of our results, invalid statistical test assumptions, reliability of measurements,  and random irrelevancies~\cite{Wohlin2012}.}

\JMRCHANGE{Since the underlying distribution of the data (i.e., code coverage achieved with test automation approaches) is not known in our context, for statistical significance, we rely on the Mann Whitney U-test, which has high \emph{statistical power} for different underlying distributions, even for a small number of samples~\cite{Shieh2006}.
Also, to let the readers draw conclusions in context about the proposed approach, we report both p-values and effect sizes.}
 
\JMRCHANGE{To avoid violating the \emph{assumptions of parametrical statistical tests}, we rely on a non-parametric test and effect size measure (i.e., Mann Whitney U-test and the Vargha and Delaney’s $A_{12}$ statistics, respectively).}

\JMRCHANGE{To ensure \emph{reliability}, our measurements (i.e., code coverage) have been collected through widely used, open-source tools.}
 
\JMRCHANGE{In our context, the only source of \emph{random irrelevancies} might be the workload of the machines used to run the experiments, which may slow down the performance of some of the tools. To mitigate this threat, in addition to rely on a Grid infrastructure with guarantees for the provided service level, we manually inspected execution logs to exclude the presence of anomalies biasing results (e.g., exceptions due to the host environment).
}

\subsubsection{Construct validity}
\JMR{3.6}{According to standard practice, we discuss \emph{construct validity} in terms of 
face, content, convergent, and predictive validity~\cite{Ralph2018}.
The constructs considered in our work are effectiveness and cost. 
Effectiveness is measured through two reflective indicators, which are target method coverage and target instruction coverage.
Cost is measured in terms of the number of inputs being generated, for reasons that were carefully discussed.
}

\JMRCHANGE{\emph{Face validity} concerns the selection of appropriate reflective indicators.
For effectiveness, we rely on method and instruction coverage, which is common practice~\cite{Choudhary-AutomatedTestInputGeneration-ASE-2015,DiMartino2020}.
For cost, we measure the number of inputs being generated. At the beginning of Section~\ref{sec:empirical}, we have discussed that, in our context, the number of inputs is a good surrogate to enable the comparison of testing cost.} 

 \JMR{2.2}{\emph{Content validity} concerns the adequacy of reflective indicators to cover the breadth of the construct.
We rely on code coverage since it has been recently shown that there is moderate to high correlation between code coverage and detection of real faults~\cite
{Kochhar2015}.
Also, code coverage is a necessary condition to uncover faults and, therefore, it remains a priority for test engineers.
Concerning the breadth of the cost construct, 
in the introduction to this Section we have discussed that a a direct and precise cost estimate 
can only be obtained with experiments involving engineers using the selected testing techniques in the field under controlled conditions, which we leave to future work.}

\JMR{3.6}{Concerning \emph{convergence}, we have computed the non-parametric Kendall's correlation coefficient, 
for all the pairs of reflective indicators, for each subject. 
Unsurprisingly, target method and target instruction coverage are highly correlated (i.e., $\tau \ge 0.7$, for all the subjects), which is expected for reflective indicators used to infer the same construct. Instead, a low correlation (i.e., $\tau < 0.35$, for all the subjects) 
is observed between inputs being triggered and target coverage, which is expected since these two reflective indicators are used for distinct constructs.}

\JMRCHANGE{To address \emph{predictive validity}, we reported statistics for all our research questions.}

\subsubsection{External validity}

To address threats to \emph{external validity} we have considered nine popular Apps, downloaded thousands of times worldwide, that have been considered in the empirical evaluation of related work. Also, for each App, we considered up to ten App versions, based on their availability, for a total of 72 App versions tested. 
The considered Apps greatly vary regarding the overall number of lines of code and updated lines between versions. Because of their diversity, we believe our subjects to be representative of the Apps landscape.

To account for randomness, we tested each App version ten times with every testing tool considered; more than the usual practice of three to five repetitions~\cite{Gu:APE:ICSE:2019}. Despite the high computational cost (17280 test execution hours, in total), this enabled us to derive solid statistical results for the comparison of  different tools.

In our experiments we considered only Android Apps, which is standard practice in most App testing research papers. 
The prevalence of Android in research papers is mostly due to its worldwide dissemination and the availability of a larger set of tools to test and analyze Android executable bytecode~\cite{Alvarez:WAMA:2019}.
In our work, the choice of relying on Android Apps enabled the comparison of \APPR with tools working for Android Apps (i.e., Monkey, APE, and DM2). However, since we do not exercise OS-specific features, results should generalize also to Apps running in different execution environments (e.g., HarmonyOS and IOS).

\section{Related Work}
\label{sec:related}

\CHANGETWO{In this section we discuss related work, which covers automated App testing tools, App regression testing techniques, incremental testing approaches, testing based on information retrieval, and test oracle automation.}

\subsection{App testing tools}

Automated App testing tools can be grouped according to the strategy adopted to generate test inputs~\cite{Linares:ICSME:2017,Amalfitano:AndroidTestingSurvey:SQJ:2018}. The most common ones are random, model-based, and evolutionary~\cite{Amalfitano:AndroidTestingSurvey:SQJ:2018}. Representative approaches of these three categories used in empirical evaluations are Monkey, Stoat~\cite{stoat17}, and Sapienz~\cite{sapienz}, respectively. Monkey has been introduced in Section~\ref{sec:empirical}. 
\FIXME{Other recent and effective approaches either rely on Q-Learning~\cite{Pan:ReinforcementLearning:2020,Koroglu2018}, execution traces~\cite{Combodroid:Wang2020}, or state cloning~\cite{Dong:20}.}

\emph{Stoat} performs stochastic model-based testing. It relies on dynamic analysis based on a weighted UI exploration strategy 
 to derive a stochastic finite state machine (FSM) of the App's GUI interactions. 
 Stoat relies on the FSM to generate test suites using an objective function that aims to maximize
code coverage, model coverage, and test diversity. The test generation process relies on Gibbs sampling to iteratively mutate and refine the FSM, based on the fitness of the generated test suite.

\emph{Sapienz} is an evolutionary approach that uses Pareto multi-objective search to automatically explore and optimise
test sequences, minimising length, while simultaneously maximising coverage and fault detection.  
Sapienz combines random fuzzing,
systematic exploration and search-based exploration.

Independent empirical evaluations performed by Choudhary et al.~\cite{Choudhary-AutomatedTestInputGeneration-ASE-2015} and Wang et al.~\cite{Wang:EmpStudy:2018} have reported that the three aforementioned testing strategies are complementary. 
Further, both studies show that the method and instruction coverage achieved by all test automation approaches are relatively low, that is below 50\%. 
Choudhary et al.~\cite{Choudhary-AutomatedTestInputGeneration-ASE-2015} note that model-based approaches complement random approaches regarding fault detection, while for code coverage, random approaches fare better.
Wang et al.~\cite{Wang:EmpStudy:2018} confirm these results. They report that random and evolutionary approaches are complementary regarding method coverage, while both evolutionary and model-based approaches complement random approaches in terms of fault detection. 
However, the validity of these findings has been weakened by recent advances in model-based approaches.
Indeed, more recent results show that model-based approaches that either integrate advanced exploration strategies (i.e., biased random in DM2) or adaptable state abstraction functions (i.e., APE~\cite{Gu:APE:ICSE:2019}) fare better than random approaches or state-of-the-art model-based approaches. APE, for example, is the most recent technique, and has been reported to perform better than Monkey, Sapienz, and Stoat.

In \emph{APE}, each window is modeled with sets of \emph{attribute paths} that univocally identify the widgets of the window.
Attribute paths resemble the AttributeValuationMaps of \APPR with the difference that \APPR considers a larger set of attributes than APE, which only accounts for type, position with respect to siblings, and appearance (e.g., text).
APE and \APPR differ for the strategy used to generate inputs (i.e., \APPR relies on the combination of static and dynamic analysis). They both rely on an adaptable state abstraction function \RF. However the state abstraction functions integrated in the two approaches present key differences.
In APE, a single \RF is defined for the whole app, while \APPR specifies one \RF for each window of the App.
Also, APE's \RF is implemented by means of a decision tree (DT), which enables APE to not rely on a predefined set of reducer functions.
The main limitation of APE is that it relies on a global abstraction function for the whole App and thus uses part of the test budget to perform \RF refinements that may be avoided by relying on static analysis (i.e., GUITrees belonging to different windows should be characterized by different abstract states).
By relying on a different \RF for every window, \APPR overcomes such limitation.
In addition, our empirical evaluation has shown that the combination of static and dynamic program analysis enables \APPR to outperform APE in terms of coverage of updated methods, while minimizing the human effort required by testing.

\JMR{1.3}{Approaches not relying on random, model-based, or evolutionary solutions 
make use of either Q-Learning~\cite{Pan:ReinforcementLearning:2020,Koroglu2018}, execution traces~\cite{Combodroid:Wang2020}, or state cloning~\cite{Dong:20}.
 \emph{Q-testing}~\cite{Pan:ReinforcementLearning:2020} relies on Q-learning (a model-free reinforcement learning algorithm~\cite{Watkins:1989}) to trigger events that enable the exploration of features that are not yet tested.  In Q-testing, an event is rewarded if it leads to an App state not visited before. A siamese neural network is used to compare states.
\emph{QBE} is a less recent approach based on Q-learning, which
relies on automated App exploratory testing to identify App states~\cite{Koroglu2018}. We did not select Q-testing~\cite{Pan:ReinforcementLearning:2020} for our empirical evaluation because there is no empirical evidence demonstrating it outperforms APE. For QBE, however, existing evidence shows it does not outperform state-of-the-art approaches.
\emph{Combodroid} works by combining test input sequences derived from execution traces~\cite{Combodroid:Wang2020}. It can work with either traces collected by automated testing tools (e.g., APE) or traces collected when end-users exercise the App under test. 
In the first case, results show that Combodroid achieves higher code coverage than APE when using more than six hours of test budget, 
which is not feasible in continuous integration contexts where test cases are always executed after code commits.
The need for execution traces collected with humans in the loop makes the second usage scenario of Combodroid inapplicable in our context. For these reasons, we did not compare \APPR with Combodroid. 
\emph{TimeMachine} implements a metaheuristic approach that relies on a pool of App states~\cite{Dong:20}. New App states are reached by triggering random events. An App state is added to the pool only if it is reached after exercising code not covered yet; App states are captured by cloning the state of the virtual machine running the App under test. When lack of progress is detected, TimeMachine resets the execution from the state with the highest fitness, which is computed by balancing the number of times the state has been visited and the number of interesting states generated from it. Similar to Combodroid, TimeMachine overcomes Monkeys, Stoat, and Sapienz when executed for more than five hours; however, as discussed above, this setting makes it inapplicable in some continuous integration contexts. 
For the reasons above, we did not consider ComboDroid, Q-testing, and TimeMachine to be suitable candidates for our empirical evaluation.}

\JMRCHANGE{Most importantly, none of the existing App testing approaches prioritize the testing of App updates. 
\textbf{\APPR is the first solution addressing App updates} by focusing on updated methods.
To efficiently test updated methods, \APPR integrates a model-based approach that combines static and dynamic program analyses. 
In the App context, static analysis is the most appropriate solution to efficiently identify test inputs because (1) Apps architecture enables the extraction of models capturing window transitions through static analysis and (2) call graph analysis enables the identification of the inputs that trigger event handlers reaching modified code. Dynamic analysis, instead, enables overcoming static analysis limitations; for this reason, \APPR associates abstract states derived through model-based exploration (i.e., dynamic analysis) to windows in EWTGs derived with static analysis.} 

\JMRTWO{1.1}{Alternative potential solutions, based on augmenting exiting approaches, can be considered to test updated Apps. For example, APE might be extended to spend more test budget on states triggering updated methods, Q-testing might be extended by increasing the reward for updated methods, ComboDroid by increasing the chance of exercising use cases related to changed classes, while TimeMachine’s fitness could be adjusted to prioritize change-related states. However, such solutions will unlikely reach updated code efficiently if they are not combined with the innovative contributions provided by \APPR, that is, procedures to combine information collected with static and dynamic program analysis.} 

\JMRTWO{1.1}{To test updated Apps, APE needs to be extended with the Phase2 and Phase3 strategies implemented by \APPR. More precisely, APE natively works by prioritizing inputs (called model actions in APE) not exercised in an App state and extending APE to prioritize target inputs based on static program analysis data (e.g., the list of target inputs generated by Extended gator, one of our contributions) may not be sufficient to efficiently test upgrades. Indeed, based on our results, when testing an updated App, it is often necessary to exercise a same Window multiple times with a same target input in order to reach the desired abstract state (this happens in \APPR{’}s Phase 2, as shown in our running example); also, it may be necessary to exercise related windows in order to cover a target method (this happens in \APPR{’}s Phase 3 and it makes \APPR more effective than related approaches, as discussed in RQ3). Considering that APE does not even track coverage information (e.g., what inputs increase coverage) this would basically lead to a re-implementation of \APPR with a different state abstraction function (i.e., the one used by APE, which, however, does not support integration with static analysis). Concerning ComboDroid, in addition to static program analysis, it would be necessary to extend it with a dynamically refined state abstraction function like the one implemented by \APPR. Indeed, for both automatically and manually derived use cases, ComboDroid relies on a static state abstraction function to transform a sequence of events (automatically generated or manually triggered by engineers) into an extended labeled transition system that is similar to \APPR{’}s DSTG. However, ComboDroid’s state abstraction function implements a transformation that is similar to the one achieved with ATUA’s abstraction level L1, that is, it ignores text content and children widgets. As shown in our running example, which is based on our case study subjects, a dynamically refined state abstraction function is necessary to reach the abstract states required to exercise updated features. Moreover, in the App upgrade context, where new features may be introduced into the App under test, the need for traces manually recorded by engineers may largely limit the benefits of test automation; indeed, to record such traces, engineers may need to exercise all the features introduced into the App under test, that is, manually perform system-level testing of the App, which is what we aim to automate with \APPR. TimeMachine, similar to APE, would also need to integrate the \APPR{’}s strategy for the selection of test inputs; an alternative is to extend \APPR to rely on TimeMachine’s time travel features to quickly reach abstract states previously visited (e.g., at the beginning of testing); we leave it for future work. Q-testing, instead, is unlikely to achieve the results obtained with ATUA’s Phase 3; indeed, without a predefined strategy to select related windows it is unlikely to automatically discover them by means of trial and error (i.e., what Q-learning does in its training phase). Indeed, as shown in our discussion for RQ3, randomly generated inputs do not enable test automation tools (i.e., Monkey, APE, and DM2) to reach some of the updated methods that need to be enabled in Windows different than the one with a target Widget.} 

\JMRTWO{1.1}{To summarize, state-of-the-art techniques are unlikely to achieve \APPR's performance by augmenting them with simple input selection strategies based on static program analysis (e.g., prioritize inputs that may trigger modified methods, based on static program analysis); also, they can’t be efficiently integrated with all the solutions provided by \APPR: (1) APE does not track code coverage, which is required to drive testing, nor has a state abstraction function been integrated with static program analysis, (2) in our context, the manual recording of traces required by ComboDroid  requires manual testing of the updated App features (i.e., what \APPR automates), (3) Q-testing is based on a paradigm (i.e., reinforcement learning) that can’t make use of a static analysis model and can’t be integrated with the three \APPR{’}s phases, (4) the time travel feature proposed by TimeMachine might be integrated into \APPR but, since it has shown to be effective, such integration may be the focus of future work.}

\subsection{App regression testing techniques}

Regression testing techniques for Apps concern the selection of events that may trigger modified code~\cite{Sharma:QADroid:ISSTA:2019}, the selection of regression test cases~\cite{Choi:DetReduce:ICSE:2018}, and the repair of existing test suites~\cite{Song:XPathRepair:2017}.
QADroid is a static analysis toolset that identifies the events (i.e., the Inputs of \APPR EWTG) that may trigger the execution of modified methods.
It implements the features of Steps 1 and 2 of \APPR, except that QADroid does not generate EWTGs.
QADroid differs from \APPR regarding the underlying static analysis framework used to identify Inputs. It relies on FlowDroid~\cite{FlowDroid}, while \APPR relies on Gator because of its capability to generate WTGs.
Also, empirical results show that Gator performs better than FlowDroid in identifying valid sequences of callbacks~\cite{Wang:FlowDroidComparison:2016}.
Different from \APPR, to select the Inputs that may trigger modified methods, QADroid performs a forward traversal of the App control flow graph obtained with Soot (i.e., it starts the traversal from event handler methods) and selects all the Inputs that reach modified methods. Consequently, QADroid cannot determine the presence of \emph{HiddenHandlers} identified by \APPR.
In addition, QADroid requires the source code of the App while \APPR works with the bytecode.
QADroid has never been applied to automated testing and cannot identify the concrete input values to be used with certain Inputs (e.g., the data to write into a TextArea).

\emph{Redroid} selects a regression test suite for an  updated version of an App~\cite{Redroid:16,Redroid:Mobisoft:2016} from an existing test suite. 
It relies on state-of-the-art static analysis procedures~\cite{Rothermel:Dejavu:1997} to perform change impact analysis and uses code coverage information to select any test case that cover modified blocks of code.
Redroid does not support the generation of a minimal test suite. 
\emph{DetReduce}~\cite{Choi:DetReduce:ICSE:2018}, instead, creates a small regression test suite for an App from a test suite generated by a model-based test automation tool.
It identifies and removes redundant method call traces and subtraces within the test suite and redundant loops within a test case. 
Redundancy is identified based on a state abstraction function that considers actionable widgets and their visible attribute values. Widgets are identified by their path in the GUITree.
Since executable test suites are often unavailable and state-of-the-art App testing tools can cover only a narrow set of modified methods, the applicability of Redroid and DetReduce remains limited. 
\JMR{1.2}{However, \emph{DetReduce} might be applied to \APPR App models to further reduce the generated test cases;
we leave to future work the integration of \APPR with DetReduce and its evaluation.
Another solution for the generation of reduced test suites is the one proposed by Jabbarvand et al.~\cite{Jabbarvand2016} in the context of energy-aware test-suite minimization. In their work, they stop and restart the execution of Monkey every 500 events; a sequence of 500 events is a test case. When the test budget is exhausted, a greedy algorithm identifies the minimal set of test cases that maximize the coverage of energy-greedy parts of the App.
Such solution might be adapted to work with \APPR and the state-of-the-art approaches considered in our experiments; for example, by restarting test generation after a predefined number of events and then by relying on the greedy algorithm to select the minimal set of test cases (e.g., up to 2000 inputs) that maximizes the coverage of updated methods. However, though such an approach might reduce the size of the generated test suites, it might not be feasible to identify a configuration that maximizes the benefits for a range of Apps. Indeed, for some Apps, 500 events might not be sufficient to reach an App state that enables exercising updated methods, especially for random approaches (e.g., if updated methods require long, specific sequences of events to be exercised), while a much larger number of events would greatly reduce the benefits of test suite reduction (e.g., for one hour budget, the test cases generated by \APPR consist of less than 2000 events, see Figure~\ref{fig:RQ3:Inputs:1:Zoom}). 
We therefore leave to future work the integration of such test suite minimization solution into \APPR.}

\CHANGED{Automated repair techniques for the GUI test scripts of mobile Apps are still preliminary~\cite{Song:XPathRepair:2017,Pan:GUIrepair:2019,Pan2020}; to repair test scripts they include strategies ranging from static program analysis~\cite{Song:XPathRepair:2017}, to model-based~\cite{Li:Atom:2017} and computer vision techniques~\cite{Pan:GUIrepair:2019,Pan2020}. Existing approaches either leave between 5\%~\cite{Pan2020} and 8\%~\cite{Song:XPathRepair:2017} of the test scripts to be manually repaired or do not preserve all the test actions (i.e., the test semantic~\cite{Li:Atom:2017}). 
Though these results show that automated GUI script repair techniques might be adopted to support the oracle automation approach we suggested for the identification of regression failures, manual intervention would still be required, as indicated in Section~\ref{sec:discussion}.}

\subsection{Incremental testing approaches}

Campos et al.
~\cite{ContinuousTestGeneration} have been the first to 
propose a technique to incrementally test the units in a software project, leading to overall higher code coverage while reducing the time spent on test generation. They apply an evolutionary approach (i.e., EvoSuite~\cite{EvoSuite}) and optimize test generation by providing more test budget to the modified portions of the code and reuse already generated test cases for seeding (i.e., they re-run all the test cases that compile). The main difference with \APPR is that they do not target the GUI testing of Apps but the unit testing of Java libraries. In our context, the reuse of existing test cases is complicated by the presence of a GUI, which is likely updated across versions and may break existing test sequences.

\subsection{Testing based on information retrieval}

In the software testing literature, \emph{information retrieval} techniques applied to the processing of program files have been integrated into fault localization~\cite{Youm:2015,Wen:2016},
test case prioritization~\cite{Saha:IRprioritization:2015}, and test input selection approaches~\cite{Toffola:MiningInputs:ASE:2017,Rau:2018}.
Concerning test input generation, TestMiner relies on information retrieval to select from a large corpus of existing test cases the input values to use in newly generated tests cases, which differs from our purpose~\cite{Toffola:MiningInputs:ASE:2017}. Poster applies a similar approach to derive sequences of Inputs for App testing~\cite{Rau:2018}. 
Different from \APPR, it relies on existing test suites developed for similar Apps.

\subsection{Test oracle automation} 

Like all the state-of-the-art approaches for App testing~\cite{Linares:ICSME:2017,Rubinov:AndroidAppTestingSurvey:2018}, \APPR does not address the \emph{oracle problem}~\cite{Barr2015}. 
\CHANGED{Oracle automation is part of our future work; however, we have presented in Section~\ref{sec:empirical} two solutions to alleviate the oracle problem in the context of App updates. 
One solution is the automated detection of functional regression faults based on the identification of unexpected changes to outputs across different software versions~\cite{Bogdan:1998,Shamshiri:2013,Pastore:2014,Gao:GUIRegressionTesting:ASE:2015}. The other solution consists in relying on crowdsourcing, a popular solution adopted by industry to reduce the costs of manual GUI testing for Apps~\cite{Minzhi:2014,ListCrowdsourcing}. Related work has shown that it is feasible for crowd workers to identify errors after visualizing the inputs and outputs of the functions under test~\cite{Pastore:CrowdOracles}; in the App context, crowd oracles may lower testing costs while test coverage is addressed by test automation.} 
In addition, \APPR could also be integrated with approaches that automatically generate in-program logical assertions 
~\cite{Jahangirova:Oracles:2016} or solutions relying on system-independent GUI oracles~\cite{Zaeem:UserInteractions:ICST:2014}. 

\subsection{Summary}

From the above, we can see that existing work lacks a model-based approach combining adaptable state abstraction functions with information retrieved from static program analysis. In \APPR, this is done to maximize the coverage of updated methods while minimizing the number of inputs required for testing. The latter is necessary to make the verification of Apps results feasible. Indeed, fully automated test oracles are currently not an option and human effort is necessary to identify both regressions and failures in new features. 
Regression testing approaches, which typically target test case selection and prioritization, are of limited applicability in this context, when test suites covering large portions of the Apps code are not available. \APPR leverages incremental testing to effectively invest the test budget and maximize the coverage of updated methods.

\section{Conclusion}
\label{sec:conclusion}

\CHANGEDOCT{State-of-the-art App testing techniques are affected by two limitations: 
limited effectiveness (i.e., low code coverage) and absence of automated oracles. To address the first limitation, given the high release frequency of Apps, we propose a solution (objective O1) to effectively focus the test budget on updated (i.e., modified and new) methods. 
In other words, within practical test execution time, we aim to maximize the coverage of updated methods and their instructions.
To address the second limitation, we aim 
(objective O2) to generate a significantly reduced set of test inputs, compared to state-of-art approaches, thus proportionally saving the corresponding human effort required to visualize test outputs or correct test scripts.}

\CHANGEDOCT{To achieve the two objectives above, we developed \APPR, an automated App testing technique that integrates multiple analysis strategies. To achieve O1, it combines static analysis, to determine the inputs that execute updated features, and random exploration, \CHANGEDNOV{to overcome the limitations
of static analysis.}
To achieve O2, it relies on  dynamically-refined state abstraction functions, to determine when distinct inputs lead to a same program state, and relies on information retrieval techniques, to identify dependencies among App features.}

\CHANGEDOCT{We performed an empirical evaluation where we compared \APPR with state-of-the-art approaches implementing testing strategies based on dynamically derived models (DM2), random exploration (Monkey), and dynamic state abstraction (APE). For our experiments, we considered practical execution time budgets of one and five hours, corresponding respectively to approximate time constraints in the context of continuous integration and overnight testing. Concerning human effort (objective O2), \APPR is the approach that generates the smallest set of inputs with the highest coverage per input.
\APPR, on average across subject Apps, saves around 
\FIXME{32.6\%} of the effort, compared to the second-best approach (DM2). Further, it exercises \FIXME{38.5\%} more instructions than DM2 per input. Differences with APE and Monkey are much larger. 
Concerning effectiveness within time budget (objective O1), on average, \APPR automatically exercises up to 70\% of updated methods and 60\% of instructions belonging to updated methods, 6\% more than the second best approach (i.e., DM2). 
These results show that the analysis strategies integrated in \APPR can drive testing towards an efficient use of the test budget (execution time and effort), thus providing clear benefits when upgrading and testing an App.
}

\begin{acks}
This project has received funding from Huawei Technologies Co., Ltd, China, and the European Research Council (ERC) under the European Union’s Horizon 2020 research and innovation programme (grant agreement No 694277).
Experiments presented in this paper were carried out using the Grid'5000 testbed, supported by a scientific interest group hosted by Inria and including CNRS, RENATER and several Universities as well as other organizations (see https://www.grid5000.fr).
Authors would like to thank Erik Derr for the initial implementation of AppDiff.
\end{acks}

\appendix

\section{\APPR Toolset}
\label{sec:toolset}

\begin{figure}[tb]
  \centering
	\includegraphics[width=14cm]{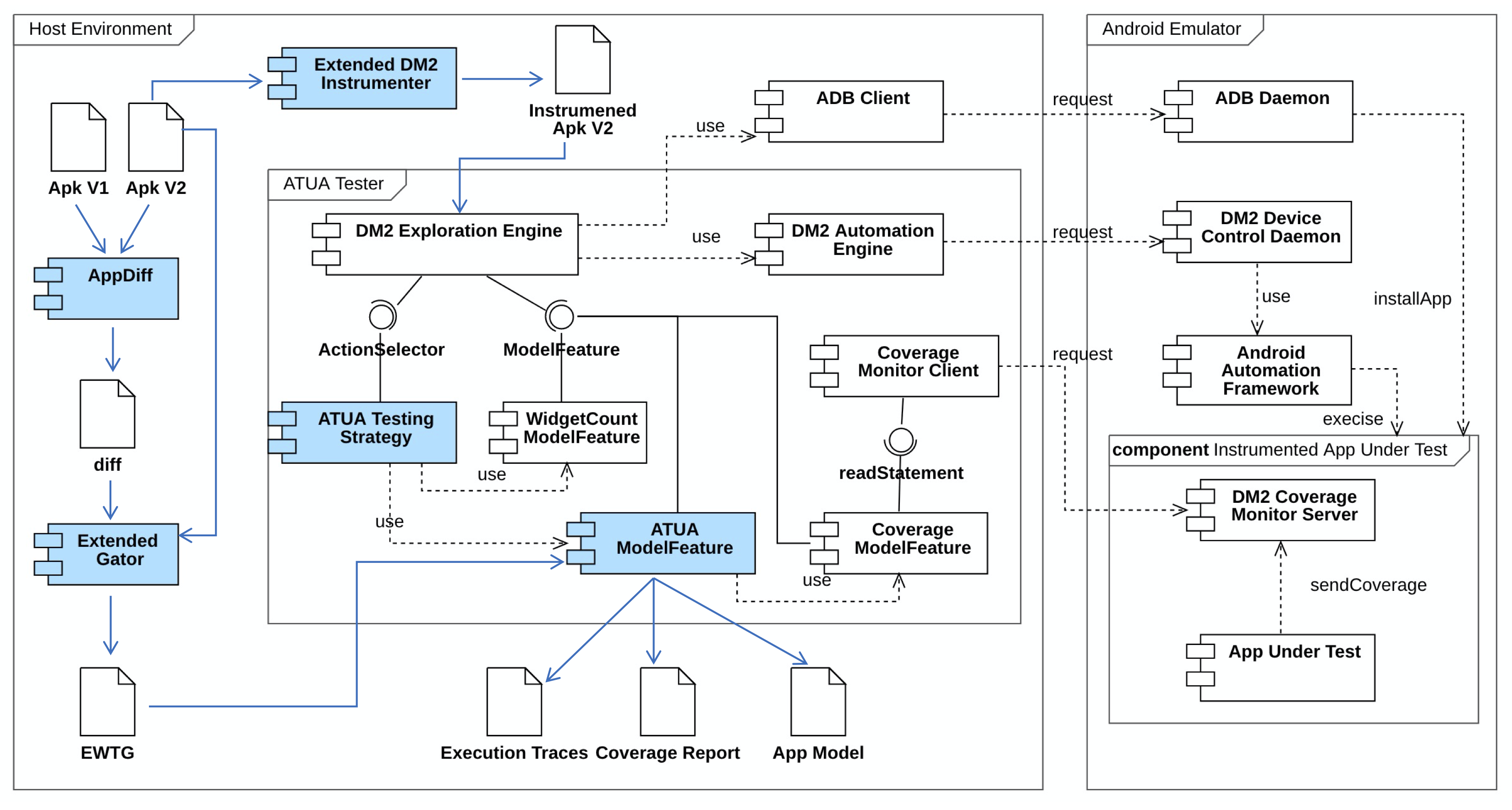}
	\footnotesize
	\textbf{Legend:}  White UML component symbols point to third-party, reused components. Blue UML component symbols highlight components developed from scratch or extended to support ATUA's features.
      \caption{Overview of the \APPR{} toolset.}
      \label{fig:toolset:architecture}
\end{figure}

\JMR{3.12}{The \APPR Toolset includes four main components: \emph{\appdiff}, which identifies the updated methods for the App under test, \emph{Extended Gator}, which generates the EWTG part of the AppModel, \emph{Extended DM2 Instrumenter}, which instruments the App under test, and \emph{\APPR Tester}, which implements the \APPR testing algorithm. The UML component diagram in Figure~\ref{fig:toolset:architecture} shows the \APPR Toolset.}

\JMRCHANGE{The features of \appdiff and {Extended Gator} have already been presented in Sections~\ref{sec:approach:identify.updated.methods} and~\ref{sec:approach:EWTG}, respectively. In this Section, we focus on the description of the {Extended DM2 Instrumenter} and the \APPR Tester.}

\JMRCHANGE{\APPR Tester has been implemented as an extension of DM2.
DM2 consists of six components (i.e., DM2 Instrumenter, DM2 Exploration Engine, DM2 Automation Engine, Coverage Monitor Client, Coverage Feature, and Widget Counting Model Feature) that  are executed on the host environment and two components (i.e., DM2 Control Device Daemon, and DM2 Coverage Monitor Server) that are deployed on the Android emulator running the App under test. The DM2 components are part of \APPR Tester, 
which automatically deploys and executes them transparently from the end-user.
\APPR Tester integrates two additional components that implement the \APPR algorithm (i.e.,  \APPR Testing Strategy and \APPR Model Feature). The integration between \APPR Tester components and DM2 is performed through the interfaces provided by DM2 (i.e., ModelFeature and ActionSelector).
\APPR Tester and DM2 rely on three additional components provided by the Android development environment: the ADB Client, the ADB Daemon, and the Android Automation Framework.}

\JMRCHANGE{The \emph{Extended DM2 Instrumenter} is used before testing to create an instrumented version of the App under test that integrates the functions required to collect code coverage. We have extended the \emph{DM2 Instrumenter} to collect method coverage information in addition to instruction coverage. Method coverage is used by \APPR to quickly determine at runtime which methods have been covered.}

\JMRCHANGE{At runtime, during testing, the \emph{DM2 Exploration Engine} acts as a controller that queries the \emph{\APPR Strategy} component, which implements the DM2 \emph{ActionSelector} interface, used by DM2 to select the next Action to trigger during testing. The \emph{\APPR Strategy} component implements the \APPR's testing algorithm. 
The DM2 Exploration Engine relies on the \emph{ADB Client} installed on the host to set up the Android emulator and deploy the App under test.
The interaction with the App under test is managed by the \emph{DM2 Automation Engine}, which sends commands to the \emph{DM2 Device Control Daemon} installed on the Android emulator. The \emph{DM2 Device Control Daemon} employs the \emph{Android Automation Framework} to execute the requested Action on the App under test and to derive the GUITree for the active Window.
After triggering an Action, the \emph{DM2 Automation Engine} receives the current GUITree, a screenshot of the Android emulator GUI, and some additional information (e.g., exception trace from Logcat) from the \emph{DM2 Device Control Daemon}. 
It then derives the GUITreeTransition performed on the GSTG and sends this information to all the registered \emph{Model Features}, including the {Widget Counting Model Feature}, the DM2's Coverage Feature, and the  \APPR App Model.
The \emph{Widget Counting Model Feature} calculates the frequency of Actions on GUI widgets, used to drive \APPR's random exploration.
The \emph{DM2 Coverage Feature} is responsible for tracking code coverage during testing. It associates to each Action the set of instructions covered when executing the Action; code coverage is provided by the \emph{Coverage Monitor Server} instrumented by DM2 within the App under test. 
The \emph{\APPR Model Feature} updates the App model consistently with the description provided in Section~\ref{sec:approach:app.model.metamodel}; for example, it implements the \APPR state abstraction mechanism. The \emph{\APPR Model Feature} is queried by the \emph{\APPR Strategy} to determine the Actions to trigger (e.g., to identify the shortest sequence of Actions required to reach a TargetWindow). 
The \emph{\APPR Model Feature} relies on the \emph{DM2 Coverage Feature} to acquire the coverage data and associate them with the App Model. At the end of the testing, the \APPR Model Feature generates ATUA's outputs (i.e., coverage report, execution traces, and the final App model). }

\bibliographystyle{ACM-Reference-Format}
\bibliography{AutAut}


\begin{thebibliography}{81}


\ifx \showCODEN    \undefined \def \showCODEN     #1{\unskip}     \fi
\ifx \showDOI      \undefined \def \showDOI       #1{#1}\fi
\ifx \showISBNx    \undefined \def \showISBNx     #1{\unskip}     \fi
\ifx \showISBNxiii \undefined \def \showISBNxiii  #1{\unskip}     \fi
\ifx \showISSN     \undefined \def \showISSN      #1{\unskip}     \fi
\ifx \showLCCN     \undefined \def \showLCCN      #1{\unskip}     \fi
\ifx \shownote     \undefined \def \shownote      #1{#1}          \fi
\ifx \showarticletitle \undefined \def \showarticletitle #1{#1}   \fi
\ifx \showURL      \undefined \def \showURL       {\relax}        \fi
\providecommand\bibfield[2]{#2}
\providecommand\bibinfo[2]{#2}
\providecommand\natexlab[1]{#1}
\providecommand\showeprint[2][]{arXiv:#2}

\bibitem[\protect\citeauthoryear{Amalfitano, Amatucci, Memon, Tramontana, and
  Fasolino}{Amalfitano et~al\mbox{.}}{2017}]%
        {Amalfitano2017}
\bibfield{author}{\bibinfo{person}{Domenico Amalfitano},
  \bibinfo{person}{Nicola Amatucci}, \bibinfo{person}{Atif~M. Memon},
  \bibinfo{person}{Porfirio Tramontana}, {and} \bibinfo{person}{Anna~Rita
  Fasolino}.} \bibinfo{year}{2017}\natexlab{}.
\newblock \showarticletitle{{A general framework for comparing automatic
  testing techniques of Android mobile apps}}.
\newblock \bibinfo{journal}{\emph{Journal of Systems and Software}}
  \bibinfo{volume}{125} (\bibinfo{year}{2017}), \bibinfo{pages}{322--343}.
\newblock
\showISSN{01641212}
\urldef\tempurl%
\url{https://doi.org/10.1016/j.jss.2016.12.017}
\showDOI{\tempurl}


\bibitem[\protect\citeauthoryear{Amalfitano, Riccio, Amatucci, Simone, and
  Fasolino}{Amalfitano et~al\mbox{.}}{2019}]%
        {AMALFITANO201995}
\bibfield{author}{\bibinfo{person}{Domenico Amalfitano},
  \bibinfo{person}{Vincenzo Riccio}, \bibinfo{person}{Nicola Amatucci},
  \bibinfo{person}{Vincenzo~De Simone}, {and} \bibinfo{person}{Anna~Rita
  Fasolino}.} \bibinfo{year}{2019}\natexlab{}.
\newblock \showarticletitle{Combining Automated GUI Exploration of Android apps
  with Capture and Replay through Machine Learning}.
\newblock \bibinfo{journal}{\emph{Information and Software Technology}}
  \bibinfo{volume}{105} (\bibinfo{year}{2019}), \bibinfo{pages}{95--116}.
\newblock
\showISSN{0950-5849}
\urldef\tempurl%
\url{https://doi.org/10.1016/j.infsof.2018.08.007}
\showDOI{\tempurl}


\bibitem[\protect\citeauthoryear{Android.com}{Android.com}{2020a}]%
        {intent}
\bibfield{author}{\bibinfo{person}{Android.com}.}
  \bibinfo{year}{2020}\natexlab{a}.
\newblock \bibinfo{title}{{Intent Resolution}}.
\newblock
  \bibinfo{howpublished}{\url{https://developer.android.com/reference/android/content/Intent}}.
\newblock
\newblock
\shownote{Last visited: 11/11/2020.}


\bibitem[\protect\citeauthoryear{Android.com}{Android.com}{2020b}]%
        {logcat}
\bibfield{author}{\bibinfo{person}{Android.com}.}
  \bibinfo{year}{2020}\natexlab{b}.
\newblock \bibinfo{title}{Logcat command line tool}.
\newblock
  \bibinfo{howpublished}{\url{https://developer.android.com/studio/command-line/logcat}}.
\newblock


\bibitem[\protect\citeauthoryear{Android.com}{Android.com}{2020c}]%
        {monkey}
\bibfield{author}{\bibinfo{person}{Android.com}.}
  \bibinfo{year}{2020}\natexlab{c}.
\newblock \bibinfo{title}{{Monkey - Android ui/application exerciser}}.
\newblock
  \bibinfo{howpublished}{\url{http://developer.android.com/tools/help/monkey.html}}.
\newblock
\newblock
\shownote{Last visited: 12/17/2018.}


\bibitem[\protect\citeauthoryear{Arcuri and Briand}{Arcuri and Briand}{2011}]%
        {Briand:Statistical:2011}
\bibfield{author}{\bibinfo{person}{Andrea Arcuri} {and} \bibinfo{person}{Lionel
  Briand}.} \bibinfo{year}{2011}\natexlab{}.
\newblock \showarticletitle{{A practical guide for using statistical tests to
  assess randomized algorithms in software engineering}}. In
  \bibinfo{booktitle}{\emph{Proceeding of the 33rd international conference on
  Software engineering - ICSE '11}}. \bibinfo{publisher}{ACM Press},
  \bibinfo{address}{New York, New York, USA}, \bibinfo{pages}{1}.
\newblock
\showISBNx{9781450304450}
\showISSN{02705257}
\urldef\tempurl%
\url{https://doi.org/10.1145/1985793.1985795}
\showDOI{\tempurl}


\bibitem[\protect\citeauthoryear{Arcuri and Briand}{Arcuri and Briand}{2014}]%
        {Briand:Hitchhiker:2014}
\bibfield{author}{\bibinfo{person}{Andrea Arcuri} {and} \bibinfo{person}{Lionel
  Briand}.} \bibinfo{year}{2014}\natexlab{}.
\newblock \showarticletitle{{A Hitchhiker's guide to statistical tests for
  assessing randomized algorithms in software engineering}}.
\newblock \bibinfo{journal}{\emph{Software Testing, Verification and
  Reliability}} \bibinfo{volume}{24}, \bibinfo{number}{3}
  (\bibinfo{year}{2014}), \bibinfo{pages}{219--250}.
\newblock
\urldef\tempurl%
\url{https://doi.org/10.1002/stvr.1486}
\showDOI{\tempurl}


\bibitem[\protect\citeauthoryear{Arzt, Rasthofer, Fritz, Bodden, Bartel, Klein,
  Le~Traon, Octeau, and McDaniel}{Arzt et~al\mbox{.}}{2014}]%
        {FlowDroid}
\bibfield{author}{\bibinfo{person}{Steven Arzt}, \bibinfo{person}{Siegfried
  Rasthofer}, \bibinfo{person}{Christian Fritz}, \bibinfo{person}{Eric Bodden},
  \bibinfo{person}{Alexandre Bartel}, \bibinfo{person}{Jacques Klein},
  \bibinfo{person}{Yves Le~Traon}, \bibinfo{person}{Damien Octeau}, {and}
  \bibinfo{person}{Patrick McDaniel}.} \bibinfo{year}{2014}\natexlab{}.
\newblock \showarticletitle{FlowDroid: Precise Context, Flow, Field,
  Object-Sensitive and Lifecycle-Aware Taint Analysis for Android Apps}.
\newblock \bibinfo{journal}{\emph{SIGPLAN Not.}} \bibinfo{volume}{49},
  \bibinfo{number}{6} (\bibinfo{date}{June} \bibinfo{year}{2014}),
  \bibinfo{pages}{259--269}.
\newblock
\showISSN{0362-1340}
\urldef\tempurl%
\url{https://doi.org/10.1145/2666356.2594299}
\showDOI{\tempurl}


\bibitem[\protect\citeauthoryear{Baek and Bae}{Baek and Bae}{2016}]%
        {Baek2016}
\bibfield{author}{\bibinfo{person}{Young-Min Baek} {and}
  \bibinfo{person}{Doo-Hwan Bae}.} \bibinfo{year}{2016}\natexlab{}.
\newblock \showarticletitle{{Automated Model-Based Android GUI Testing Using
  Multi-Level GUI Comparison Criteria}}. In
  \bibinfo{booktitle}{\emph{Proceedings of the 31st IEEE/ACM International
  Conference on Automated Software Engineering}} \emph{(\bibinfo{series}{ASE
  2016})}. \bibinfo{publisher}{Association for Computing Machinery},
  \bibinfo{address}{New York, NY, USA}, \bibinfo{pages}{238--249}.
\newblock
\showISBNx{9781450338455}
\urldef\tempurl%
\url{https://doi.org/10.1145/2970276.2970313}
\showDOI{\tempurl}


\bibitem[\protect\citeauthoryear{Balouek, Carpen~Amarie, Charrier, Desprez,
  Jeannot, Jeanvoine, L{\`e}bre, Margery, Niclausse, Nussbaum, Richard,
  P{\'e}rez, Quesnel, Rohr, and Sarzyniec}{Balouek et~al\mbox{.}}{2013}]%
        {grid5000}
\bibfield{author}{\bibinfo{person}{Daniel Balouek}, \bibinfo{person}{Alexandra
  Carpen~Amarie}, \bibinfo{person}{Ghislain Charrier},
  \bibinfo{person}{Fr{\'e}d{\'e}ric Desprez}, \bibinfo{person}{Emmanuel
  Jeannot}, \bibinfo{person}{Emmanuel Jeanvoine}, \bibinfo{person}{Adrien
  L{\`e}bre}, \bibinfo{person}{David Margery}, \bibinfo{person}{Nicolas
  Niclausse}, \bibinfo{person}{Lucas Nussbaum}, \bibinfo{person}{Olivier
  Richard}, \bibinfo{person}{Christian P{\'e}rez}, \bibinfo{person}{Flavien
  Quesnel}, \bibinfo{person}{Cyril Rohr}, {and} \bibinfo{person}{Luc
  Sarzyniec}.} \bibinfo{year}{2013}\natexlab{}.
\newblock \showarticletitle{Adding Virtualization Capabilities to the
  {Grid'5000} Testbed}.
\newblock In \bibinfo{booktitle}{\emph{Cloud Computing and Services Science}},
  \bibfield{editor}{\bibinfo{person}{Ivan~I. Ivanov}, \bibinfo{person}{Marten
  van Sinderen}, \bibinfo{person}{Frank Leymann}, {and} \bibinfo{person}{Tony
  Shan}} (Eds.). \bibinfo{series}{Communications in Computer and Information
  Science}, Vol.~\bibinfo{volume}{367}. \bibinfo{publisher}{Springer
  International Publishing}, \bibinfo{pages}{3--20}.
\newblock
\showISBNx{978-3-319-04518-4}
\urldef\tempurl%
\url{https://doi.org/10.1007/978-3-319-04519-1\_1}
\showDOI{\tempurl}


\bibitem[\protect\citeauthoryear{Barr, Harman, McMinn, Shahbaz, and Yoo}{Barr
  et~al\mbox{.}}{2015}]%
        {Barr2015}
\bibfield{author}{\bibinfo{person}{Earl~T Barr}, \bibinfo{person}{Mark Harman},
  \bibinfo{person}{Phil McMinn}, \bibinfo{person}{Muzammil Shahbaz}, {and}
  \bibinfo{person}{Shin Yoo}.} \bibinfo{year}{2015}\natexlab{}.
\newblock \showarticletitle{The Oracle Problem in Software Testing: A Survey}.
\newblock \bibinfo{journal}{\emph{IEEE Transactions on Software Engineering}}
  \bibinfo{volume}{41}, \bibinfo{number}{5} (\bibinfo{year}{2015}),
  \bibinfo{pages}{507--525}.
\newblock


\bibitem[\protect\citeauthoryear{Borges~Jr., Hotzkow, and Zeller}{Borges~Jr.
  et~al\mbox{.}}{2018}]%
        {Borges-Droidmate2-ASE-2018}
\bibfield{author}{\bibinfo{person}{Nataniel~P. Borges~Jr.},
  \bibinfo{person}{Jenny Hotzkow}, {and} \bibinfo{person}{Andreas Zeller}.}
  \bibinfo{year}{2018}\natexlab{}.
\newblock \showarticletitle{DroidMate-2: A Platform for Android Test
  Generation}. In \bibinfo{booktitle}{\emph{Proceedings of the 33rd ACM/IEEE
  International Conference on Automated Software Engineering}}
  \emph{(\bibinfo{series}{ASE 2018})}. \bibinfo{publisher}{ACM},
  \bibinfo{address}{New York, NY, USA}, \bibinfo{pages}{916--919}.
\newblock
\showISBNx{978-1-4503-5937-5}
\urldef\tempurl%
\url{https://doi.org/10.1145/3238147.3240479}
\showDOI{\tempurl}


\bibitem[\protect\citeauthoryear{Calciati, Kuznetsov, Bai, and Gorla}{Calciati
  et~al\mbox{.}}{2018}]%
        {Calciati:NewAppReseases:MSR:2018}
\bibfield{author}{\bibinfo{person}{Paolo Calciati}, \bibinfo{person}{Konstantin
  Kuznetsov}, \bibinfo{person}{Xue Bai}, {and} \bibinfo{person}{Alessandra
  Gorla}.} \bibinfo{year}{2018}\natexlab{}.
\newblock \showarticletitle{What Did Really Change with the New Release of the
  App?}. In \bibinfo{booktitle}{\emph{Proceedings of the 15th International
  Conference on Mining Software Repositories}} \emph{(\bibinfo{series}{MSR
  '18})}. \bibinfo{publisher}{Association for Computing Machinery},
  \bibinfo{address}{New York, NY, USA}, \bibinfo{pages}{142--152}.
\newblock
\showISBNx{9781450357166}
\urldef\tempurl%
\url{https://doi.org/10.1145/3196398.3196449}
\showDOI{\tempurl}


\bibitem[\protect\citeauthoryear{Campos, Arcuri, Fraser, and Abreu}{Campos
  et~al\mbox{.}}{2014}]%
        {ContinuousTestGeneration}
\bibfield{author}{\bibinfo{person}{Jos\'{e} Campos}, \bibinfo{person}{Andrea
  Arcuri}, \bibinfo{person}{Gordon Fraser}, {and} \bibinfo{person}{Rui Abreu}.}
  \bibinfo{year}{2014}\natexlab{}.
\newblock \showarticletitle{Continuous Test Generation: Enhancing Continuous
  Integration with Automated Test Generation}. In
  \bibinfo{booktitle}{\emph{Proceedings of the 29th ACM/IEEE International
  Conference on Automated Software Engineering}} \emph{(\bibinfo{series}{ASE
  '14})}. \bibinfo{publisher}{Association for Computing Machinery},
  \bibinfo{address}{New York, NY, USA}, \bibinfo{pages}{55--66}.
\newblock
\showISBNx{9781450330138}
\urldef\tempurl%
\url{https://doi.org/10.1145/2642937.2643002}
\showDOI{\tempurl}


\bibitem[\protect\citeauthoryear{Choi, Sen, Necula, and Wang}{Choi
  et~al\mbox{.}}{2018}]%
        {Choi:DetReduce:ICSE:2018}
\bibfield{author}{\bibinfo{person}{Wontae Choi}, \bibinfo{person}{Koushik Sen},
  \bibinfo{person}{George Necula}, {and} \bibinfo{person}{Wenyu Wang}.}
  \bibinfo{year}{2018}\natexlab{}.
\newblock \showarticletitle{DetReduce: Minimizing Android GUI Test Suites for
  Regression Testing}. In \bibinfo{booktitle}{\emph{Proceedings of the 40th
  International Conference on Software Engineering}}
  \emph{(\bibinfo{series}{ICSE '18})}. \bibinfo{publisher}{Association for
  Computing Machinery}, \bibinfo{address}{New York, NY, USA},
  \bibinfo{pages}{445--455}.
\newblock
\showISBNx{9781450356381}
\urldef\tempurl%
\url{https://doi.org/10.1145/3180155.3180173}
\showDOI{\tempurl}


\bibitem[\protect\citeauthoryear{Choudhary, Gorla, and Orso}{Choudhary
  et~al\mbox{.}}{2015}]%
        {Choudhary-AutomatedTestInputGeneration-ASE-2015}
\bibfield{author}{\bibinfo{person}{Shauvik~Roy Choudhary},
  \bibinfo{person}{Alessandra Gorla}, {and} \bibinfo{person}{Alessandro Orso}.}
  \bibinfo{year}{2015}\natexlab{}.
\newblock \showarticletitle{Automated Test Input Generation for Android: Are We
  There Yet? (E)}. In \bibinfo{booktitle}{\emph{Proceedings of the 2015 30th
  IEEE/ACM International Conference on Automated Software Engineering (ASE)}}
  \emph{(\bibinfo{series}{ASE '15})}. \bibinfo{publisher}{IEEE Computer
  Society}, \bibinfo{address}{Washington, DC, USA}, \bibinfo{pages}{429--440}.
\newblock
\showISBNx{978-1-5090-0025-8}
\urldef\tempurl%
\url{https://doi.org/10.1109/ASE.2015.89}
\showDOI{\tempurl}


\bibitem[\protect\citeauthoryear{Deloitte}{Deloitte}{[n.d.]}]%
        {Deloitte:US:2018}
\bibfield{author}{\bibinfo{person}{Deloitte}.}
  \bibinfo{year}{[n.d.]}\natexlab{}.
\newblock \bibinfo{title}{{2018 Global Mobile Consumer Survey: US Edition}}.
\newblock
  \bibinfo{howpublished}{\url{https://www2.deloitte.com/content/dam/Deloitte/us/Documents/technology-media-telecommunications/us-tmt-global-mobile-consumer-survey-exec-summary-2018.pdf}}.
\newblock


\bibitem[\protect\citeauthoryear{Derr, Bugiel, Fahl, Acar, and Backes}{Derr
  et~al\mbox{.}}{2017}]%
        {derr:ccs17}
\bibfield{author}{\bibinfo{person}{Erik Derr}, \bibinfo{person}{Sven Bugiel},
  \bibinfo{person}{Sascha Fahl}, \bibinfo{person}{Yasemin Acar}, {and}
  \bibinfo{person}{Michael Backes}.} \bibinfo{year}{2017}\natexlab{}.
\newblock \showarticletitle{Keep me Updated: An Empirical Study of Third-Party
  Library Updatability on Android}. In \bibinfo{booktitle}{\emph{Proceedings of
  the 24th ACM Conference on Computer and Communication Security (CCS 2017)}}.
  \bibinfo{publisher}{ACM}.
\newblock


\bibitem[\protect\citeauthoryear{{Di Martino}, Fasolino, Starace, and
  Tramontana}{{Di Martino} et~al\mbox{.}}{2020}]%
        {DiMartino2020}
\bibfield{author}{\bibinfo{person}{Sergio {Di Martino}},
  \bibinfo{person}{Anna~Rita Fasolino}, \bibinfo{person}{Luigi Libero~Lucio
  Starace}, {and} \bibinfo{person}{Porfirio Tramontana}.}
  \bibinfo{year}{2020}\natexlab{}.
\newblock \showarticletitle{{Comparing the effectiveness of capture and replay
  against automatic input generation for Android graphical user interface
  testing}}.
\newblock \bibinfo{journal}{\emph{Software Testing Verification and
  Reliability}} (\bibinfo{year}{2020}), \bibinfo{pages}{1--27}.
\newblock
\showISSN{10991689}
\urldef\tempurl%
\url{https://doi.org/10.1002/stvr.1754}
\showDOI{\tempurl}


\bibitem[\protect\citeauthoryear{Do, Yang, Che, Hui, and Ridgeway}{Do
  et~al\mbox{.}}{2016a}]%
        {Redroid:Mobisoft:2016}
\bibfield{author}{\bibinfo{person}{Quan Do}, \bibinfo{person}{Guowei Yang},
  \bibinfo{person}{Meiru Che}, \bibinfo{person}{Darren Hui}, {and}
  \bibinfo{person}{Jefferson Ridgeway}.} \bibinfo{year}{2016}\natexlab{a}.
\newblock \showarticletitle{Regression Test Selection for Android
  Applications}. In \bibinfo{booktitle}{\emph{Proceedings of the International
  Conference on Mobile Software Engineering and Systems}}
  \emph{(\bibinfo{series}{MOBILESoft '16})}. \bibinfo{publisher}{Association
  for Computing Machinery}, \bibinfo{address}{New York, NY, USA},
  \bibinfo{pages}{27--28}.
\newblock
\showISBNx{9781450341783}
\urldef\tempurl%
\url{https://doi.org/10.1145/2897073.2897127}
\showDOI{\tempurl}


\bibitem[\protect\citeauthoryear{Do, Yang, Che, Hui, and Ridgeway}{Do
  et~al\mbox{.}}{2016b}]%
        {Redroid:16}
\bibfield{author}{\bibinfo{person}{Quan Chau~Dong Do}, \bibinfo{person}{Guowei
  Yang}, \bibinfo{person}{Meiru Che}, \bibinfo{person}{Darren Hui}, {and}
  \bibinfo{person}{Jefferson Ridgeway}.} \bibinfo{year}{2016}\natexlab{b}.
\newblock \showarticletitle{Redroid: A Regression Test Selection Approach for
  Android Applications}. In \bibinfo{booktitle}{\emph{Procedings of the 28th
  International Conference on Software Engineering and Knowledge Engineering}}
  \emph{(\bibinfo{series}{SEKE 2016})}. \bibinfo{pages}{486--491}.
\newblock
\urldef\tempurl%
\url{https://doi.org/10.18293/SEKE2016-223}
\showDOI{\tempurl}


\bibitem[\protect\citeauthoryear{Dom\'{\i}nguez-\'{A}lvarez and
  Gorla}{Dom\'{\i}nguez-\'{A}lvarez and Gorla}{2019}]%
        {Alvarez:WAMA:2019}
\bibfield{author}{\bibinfo{person}{Daniel Dom\'{\i}nguez-\'{A}lvarez} {and}
  \bibinfo{person}{Alessandra Gorla}.} \bibinfo{year}{2019}\natexlab{}.
\newblock \showarticletitle{Release Practices for IOS and Android Apps}. In
  \bibinfo{booktitle}{\emph{Proceedings of the 3rd ACM SIGSOFT International
  Workshop on App Market Analytics}} \emph{(\bibinfo{series}{WAMA 2019})}.
  \bibinfo{publisher}{Association for Computing Machinery},
  \bibinfo{address}{New York, NY, USA}, \bibinfo{pages}{15--18}.
\newblock
\showISBNx{9781450368582}
\urldef\tempurl%
\url{https://doi.org/10.1145/3340496.3342762}
\showDOI{\tempurl}


\bibitem[\protect\citeauthoryear{Dong, B\"{o}hme, Cojocaru, and
  Roychoudhury}{Dong et~al\mbox{.}}{2020}]%
        {Dong:20}
\bibfield{author}{\bibinfo{person}{Zhen Dong}, \bibinfo{person}{Marcel
  B\"{o}hme}, \bibinfo{person}{Lucia Cojocaru}, {and} \bibinfo{person}{Abhik
  Roychoudhury}.} \bibinfo{year}{2020}\natexlab{}.
\newblock \showarticletitle{Time-Travel Testing of Android Apps}. In
  \bibinfo{booktitle}{\emph{Proceedings of the ACM/IEEE 42nd International
  Conference on Software Engineering}} \emph{(\bibinfo{series}{ICSE '20})}.
  \bibinfo{publisher}{Association for Computing Machinery},
  \bibinfo{address}{New York, NY, USA}, \bibinfo{pages}{481–492}.
\newblock
\showISBNx{9781450371216}
\urldef\tempurl%
\url{https://doi.org/10.1145/3377811.3380402}
\showDOI{\tempurl}


\bibitem[\protect\citeauthoryear{Feldt and Magazinius}{Feldt and
  Magazinius}{2010}]%
        {Feldt2010}
\bibfield{author}{\bibinfo{person}{Robert Feldt} {and} \bibinfo{person}{Ana
  Magazinius}.} \bibinfo{year}{2010}\natexlab{}.
\newblock \showarticletitle{{Validity threats in empirical software engineering
  research - An initial survey}}.
\newblock \bibinfo{journal}{\emph{SEKE 2010 - Proceedings of the 22nd
  International Conference on Software Engineering and Knowledge Engineering}}
  (\bibinfo{year}{2010}), \bibinfo{pages}{374--379}.
\newblock
\showISBNx{1891706268}


\bibitem[\protect\citeauthoryear{Fraser and Arcuri}{Fraser and Arcuri}{2011}]%
        {EvoSuite}
\bibfield{author}{\bibinfo{person}{Gordon Fraser} {and} \bibinfo{person}{Andrea
  Arcuri}.} \bibinfo{year}{2011}\natexlab{}.
\newblock \showarticletitle{EvoSuite: Automatic Test Suite Generation for
  Object-Oriented Software}. In \bibinfo{booktitle}{\emph{Proceedings of the
  19th ACM SIGSOFT Symposium and the 13th European Conference on Foundations of
  Software Engineering}} \emph{(\bibinfo{series}{ESEC/FSE '11})}.
  \bibinfo{publisher}{Association for Computing Machinery},
  \bibinfo{address}{New York, NY, USA}, \bibinfo{pages}{416--419}.
\newblock
\showISBNx{9781450304436}
\urldef\tempurl%
\url{https://doi.org/10.1145/2025113.2025179}
\showDOI{\tempurl}


\bibitem[\protect\citeauthoryear{{Gao}, {Fang}, and {Memon}}{{Gao}
  et~al\mbox{.}}{2015}]%
        {Gao:GUIRegressionTesting:ASE:2015}
\bibfield{author}{\bibinfo{person}{Z. {Gao}}, \bibinfo{person}{C. {Fang}},
  {and} \bibinfo{person}{A.~M. {Memon}}.} \bibinfo{year}{2015}\natexlab{}.
\newblock \showarticletitle{Pushing the limits on automation in GUI regression
  testing}. In \bibinfo{booktitle}{\emph{2015 IEEE 26th International Symposium
  on Software Reliability Engineering (ISSRE)}}. \bibinfo{pages}{565--575}.
\newblock
\showISSN{null}
\urldef\tempurl%
\url{https://doi.org/10.1109/ISSRE.2015.7381848}
\showDOI{\tempurl}


\bibitem[\protect\citeauthoryear{Gazzola, Mariani, Pastore, and Pezzè}{Gazzola
  et~al\mbox{.}}{2017}]%
        {Gazzola}
\bibfield{author}{\bibinfo{person}{Luca Gazzola}, \bibinfo{person}{Leonardo
  Mariani}, \bibinfo{person}{Fabrizio Pastore}, {and} \bibinfo{person}{Mauro
  Pezzè}.} \bibinfo{year}{2017}\natexlab{}.
\newblock \showarticletitle{An Exploratory Study of Field Failures}. In
  \bibinfo{booktitle}{\emph{2017 IEEE 28th International Symposium on Software
  Reliability Engineering (ISSRE)}}. \bibinfo{pages}{67--77}.
\newblock
\urldef\tempurl%
\url{https://doi.org/10.1109/ISSRE.2017.10}
\showDOI{\tempurl}


\bibitem[\protect\citeauthoryear{Gu}{Gu}{[n.d.]}]%
        {MiniTracing}
\bibfield{author}{\bibinfo{person}{Tianxiao Gu}.}
  \bibinfo{year}{[n.d.]}\natexlab{}.
\newblock \bibinfo{title}{{MiniTracing}, {APE} coverage tool}.
\newblock
  \bibinfo{howpublished}{\url{http://gutianxiao.com/ape/install-mini-tracing}}.
\newblock
\newblock
\shownote{Last visited: 07/07/2020.}


\bibitem[\protect\citeauthoryear{Gu, Sun, Ma, Cao, Xu, Yao, Zhang, Lu, and
  Su}{Gu et~al\mbox{.}}{2019}]%
        {Gu:APE:ICSE:2019}
\bibfield{author}{\bibinfo{person}{Tianxiao Gu}, \bibinfo{person}{Chengnian
  Sun}, \bibinfo{person}{Xiaoxing Ma}, \bibinfo{person}{Chun Cao},
  \bibinfo{person}{Chang Xu}, \bibinfo{person}{Yuan Yao},
  \bibinfo{person}{Qirun Zhang}, \bibinfo{person}{Jian Lu}, {and}
  \bibinfo{person}{Zhendong Su}.} \bibinfo{year}{2019}\natexlab{}.
\newblock \showarticletitle{Practical GUI Testing of Android Applications via
  Model Abstraction and Refinement}. In \bibinfo{booktitle}{\emph{Proceedings
  of the 41st International Conference on Software Engineering}}
  \emph{(\bibinfo{series}{ICSE '19})}. \bibinfo{publisher}{IEEE Press},
  \bibinfo{pages}{269--280}.
\newblock
\urldef\tempurl%
\url{https://doi.org/10.1109/ICSE.2019.00042}
\showDOI{\tempurl}


\bibitem[\protect\citeauthoryear{Hammoudi, Rothermel, and Stocco}{Hammoudi
  et~al\mbox{.}}{2016}]%
        {Hammoudi:2016}
\bibfield{author}{\bibinfo{person}{Mouna Hammoudi}, \bibinfo{person}{Gregg
  Rothermel}, {and} \bibinfo{person}{Andrea Stocco}.}
  \bibinfo{year}{2016}\natexlab{}.
\newblock \showarticletitle{WATERFALL: An Incremental Approach for Repairing
  Record-Replay Tests of Web Applications}. In
  \bibinfo{booktitle}{\emph{Proceedings of the 2016 24th ACM SIGSOFT
  International Symposium on Foundations of Software Engineering}}
  \emph{(\bibinfo{series}{FSE 2016})}. \bibinfo{publisher}{Association for
  Computing Machinery}, \bibinfo{address}{New York, NY, USA},
  \bibinfo{pages}{751--762}.
\newblock
\showISBNx{9781450342186}
\urldef\tempurl%
\url{https://doi.org/10.1145/2950290.2950294}
\showDOI{\tempurl}


\bibitem[\protect\citeauthoryear{Inc.}{Inc.}{2020}]%
        {Animator:online}
\bibfield{author}{\bibinfo{person}{Google Inc.}}
  \bibinfo{year}{2020}\natexlab{}.
\newblock \bibinfo{title}{Animator | Android Developers}.
\newblock
  \bibinfo{howpublished}{\url{https://developer.android.com/reference/kotlin/android/animation/Animator}}.
\newblock
\newblock
\shownote{Last visited: 11/11/2020.}


\bibitem[\protect\citeauthoryear{{INRIA}}{{INRIA}}{[n.d.]}]%
        {grid5000Web}
\bibfield{author}{\bibinfo{person}{{CNRS} {INRIA}}.}
  \bibinfo{year}{[n.d.]}\natexlab{}.
\newblock \bibinfo{title}{{Grid5000} infrastructure}.
\newblock \bibinfo{howpublished}{\url{https://www.grid5000.fr}}.
\newblock
\newblock
\shownote{Last visited: 07/07/2020.}


\bibitem[\protect\citeauthoryear{Jabbarvand, Sadeghi, Bagheri, and
  Malek}{Jabbarvand et~al\mbox{.}}{2016}]%
        {Jabbarvand2016}
\bibfield{author}{\bibinfo{person}{Reyhaneh Jabbarvand},
  \bibinfo{person}{Alireza Sadeghi}, \bibinfo{person}{Hamid Bagheri}, {and}
  \bibinfo{person}{Sam Malek}.} \bibinfo{year}{2016}\natexlab{}.
\newblock \showarticletitle{{Energy-aware test-suite minimization for android
  apps}}.
\newblock \bibinfo{journal}{\emph{ISSTA 2016 - Proceedings of the 25th
  International Symposium on Software Testing and Analysis}}
  (\bibinfo{year}{2016}), \bibinfo{pages}{425--436}.
\newblock
\showISBNx{9781450343909}
\urldef\tempurl%
\url{https://doi.org/10.1145/2931037.2931067}
\showDOI{\tempurl}


\bibitem[\protect\citeauthoryear{Jahangirova, Clark, Harman, and
  Tonella}{Jahangirova et~al\mbox{.}}{2016}]%
        {Jahangirova:Oracles:2016}
\bibfield{author}{\bibinfo{person}{Gunel Jahangirova}, \bibinfo{person}{David
  Clark}, \bibinfo{person}{Mark Harman}, {and} \bibinfo{person}{Paolo
  Tonella}.} \bibinfo{year}{2016}\natexlab{}.
\newblock \showarticletitle{Test Oracle Assessment and Improvement}. In
  \bibinfo{booktitle}{\emph{Proceedings of the 25th International Symposium on
  Software Testing and Analysis}} \emph{(\bibinfo{series}{ISSTA 2016})}.
  \bibinfo{publisher}{Association for Computing Machinery},
  \bibinfo{address}{New York, NY, USA}, \bibinfo{pages}{247--258}.
\newblock
\showISBNx{9781450343909}
\urldef\tempurl%
\url{https://doi.org/10.1145/2931037.2931062}
\showDOI{\tempurl}


\bibitem[\protect\citeauthoryear{Jamendo}{Jamendo}{2020}]%
        {Jamendo}
\bibfield{author}{\bibinfo{person}{Jamendo}.} \bibinfo{year}{2020}\natexlab{}.
\newblock \bibinfo{title}{Music Streaming App}.
\newblock \bibinfo{howpublished}{\url{https://www.jamendo.com/}}.
\newblock


\bibitem[\protect\citeauthoryear{Kochhar, Thung, and Lo}{Kochhar
  et~al\mbox{.}}{2015}]%
        {Kochhar2015}
\bibfield{author}{\bibinfo{person}{Pavneet~Singh Kochhar},
  \bibinfo{person}{Ferdian Thung}, {and} \bibinfo{person}{David Lo}.}
  \bibinfo{year}{2015}\natexlab{}.
\newblock \showarticletitle{{Code coverage and test suite effectiveness:
  Empirical study with real bugs in large systems}}.
\newblock \bibinfo{journal}{\emph{2015 IEEE 22nd International Conference on
  Software Analysis, Evolution, and Reengineering, SANER 2015 - Proceedings}}
  (\bibinfo{year}{2015}), \bibinfo{pages}{560--564}.
\newblock
\showISBNx{9781479984695}
\urldef\tempurl%
\url{https://doi.org/10.1109/SANER.2015.7081877}
\showDOI{\tempurl}


\bibitem[\protect\citeauthoryear{Korel and Al-Yami}{Korel and Al-Yami}{1998}]%
        {Bogdan:1998}
\bibfield{author}{\bibinfo{person}{Bogdan Korel} {and} \bibinfo{person}{Ali~M.
  Al-Yami}.} \bibinfo{year}{1998}\natexlab{}.
\newblock \showarticletitle{Automated Regression Test Generation}. In
  \bibinfo{booktitle}{\emph{Proceedings of the 1998 ACM SIGSOFT International
  Symposium on Software Testing and Analysis}} \emph{(\bibinfo{series}{ISSTA
  '98})}. \bibinfo{publisher}{Association for Computing Machinery},
  \bibinfo{address}{New York, NY, USA}, \bibinfo{pages}{143--152}.
\newblock
\showISBNx{0897919718}
\urldef\tempurl%
\url{https://doi.org/10.1145/271771.271803}
\showDOI{\tempurl}


\bibitem[\protect\citeauthoryear{Koroglu, Sen, Muslu, Mete, Ulker, Tanriverdi,
  and Donmez}{Koroglu et~al\mbox{.}}{2018}]%
        {Koroglu2018}
\bibfield{author}{\bibinfo{person}{Yavuz Koroglu}, \bibinfo{person}{Alper Sen},
  \bibinfo{person}{Ozlem Muslu}, \bibinfo{person}{Yunus Mete},
  \bibinfo{person}{Ceyda Ulker}, \bibinfo{person}{Tolga Tanriverdi}, {and}
  \bibinfo{person}{Yunus Donmez}.} \bibinfo{year}{2018}\natexlab{}.
\newblock \showarticletitle{{QBE: QLearning-Based Exploration of Android
  Applications}}. In \bibinfo{booktitle}{\emph{2018 IEEE 11th International
  Conference on Software Testing, Verification and Validation (ICST)}}.
  \bibinfo{publisher}{IEEE}, \bibinfo{pages}{105--115}.
\newblock
\showISBNx{978-1-5386-5012-7}
\urldef\tempurl%
\url{https://doi.org/10.1109/ICST.2018.00020}
\showDOI{\tempurl}


\bibitem[\protect\citeauthoryear{{Kuznetsov}, {Avdiienko}, {Gorla}, and
  {Zeller}}{{Kuznetsov} et~al\mbox{.}}{2018}]%
        {Kuznetsov:2018}
\bibfield{author}{\bibinfo{person}{K. {Kuznetsov}}, \bibinfo{person}{V.
  {Avdiienko}}, \bibinfo{person}{A. {Gorla}}, {and} \bibinfo{person}{A.
  {Zeller}}.} \bibinfo{year}{2018}\natexlab{}.
\newblock \showarticletitle{Analyzing the User Interface of Android Apps}. In
  \bibinfo{booktitle}{\emph{2018 IEEE/ACM 5th International Conference on
  Mobile Software Engineering and Systems (MOBILESoft)}}.
  \bibinfo{pages}{84--87}.
\newblock


\bibitem[\protect\citeauthoryear{{Li}, {Chang}, {Wang}, {Huang}, {Pei}, {Wang},
  and {Li}}{{Li} et~al\mbox{.}}{2017}]%
        {Li:Atom:2017}
\bibfield{author}{\bibinfo{person}{X. {Li}}, \bibinfo{person}{N. {Chang}},
  \bibinfo{person}{Y. {Wang}}, \bibinfo{person}{H. {Huang}},
  \bibinfo{person}{Y. {Pei}}, \bibinfo{person}{L. {Wang}}, {and}
  \bibinfo{person}{X. {Li}}.} \bibinfo{year}{2017}\natexlab{}.
\newblock \showarticletitle{ATOM: Automatic Maintenance of GUI Test Scripts for
  Evolving Mobile Applications}. In \bibinfo{booktitle}{\emph{2017 IEEE
  International Conference on Software Testing, Verification and Validation
  (ICST)}}. \bibinfo{pages}{161--171}.
\newblock


\bibitem[\protect\citeauthoryear{Linares-Vasquez, Moran, and
  Poshyvanyk}{Linares-Vasquez et~al\mbox{.}}{2017}]%
        {Linares:ICSME:2017}
\bibfield{author}{\bibinfo{person}{M. Linares-Vasquez}, \bibinfo{person}{K.
  Moran}, {and} \bibinfo{person}{D. Poshyvanyk}.}
  \bibinfo{year}{2017}\natexlab{}.
\newblock \showarticletitle{Continuous, Evolutionary and Large-Scale: A New
  Perspective for Automated Mobile App Testing}. In
  \bibinfo{booktitle}{\emph{2017 IEEE International Conference on Software
  Maintenance and Evolution (ICSME)}}. \bibinfo{pages}{399--410}.
\newblock
\urldef\tempurl%
\url{https://doi.org/10.1109/ICSME.2017.27}
\showDOI{\tempurl}


\bibitem[\protect\citeauthoryear{Manning, Raghavan, and Sch\"{u}tze}{Manning
  et~al\mbox{.}}{2008}]%
        {IRbook}
\bibfield{author}{\bibinfo{person}{Christopher~D. Manning},
  \bibinfo{person}{Prabhakar Raghavan}, {and} \bibinfo{person}{Hinrich
  Sch\"{u}tze}.} \bibinfo{year}{2008}\natexlab{}.
\newblock \bibinfo{booktitle}{\emph{Introduction to Information Retrieval}}.
\newblock \bibinfo{publisher}{Cambridge University Press},
  \bibinfo{address}{USA}.
\newblock
\showISBNx{0521865719}


\bibitem[\protect\citeauthoryear{Mao, Harman, and Jia}{Mao
  et~al\mbox{.}}{2016}]%
        {sapienz}
\bibfield{author}{\bibinfo{person}{Ke Mao}, \bibinfo{person}{Mark Harman},
  {and} \bibinfo{person}{Yue Jia}.} \bibinfo{year}{2016}\natexlab{}.
\newblock \showarticletitle{Sapienz: Multi-objective Automated Testing for
  Android Applications}. In \bibinfo{booktitle}{\emph{Proceedings of the 25th
  International Symposium on Software Testing and Analysis}}
  \emph{(\bibinfo{series}{ISSTA 2016})}. \bibinfo{publisher}{ACM},
  \bibinfo{address}{New York, NY, USA}, \bibinfo{pages}{94--105}.
\newblock
\showISBNx{978-1-4503-4390-9}
\urldef\tempurl%
\url{https://doi.org/10.1145/2931037.2931054}
\showDOI{\tempurl}


\bibitem[\protect\citeauthoryear{Mcilroy, Ali, and Hassan}{Mcilroy
  et~al\mbox{.}}{2016}]%
        {Mcilroy:FrequentlyUpdatedApps:ESE:2016}
\bibfield{author}{\bibinfo{person}{Stuart Mcilroy}, \bibinfo{person}{Nasir
  Ali}, {and} \bibinfo{person}{Ahmed~E. Hassan}.}
  \bibinfo{year}{2016}\natexlab{}.
\newblock \showarticletitle{Fresh Apps: An Empirical Study of
  Frequently-Updated Mobile Apps in the Google Play Store}.
\newblock \bibinfo{journal}{\emph{Empirical Softw. Engg.}}
  \bibinfo{volume}{21}, \bibinfo{number}{3} (\bibinfo{date}{June}
  \bibinfo{year}{2016}), \bibinfo{pages}{1346--1370}.
\newblock
\showISSN{1382-3256}
\urldef\tempurl%
\url{https://doi.org/10.1007/s10664-015-9388-2}
\showDOI{\tempurl}


\bibitem[\protect\citeauthoryear{Ngo, Pastore, and Briand}{Ngo
  et~al\mbox{.}}{2020}]%
        {replicability}
\bibfield{author}{\bibinfo{person}{Chanh~Duc Ngo}, \bibinfo{person}{Fabrizio
  Pastore}, {and} \bibinfo{person}{Lionel Briand}.}
  \bibinfo{year}{2020}\natexlab{}.
\newblock \bibinfo{title}{ATUA toolset and replicability package}.
\newblock \bibinfo{howpublished}{\url{https://github.com/SNTSVV/ATUA/}}.
\newblock
\urldef\tempurl%
\url{https://doi.org/10.5281/zenodo.5734090}
\showDOI{\tempurl}
\newblock
\shownote{Last visited: 30/11/2021.}


\bibitem[\protect\citeauthoryear{Pan, Huang, Wang, Zhang, and Li}{Pan
  et~al\mbox{.}}{2020a}]%
        {Pan:ReinforcementLearning:2020}
\bibfield{author}{\bibinfo{person}{Minxue Pan}, \bibinfo{person}{An Huang},
  \bibinfo{person}{Guoxin Wang}, \bibinfo{person}{Tian Zhang}, {and}
  \bibinfo{person}{Xuandong Li}.} \bibinfo{year}{2020}\natexlab{a}.
\newblock \showarticletitle{{Reinforcement learning based curiosity-driven
  testing of Android applications}}.
\newblock \bibinfo{journal}{\emph{ISSTA 2020 - Proceedings of the 29th ACM
  SIGSOFT International Symposium on Software Testing and Analysis}}
  (\bibinfo{year}{2020}), \bibinfo{pages}{153--164}.
\newblock
\showISBNx{9781450380089}
\urldef\tempurl%
\url{https://doi.org/10.1145/3395363.3397354}
\showDOI{\tempurl}


\bibitem[\protect\citeauthoryear{{Pan}, {Xu}, {Pei}, {Li}, {Zhang}, and
  {Li}}{{Pan} et~al\mbox{.}}{2019}]%
        {Pan:GUIrepair:2019}
\bibfield{author}{\bibinfo{person}{M. {Pan}}, \bibinfo{person}{T. {Xu}},
  \bibinfo{person}{Y. {Pei}}, \bibinfo{person}{Z. {Li}}, \bibinfo{person}{T.
  {Zhang}}, {and} \bibinfo{person}{X. {Li}}.} \bibinfo{year}{2019}\natexlab{}.
\newblock \showarticletitle{GUI-Guided Repair of Mobile Test Scripts}. In
  \bibinfo{booktitle}{\emph{2019 IEEE/ACM 41st International Conference on
  Software Engineering: Companion Proceedings (ICSE-Companion)}}.
  \bibinfo{pages}{326--327}.
\newblock


\bibitem[\protect\citeauthoryear{Pan, Xu, Pei, Li, Zhang, and Li}{Pan
  et~al\mbox{.}}{2020b}]%
        {Pan2020}
\bibfield{author}{\bibinfo{person}{Minxue Pan}, \bibinfo{person}{Tongtong Xu},
  \bibinfo{person}{Yu Pei}, \bibinfo{person}{Zhong Li}, \bibinfo{person}{Tian
  Zhang}, {and} \bibinfo{person}{Xuandong Li}.}
  \bibinfo{year}{2020}\natexlab{b}.
\newblock \showarticletitle{{GUI-Guided Test Script Repair for Mobile Apps}}.
\newblock \bibinfo{journal}{\emph{IEEE Transactions on Software Engineering}}
  \bibinfo{volume}{5589}, \bibinfo{number}{c} (\bibinfo{year}{2020}),
  \bibinfo{pages}{1--1}.
\newblock
\showISSN{0098-5589}
\urldef\tempurl%
\url{https://doi.org/10.1109/tse.2020.3007664}
\showDOI{\tempurl}


\bibitem[\protect\citeauthoryear{{Pastore}, {Mariani}, and {Fraser}}{{Pastore}
  et~al\mbox{.}}{2013}]%
        {Pastore:CrowdOracles}
\bibfield{author}{\bibinfo{person}{F. {Pastore}}, \bibinfo{person}{L.
  {Mariani}}, {and} \bibinfo{person}{G. {Fraser}}.}
  \bibinfo{year}{2013}\natexlab{}.
\newblock \showarticletitle{CrowdOracles: Can the Crowd Solve the Oracle
  Problem?}. In \bibinfo{booktitle}{\emph{2013 IEEE Sixth International
  Conference on Software Testing, Verification and Validation}}.
  \bibinfo{pages}{342--351}.
\newblock


\bibitem[\protect\citeauthoryear{Pastore, Mariani, Goffi, Oriol, and
  Wahler}{Pastore et~al\mbox{.}}{2012}]%
        {Pastore:ISSRE:12}
\bibfield{author}{\bibinfo{person}{Fabrizio Pastore}, \bibinfo{person}{Leonardo
  Mariani}, \bibinfo{person}{Alberto Goffi}, \bibinfo{person}{Manuel Oriol},
  {and} \bibinfo{person}{Michael Wahler}.} \bibinfo{year}{2012}\natexlab{}.
\newblock \showarticletitle{Dynamic Analysis of Upgrades in C/C++ Software}. In
  \bibinfo{booktitle}{\emph{Proceedings of the 2012 IEEE 23rd International
  Symposium on Software Reliability Engineering}} \emph{(\bibinfo{series}{ISSRE
  '12})}. \bibinfo{publisher}{IEEE Computer Society}, \bibinfo{address}{USA},
  \bibinfo{pages}{91--100}.
\newblock
\showISBNx{9780769548883}
\urldef\tempurl%
\url{https://doi.org/10.1109/ISSRE.2012.9}
\showDOI{\tempurl}


\bibitem[\protect\citeauthoryear{Pastore, Mariani, Hyv\"{a}rinen, Fedyukovich,
  Sharygina, Sehestedt, and Muhammad}{Pastore et~al\mbox{.}}{2014}]%
        {Pastore:2014}
\bibfield{author}{\bibinfo{person}{Fabrizio Pastore}, \bibinfo{person}{Leonardo
  Mariani}, \bibinfo{person}{Antti E.~J. Hyv\"{a}rinen},
  \bibinfo{person}{Grigory Fedyukovich}, \bibinfo{person}{Natasha Sharygina},
  \bibinfo{person}{Stephan Sehestedt}, {and} \bibinfo{person}{Ali Muhammad}.}
  \bibinfo{year}{2014}\natexlab{}.
\newblock \showarticletitle{Verification-Aided Regression Testing}. In
  \bibinfo{booktitle}{\emph{Proceedings of the 2014 International Symposium on
  Software Testing and Analysis}} \emph{(\bibinfo{series}{ISSTA 2014})}.
  \bibinfo{publisher}{Association for Computing Machinery},
  \bibinfo{address}{New York, NY, USA}, \bibinfo{pages}{37--48}.
\newblock
\showISBNx{9781450326452}
\urldef\tempurl%
\url{https://doi.org/10.1145/2610384.2610387}
\showDOI{\tempurl}


\bibitem[\protect\citeauthoryear{Phillips and Hardy}{Phillips and
  Hardy}{2013}]%
        {Phillips:AndroidProgramming}
\bibfield{author}{\bibinfo{person}{Bill Phillips} {and} \bibinfo{person}{Brian
  Hardy}.} \bibinfo{year}{2013}\natexlab{}.
\newblock \bibinfo{booktitle}{\emph{Android Programming: The Big Nerd Ranch
  Guide} (\bibinfo{edition}{1st} ed.)}.
\newblock \bibinfo{publisher}{Big Nerd Ranch}.
\newblock
\showISBNx{0321804333}


\bibitem[\protect\citeauthoryear{Ralph and Tempero}{Ralph and Tempero}{2018}]%
        {Ralph2018}
\bibfield{author}{\bibinfo{person}{Paul Ralph} {and} \bibinfo{person}{Ewan
  Tempero}.} \bibinfo{year}{2018}\natexlab{}.
\newblock \showarticletitle{{Construct Validity in Software Engineering
  Research and Software Metrics}}. In \bibinfo{booktitle}{\emph{Proceedings of
  the 22nd International Conference on Evaluation and Assessment in Software
  Engineering 2018}}, Vol.~\bibinfo{volume}{Part F1377}.
  \bibinfo{publisher}{ACM}, \bibinfo{address}{New York, NY, USA},
  \bibinfo{pages}{13--23}.
\newblock
\showISBNx{9781450364034}
\urldef\tempurl%
\url{https://doi.org/10.1145/3210459.3210461}
\showDOI{\tempurl}


\bibitem[\protect\citeauthoryear{Rau, Hotzkow, and Zeller}{Rau
  et~al\mbox{.}}{2018}]%
        {Rau:2018}
\bibfield{author}{\bibinfo{person}{Andreas Rau}, \bibinfo{person}{Jenny
  Hotzkow}, {and} \bibinfo{person}{Andreas Zeller}.}
  \bibinfo{year}{2018}\natexlab{}.
\newblock \showarticletitle{Efficient GUI Test Generation by Learning from
  Tests of Other Apps}. In \bibinfo{booktitle}{\emph{Proceedings of the 40th
  International Conference on Software Engineering: Companion Proceeedings}}
  \emph{(\bibinfo{series}{ICSE '18})}. \bibinfo{publisher}{Association for
  Computing Machinery}, \bibinfo{address}{New York, NY, USA},
  \bibinfo{pages}{370--371}.
\newblock
\showISBNx{9781450356633}
\urldef\tempurl%
\url{https://doi.org/10.1145/3183440.3195014}
\showDOI{\tempurl}


\bibitem[\protect\citeauthoryear{Rothermel and Harrold}{Rothermel and
  Harrold}{1997}]%
        {Rothermel:Dejavu:1997}
\bibfield{author}{\bibinfo{person}{Gregg Rothermel} {and}
  \bibinfo{person}{Mary~Jean Harrold}.} \bibinfo{year}{1997}\natexlab{}.
\newblock \showarticletitle{A Safe, Efficient Regression Test Selection
  Technique}.
\newblock \bibinfo{journal}{\emph{ACM Trans. Softw. Eng. Methodol.}}
  \bibinfo{volume}{6}, \bibinfo{number}{2} (\bibinfo{date}{April}
  \bibinfo{year}{1997}), \bibinfo{pages}{173--210}.
\newblock
\showISSN{1049-331X}
\urldef\tempurl%
\url{https://doi.org/10.1145/248233.248262}
\showDOI{\tempurl}


\bibitem[\protect\citeauthoryear{Rountev and Yan}{Rountev and Yan}{2014}]%
        {Rountev:2014}
\bibfield{author}{\bibinfo{person}{Atanas Rountev} {and}
  \bibinfo{person}{Dacong Yan}.} \bibinfo{year}{2014}\natexlab{}.
\newblock \showarticletitle{Static Reference Analysis for GUI Objects in
  Android Software}. In \bibinfo{booktitle}{\emph{Proceedings of Annual
  IEEE/ACM International Symposium on Code Generation and Optimization}}
  \emph{(\bibinfo{series}{CGO '14})}. \bibinfo{publisher}{Association for
  Computing Machinery}, \bibinfo{address}{New York, NY, USA},
  \bibinfo{pages}{143--153}.
\newblock
\showISBNx{9781450326704}
\urldef\tempurl%
\url{https://doi.org/10.1145/2581122.2544159}
\showDOI{\tempurl}


\bibitem[\protect\citeauthoryear{{Rubinov} and {Baresi}}{{Rubinov} and
  {Baresi}}{2018}]%
        {Rubinov:AndroidAppTestingSurvey:2018}
\bibfield{author}{\bibinfo{person}{K. {Rubinov}} {and} \bibinfo{person}{L.
  {Baresi}}.} \bibinfo{year}{2018}\natexlab{}.
\newblock \showarticletitle{What Are We Missing When Testing Our Android Apps?}
\newblock \bibinfo{journal}{\emph{Computer}} \bibinfo{volume}{51},
  \bibinfo{number}{4} (\bibinfo{date}{April} \bibinfo{year}{2018}),
  \bibinfo{pages}{60--68}.
\newblock
\showISSN{1558-0814}
\urldef\tempurl%
\url{https://doi.org/10.1109/MC.2018.2141024}
\showDOI{\tempurl}


\bibitem[\protect\citeauthoryear{Ryder and Tip}{Ryder and Tip}{2001}]%
        {Ryder:2001}
\bibfield{author}{\bibinfo{person}{Barbara~G. Ryder} {and}
  \bibinfo{person}{Frank Tip}.} \bibinfo{year}{2001}\natexlab{}.
\newblock \showarticletitle{Change Impact Analysis for Object-Oriented
  Programs}. In \bibinfo{booktitle}{\emph{Proceedings of the 2001 ACM
  SIGPLAN-SIGSOFT Workshop on Program Analysis for Software Tools and
  Engineering}} \emph{(\bibinfo{series}{PASTE '01})}.
  \bibinfo{publisher}{Association for Computing Machinery},
  \bibinfo{address}{New York, NY, USA}, \bibinfo{pages}{46--53}.
\newblock
\showISBNx{1581134134}
\urldef\tempurl%
\url{https://doi.org/10.1145/379605.379661}
\showDOI{\tempurl}


\bibitem[\protect\citeauthoryear{Saha, Zhang, Khurshid, and Perry}{Saha
  et~al\mbox{.}}{2015}]%
        {Saha:IRprioritization:2015}
\bibfield{author}{\bibinfo{person}{Ripon~K. Saha}, \bibinfo{person}{Lingming
  Zhang}, \bibinfo{person}{Sarfraz Khurshid}, {and} \bibinfo{person}{Dewayne~E.
  Perry}.} \bibinfo{year}{2015}\natexlab{}.
\newblock \showarticletitle{An Information Retrieval Approach for Regression
  Test Prioritization Based on Program Changes}. In
  \bibinfo{booktitle}{\emph{Proceedings of the 37th International Conference on
  Software Engineering - Volume 1}} \emph{(\bibinfo{series}{ICSE '15})}.
  \bibinfo{publisher}{IEEE Press}, \bibinfo{pages}{268--279}.
\newblock
\showISBNx{9781479919345}


\bibitem[\protect\citeauthoryear{{Shamshiri}, {Fraser}, {Mcminn}, and
  {Orso}}{{Shamshiri} et~al\mbox{.}}{2013}]%
        {Shamshiri:2013}
\bibfield{author}{\bibinfo{person}{S. {Shamshiri}}, \bibinfo{person}{G.
  {Fraser}}, \bibinfo{person}{P. {Mcminn}}, {and} \bibinfo{person}{A. {Orso}}.}
  \bibinfo{year}{2013}\natexlab{}.
\newblock \showarticletitle{Search-Based Propagation of Regression Faults in
  Automated Regression Testing}. In \bibinfo{booktitle}{\emph{2013 IEEE Sixth
  International Conference on Software Testing, Verification and Validation
  Workshops}}. \bibinfo{pages}{396--399}.
\newblock
\showISSN{null}
\urldef\tempurl%
\url{https://doi.org/10.1109/ICSTW.2013.51}
\showDOI{\tempurl}


\bibitem[\protect\citeauthoryear{Sharma and Nasre}{Sharma and Nasre}{2019}]%
        {Sharma:QADroid:ISSTA:2019}
\bibfield{author}{\bibinfo{person}{Aman Sharma} {and} \bibinfo{person}{Rupesh
  Nasre}.} \bibinfo{year}{2019}\natexlab{}.
\newblock \showarticletitle{QADroid: Regression Event Selection for Android
  Applications}. In \bibinfo{booktitle}{\emph{Proceedings of the 28th ACM
  SIGSOFT International Symposium on Software Testing and Analysis}}
  \emph{(\bibinfo{series}{ISSTA 2019})}. \bibinfo{publisher}{Association for
  Computing Machinery}, \bibinfo{address}{New York, NY, USA},
  \bibinfo{pages}{66--77}.
\newblock
\showISBNx{9781450362245}
\urldef\tempurl%
\url{https://doi.org/10.1145/3293882.3330550}
\showDOI{\tempurl}


\bibitem[\protect\citeauthoryear{Shieh, Jan, and Randles}{Shieh
  et~al\mbox{.}}{2006}]%
        {Shieh2006}
\bibfield{author}{\bibinfo{person}{Gwowen Shieh}, \bibinfo{person}{Show~Li
  Jan}, {and} \bibinfo{person}{Ronald~H. Randles}.}
  \bibinfo{year}{2006}\natexlab{}.
\newblock \showarticletitle{{On power and sample size determinations for the
  Wilcoxon-Mann-Whitney test}}.
\newblock \bibinfo{journal}{\emph{Journal of Nonparametric Statistics}}
  \bibinfo{volume}{18}, \bibinfo{number}{1} (\bibinfo{year}{2006}),
  \bibinfo{pages}{33--43}.
\newblock
\showISSN{10485252}
\urldef\tempurl%
\url{https://doi.org/10.1080/10485250500473099}
\showDOI{\tempurl}


\bibitem[\protect\citeauthoryear{{Song}, {Xu}, and {Xu}}{{Song}
  et~al\mbox{.}}{2017}]%
        {Song:XPathRepair:2017}
\bibfield{author}{\bibinfo{person}{F. {Song}}, \bibinfo{person}{Z. {Xu}}, {and}
  \bibinfo{person}{F. {Xu}}.} \bibinfo{year}{2017}\natexlab{}.
\newblock \showarticletitle{An XPath-Based Approach to Reusing Test Scripts for
  Android Applications}. In \bibinfo{booktitle}{\emph{2017 14th Web Information
  Systems and Applications Conference (WISA)}}. \bibinfo{pages}{143--148}.
\newblock


\bibitem[\protect\citeauthoryear{{STH blog editors}}{{STH blog
  editors}}{[n.d.]}]%
        {ListCrowdsourcing}
\bibfield{author}{\bibinfo{person}{{STH blog editors}}.}
  \bibinfo{year}{[n.d.]}\natexlab{}.
\newblock \bibinfo{title}{{10 Most Popular Crowdsourced Testing Companies in
  2020}}.
\newblock
  \bibinfo{howpublished}{\url{https://www.softwaretestinghelp.com/crowdsourced-testing-companies/}}.
\newblock
\newblock
\shownote{Last visited: 07/07/2020.}


\bibitem[\protect\citeauthoryear{Su, Meng, Chen, Wu, Yang, Yao, Pu, Liu, and
  Su}{Su et~al\mbox{.}}{2017}]%
        {stoat17}
\bibfield{author}{\bibinfo{person}{Ting Su}, \bibinfo{person}{Guozhu Meng},
  \bibinfo{person}{Yuting Chen}, \bibinfo{person}{Ke Wu},
  \bibinfo{person}{Weiming Yang}, \bibinfo{person}{Yao Yao},
  \bibinfo{person}{Geguang Pu}, \bibinfo{person}{Yang Liu}, {and}
  \bibinfo{person}{Zhendong Su}.} \bibinfo{year}{2017}\natexlab{}.
\newblock \showarticletitle{Guided, Stochastic Model-based GUI Testing of
  Android Apps}. In \bibinfo{booktitle}{\emph{Proceedings of the 2017 11th
  Joint Meeting on Foundations of Software Engineering}}
  \emph{(\bibinfo{series}{ESEC/FSE 2017})}. \bibinfo{publisher}{ACM},
  \bibinfo{address}{New York, NY, USA}, \bibinfo{pages}{245--256}.
\newblock
\showISBNx{978-1-4503-5105-8}
\urldef\tempurl%
\url{https://doi.org/10.1145/3106237.3106298}
\showDOI{\tempurl}


\bibitem[\protect\citeauthoryear{{Toffola}, {Staicu}, and {Pradel}}{{Toffola}
  et~al\mbox{.}}{2017}]%
        {Toffola:MiningInputs:ASE:2017}
\bibfield{author}{\bibinfo{person}{L.~D. {Toffola}}, \bibinfo{person}{C.
  {Staicu}}, {and} \bibinfo{person}{M. {Pradel}}.}
  \bibinfo{year}{2017}\natexlab{}.
\newblock \showarticletitle{Saying 'Hi!' is not enough: Mining inputs for
  effective test generation}. In \bibinfo{booktitle}{\emph{2017 32nd IEEE/ACM
  International Conference on Automated Software Engineering (ASE)}}.
  \bibinfo{pages}{44--49}.
\newblock
\showISSN{null}
\urldef\tempurl%
\url{https://doi.org/10.1109/ASE.2017.8115617}
\showDOI{\tempurl}


\bibitem[\protect\citeauthoryear{Tramontana, Amalfitano, Amatucci, and
  Fasolino}{Tramontana et~al\mbox{.}}{2019}]%
        {Amalfitano:AndroidTestingSurvey:SQJ:2018}
\bibfield{author}{\bibinfo{person}{Porfirio Tramontana},
  \bibinfo{person}{Domenico Amalfitano}, \bibinfo{person}{Nicola Amatucci},
  {and} \bibinfo{person}{Anna~Rita Fasolino}.} \bibinfo{year}{2019}\natexlab{}.
\newblock \showarticletitle{Automated functional testing of mobile
  applications: a systematic mapping study}.
\newblock \bibinfo{journal}{\emph{Software Quality Journal}}
  \bibinfo{volume}{27}, \bibinfo{number}{1} (\bibinfo{year}{2019}),
  \bibinfo{pages}{149--201}.
\newblock
\showISBNx{1573-1367}
\urldef\tempurl%
\url{https://doi.org/10.1007/s11219-018-9418-6}
\showDOI{\tempurl}


\bibitem[\protect\citeauthoryear{Vargha and Delaney}{Vargha and
  Delaney}{2000}]%
        {VDA}
\bibfield{author}{\bibinfo{person}{András Vargha} {and}
  \bibinfo{person}{Harold~D. Delaney}.} \bibinfo{year}{2000}\natexlab{}.
\newblock \showarticletitle{A Critique and Improvement of the CL Common
  Language Effect Size Statistics of McGraw and Wong}.
\newblock \bibinfo{journal}{\emph{Journal of Educational and Behavioral
  Statistics}} \bibinfo{volume}{25}, \bibinfo{number}{2}
  (\bibinfo{year}{2000}), \bibinfo{pages}{101--132}.
\newblock
\urldef\tempurl%
\url{https://doi.org/10.3102/10769986025002101}
\showDOI{\tempurl}
\showeprint{https://doi.org/10.3102/10769986025002101}


\bibitem[\protect\citeauthoryear{Wang, Jiang, Xu, Cao, Ma, and Lu}{Wang
  et~al\mbox{.}}{2020}]%
        {Combodroid:Wang2020}
\bibfield{author}{\bibinfo{person}{Jue Wang}, \bibinfo{person}{Yanyan Jiang},
  \bibinfo{person}{Chang Xu}, \bibinfo{person}{Chun Cao},
  \bibinfo{person}{Xiaoxing Ma}, {and} \bibinfo{person}{Jian Lu}.}
  \bibinfo{year}{2020}\natexlab{}.
\newblock \showarticletitle{{Combodroid: Generating high-quality test inputs
  for android apps via use case combinations}}.
\newblock \bibinfo{journal}{\emph{Proceedings - International Conference on
  Software Engineering}} (\bibinfo{year}{2020}), \bibinfo{pages}{469--480}.
\newblock
\showISBNx{9781450371216}
\showISSN{02705257}
\urldef\tempurl%
\url{https://doi.org/10.1145/3377811.3380382}
\showDOI{\tempurl}


\bibitem[\protect\citeauthoryear{Wang, Li, Yang, Cao, Zhang, Deng, and
  Xie}{Wang et~al\mbox{.}}{2018}]%
        {Wang:EmpStudy:2018}
\bibfield{author}{\bibinfo{person}{Wenyu Wang}, \bibinfo{person}{Dengfeng Li},
  \bibinfo{person}{Wei Yang}, \bibinfo{person}{Yurui Cao},
  \bibinfo{person}{Zhenwen Zhang}, \bibinfo{person}{Yuetang Deng}, {and}
  \bibinfo{person}{Tao Xie}.} \bibinfo{year}{2018}\natexlab{}.
\newblock \showarticletitle{An Empirical Study of Android Test Generation Tools
  in Industrial Cases}. In \bibinfo{booktitle}{\emph{Proceedings of the 33rd
  ACM/IEEE International Conference on Automated Software Engineering}}
  \emph{(\bibinfo{series}{ASE 2018})}. \bibinfo{publisher}{ACM},
  \bibinfo{address}{New York, NY, USA}, \bibinfo{pages}{738--748}.
\newblock
\showISBNx{978-1-4503-5937-5}
\urldef\tempurl%
\url{https://doi.org/10.1145/3238147.3240465}
\showDOI{\tempurl}


\bibitem[\protect\citeauthoryear{Wang, Zhang, and Rountev}{Wang
  et~al\mbox{.}}{2016}]%
        {Wang:FlowDroidComparison:2016}
\bibfield{author}{\bibinfo{person}{Yan Wang}, \bibinfo{person}{Hailong Zhang},
  {and} \bibinfo{person}{Atanas Rountev}.} \bibinfo{year}{2016}\natexlab{}.
\newblock \showarticletitle{On the Unsoundness of Static Analysis for Android
  GUIs}. In \bibinfo{booktitle}{\emph{Proceedings of the 5th ACM SIGPLAN
  International Workshop on State Of the Art in Program Analysis}}
  \emph{(\bibinfo{series}{SOAP 2016})}. \bibinfo{publisher}{Association for
  Computing Machinery}, \bibinfo{address}{New York, NY, USA},
  \bibinfo{pages}{18--23}.
\newblock
\showISBNx{9781450343855}
\urldef\tempurl%
\url{https://doi.org/10.1145/2931021.2931026}
\showDOI{\tempurl}


\bibitem[\protect\citeauthoryear{Watkins}{Watkins}{1989}]%
        {Watkins:1989}
\bibfield{author}{\bibinfo{person}{Christopher John Cornish~Hellaby Watkins}.}
  \bibinfo{year}{1989}\natexlab{}.
\newblock \emph{\bibinfo{title}{Learning from Delayed Rewards}}.
\newblock \bibinfo{thesistype}{Ph.D. Dissertation}. \bibinfo{school}{King's
  College}, \bibinfo{address}{Cambridge, UK}.
\newblock


\bibitem[\protect\citeauthoryear{{Wen}, {Wu}, and {Cheung}}{{Wen}
  et~al\mbox{.}}{2016}]%
        {Wen:2016}
\bibfield{author}{\bibinfo{person}{M. {Wen}}, \bibinfo{person}{R. {Wu}}, {and}
  \bibinfo{person}{S. {Cheung}}.} \bibinfo{year}{2016}\natexlab{}.
\newblock \showarticletitle{Locus: Locating bugs from software changes}. In
  \bibinfo{booktitle}{\emph{2016 31st IEEE/ACM International Conference on
  Automated Software Engineering (ASE)}}. \bibinfo{pages}{262--273}.
\newblock


\bibitem[\protect\citeauthoryear{Wohlin, Runeson, H{\"{o}}st, Ohlsson, Regnell,
  and Wessl{\'{e}}n}{Wohlin et~al\mbox{.}}{2012}]%
        {Wohlin2012}
\bibfield{author}{\bibinfo{person}{Claes Wohlin}, \bibinfo{person}{Per
  Runeson}, \bibinfo{person}{Martin H{\"{o}}st}, \bibinfo{person}{Magnus~C.
  Ohlsson}, \bibinfo{person}{Bj{\"{o}}rn Regnell}, {and}
  \bibinfo{person}{Anders Wessl{\'{e}}n}.} \bibinfo{year}{2012}\natexlab{}.
\newblock \bibinfo{booktitle}{\emph{{Experimentation in software
  engineering}}}. Vol.~\bibinfo{volume}{9783642290}.
\newblock 1--236 pages.
\newblock
\showISBNx{9783642290442}
\urldef\tempurl%
\url{https://doi.org/10.1007/978-3-642-29044-2}
\showDOI{\tempurl}


\bibitem[\protect\citeauthoryear{Wu, Zhang, Wang, and Rountev}{Wu
  et~al\mbox{.}}{2019}]%
        {Wu:Gator:Leak:2019}
\bibfield{author}{\bibinfo{person}{Haowei Wu}, \bibinfo{person}{Hailong Zhang},
  \bibinfo{person}{Yan Wang}, {and} \bibinfo{person}{Atanas Rountev}.}
  \bibinfo{year}{2019}\natexlab{}.
\newblock \showarticletitle{Sentinel: generating GUI tests for sensor leaks in
  Android and Android wear apps}.
\newblock \bibinfo{journal}{\emph{Software Quality Journal}}
  (\bibinfo{year}{2019}).
\newblock
\showISBNx{1573-1367}
\urldef\tempurl%
\url{https://doi.org/10.1007/s11219-019-09484-z}
\showDOI{\tempurl}


\bibitem[\protect\citeauthoryear{Yan, Sun, and Liu}{Yan et~al\mbox{.}}{2014}]%
        {Minzhi:2014}
\bibfield{author}{\bibinfo{person}{Minzhi Yan}, \bibinfo{person}{Hailong Sun},
  {and} \bibinfo{person}{Xudong Liu}.} \bibinfo{year}{2014}\natexlab{}.
\newblock \showarticletitle{ITest: Testing Software with Mobile Crowdsourcing}.
  In \bibinfo{booktitle}{\emph{Proceedings of the 1st International Workshop on
  Crowd-Based Software Development Methods and Technologies}}
  \emph{(\bibinfo{series}{CrowdSoft 2014})}. \bibinfo{publisher}{Association
  for Computing Machinery}, \bibinfo{address}{New York, NY, USA},
  \bibinfo{pages}{19--24}.
\newblock
\showISBNx{9781450332248}
\urldef\tempurl%
\url{https://doi.org/10.1145/2666539.2666569}
\showDOI{\tempurl}


\bibitem[\protect\citeauthoryear{Yang, Wu, Zhang, Wang, Swaminathan, Yan, and
  Rountev}{Yang et~al\mbox{.}}{2018}]%
        {yang-jase18}
\bibfield{author}{\bibinfo{person}{Shengqian Yang}, \bibinfo{person}{Haowei
  Wu}, \bibinfo{person}{Hailong Zhang}, \bibinfo{person}{Yan Wang},
  \bibinfo{person}{Chandrasekar Swaminathan}, \bibinfo{person}{Dacong Yan},
  {and} \bibinfo{person}{Atanas Rountev}.} \bibinfo{year}{2018}\natexlab{}.
\newblock \showarticletitle{Static Window Transition Graphs for {Android}}.
\newblock \bibinfo{journal}{\emph{International Journal of Automated Software
  Engineering}} \bibinfo{volume}{25}, \bibinfo{number}{4} (\bibinfo{date}{Dec.}
  \bibinfo{year}{2018}), \bibinfo{pages}{833--873}.
\newblock


\bibitem[\protect\citeauthoryear{Yang, Zhang, Wu, Wang, Yan, and Rountev}{Yang
  et~al\mbox{.}}{2015}]%
        {yang-ase15}
\bibfield{author}{\bibinfo{person}{Shengqian Yang}, \bibinfo{person}{Hailong
  Zhang}, \bibinfo{person}{Haowei Wu}, \bibinfo{person}{Yan Wang},
  \bibinfo{person}{Dacong Yan}, {and} \bibinfo{person}{Atanas Rountev}.}
  \bibinfo{year}{2015}\natexlab{}.
\newblock \showarticletitle{Static Window Transition Graphs for {A}ndroid}. In
  \bibinfo{booktitle}{\emph{IEEE/ACM International Conference on Automated
  Software Engineering}}. \bibinfo{pages}{658--668}.
\newblock


\bibitem[\protect\citeauthoryear{Yang, Prasad, and Xie}{Yang
  et~al\mbox{.}}{2013}]%
        {Yang13}
\bibfield{author}{\bibinfo{person}{Wei Yang}, \bibinfo{person}{Mukul~R Prasad},
  {and} \bibinfo{person}{Tao Xie}.} \bibinfo{year}{2013}\natexlab{}.
\newblock \showarticletitle{{A Grey-Box Approach for Automated GUI-Model
  Generation of Mobile Applications}}. In \bibinfo{booktitle}{\emph{Fundamental
  Approaches to Software Engineering}},
  \bibfield{editor}{\bibinfo{person}{Vittorio Cortellessa} {and}
  \bibinfo{person}{D{\'{a}}niel Varr{\'{o}}}} (Eds.).
  \bibinfo{publisher}{Springer Berlin Heidelberg}, \bibinfo{address}{Berlin,
  Heidelberg}, \bibinfo{pages}{250--265}.
\newblock
\showISBNx{978-3-642-37057-1}


\bibitem[\protect\citeauthoryear{{Youm}, {Ahn}, {Kim}, and {Lee}}{{Youm}
  et~al\mbox{.}}{2015}]%
        {Youm:2015}
\bibfield{author}{\bibinfo{person}{K.~C. {Youm}}, \bibinfo{person}{J. {Ahn}},
  \bibinfo{person}{J. {Kim}}, {and} \bibinfo{person}{E. {Lee}}.}
  \bibinfo{year}{2015}\natexlab{}.
\newblock \showarticletitle{Bug Localization Based on Code Change Histories and
  Bug Reports}. In \bibinfo{booktitle}{\emph{2015 Asia-Pacific Software
  Engineering Conference (APSEC)}}. \bibinfo{pages}{190--197}.
\newblock


\bibitem[\protect\citeauthoryear{{Zaeem}, {Prasad}, and {Khurshid}}{{Zaeem}
  et~al\mbox{.}}{2014}]%
        {Zaeem:UserInteractions:ICST:2014}
\bibfield{author}{\bibinfo{person}{R.~N. {Zaeem}}, \bibinfo{person}{M.~R.
  {Prasad}}, {and} \bibinfo{person}{S. {Khurshid}}.}
  \bibinfo{year}{2014}\natexlab{}.
\newblock \showarticletitle{Automated Generation of Oracles for Testing
  User-Interaction Features of Mobile Apps}. In \bibinfo{booktitle}{\emph{2014
  IEEE Seventh International Conference on Software Testing, Verification and
  Validation}}. \bibinfo{pages}{183--192}.
\newblock
\showISSN{2159-4848}
\urldef\tempurl%
\url{https://doi.org/10.1109/ICST.2014.31}
\showDOI{\tempurl}


\end{thebibliography}

\end{document}